\documentclass[10pt,aps,prd,twocolumn,showpacs,preprintnumbers,nofootinbib,bibnotes]{revtex4-1}

\usepackage{graphicx,amsmath,amssymb,hyperref,natbib,slashed}
\usepackage{soul} %For strikethroughs \st{}
\usepackage{cancel}
\usepackage[dvipsnames]{xcolor}
\usepackage[export]{adjustbox}
\usepackage{float}
\usepackage{wasysym}
\usepackage{slashed}
\usepackage[utf8]{inputenc}
\usepackage[T1]{fontenc}
\usepackage{multirow}
\usepackage{pifont}
\usepackage{array}
\usepackage{color, colortbl}
\definecolor{Gray}{gray}{0.8}

\newcolumntype{L}[1]{>{\raggedright\let\newline\\\arraybackslash\hspace{0pt}}m{#1}}
\newcolumntype{C}[1]{>{\centering\let\newline\\\arraybackslash\hspace{0pt}}m{#1}}
\newcolumntype{R}[1]{>{\raggedleft\let\newline\\\arraybackslash\hspace{0pt}}m{#1}}

\mathchardef\mhyphen="2D
\newcommand{\ds}{{\sf DarkSUSY}}

\newcommand{\be}{\begin{equation}}
\newcommand{\ee}{\end{equation}}
\newcommand{\bea}{\begin{eqnarray}}
\newcommand{\eea}{\end{eqnarray}}

\begin{document}

\title{Suppressing structure formation at dwarf galaxy scales and below:\\
late kinetic decoupling as a compelling alternative to warm dark matter}

\author{Torsten Bringmann} 
\email{torsten.bringmann@fys.uio.no}
\affiliation{Department of Physics, University of Oslo, Box 1048, N-0371 Oslo, Norway}

\author{H\aa vard Tveit Ihle}
\email{h.t.ihle@astro.uio.no}
\affiliation{Institute of Theoretical Astrophysics, University of Oslo, N-0315 Oslo, Norway}

\author{J\"orn Kersten}
\email{joern.kersten@uib.no}
\affiliation{University of Bergen, Institute for Physics and Technology, Postboks 7803, N-5020 Bergen, Norway}

\author{Parampreet Walia}
\email{p.s.walia@fys.uio.no}
\affiliation{Department of Physics, University of Oslo, Box~1048, N-0371 Oslo, Norway
\vspace*{0.2cm}
}

\date{November 30, 2016}

\begin{abstract}\noindent
Warm dark matter cosmologies have been widely studied as an alternative to
the cold dark matter paradigm, the characteristic feature being a suppression
of structure formation on small cosmological scales. A very similar situation occurs
if standard cold dark matter particles are kept in local thermal equilibrium with a, possibly
dark, relativistic species until the universe has cooled down to keV temperatures.
We perform a systematic phenomenological study of this possibility, and classify all
minimal models containing dark matter and an arbitrary radiation component that allow
such a late kinetic  decoupling. We recover explicit cases recently
discussed in the literature and identify new classes of examples that are very interesting from
a model-building point of view. In some of these models dark matter is inevitably
self-interacting, which is remarkable in view of recent observational support for this
possibility.
Hence, dark matter models featuring late kinetic decoupling
have the potential not only to alleviate the missing satellites problem but also
to address other problems of the cosmological concordance model 
on small scales, in particular the cusp-core and 
too-big-too-fail problems, in some cases without invoking any additional input.

\end{abstract}

\maketitle

%%%%%%%%%%%%%%%%%%%%%%%%%%%%%%%%%%%%%%%%%%%%
\section{Introduction}
%%%%%%%%%%%%%%%%%%%%%%%%%%%%%%%%%%%%%%%%%%%%
Dark matter (DM) is about five times as abundant as ordinary matter \cite{Ade:2015xua} 
and known to be the dominant driver of cosmological structure formation. The $\Lambda$CDM 
cosmological concordance model, which treats DM as a completely cold and 
collisionless component in the cosmic energy budget, is remarkably successful in 
describing the large scale structure of the universe 
\cite{Springel:2005mi,Vogelsberger:2014kha}. 
At galactic scales and below, on the other hand, the observational situation is 
less clear and leaves considerable room for various new physics effects leading to
deviations from the standard scenario. 
Several observations at  such scales have even been claimed to be in tension
with the expectations within the $\Lambda$CDM paradigm 
\cite{deBlok:1997zlw,Klypin:1999uc,Moore:1999nt,Zavala:2009ms,Oh:2010ea,
Papastergis:2011xe, BoylanKolchin:2011de, Walker:2011zu, Pawlowski:2013kpa,
Klypin:2014ira,Papastergis:2014aba,Oman:2015xda,Massey:2015dkw}, 
see Ref.~\cite{Vogelsberger:2015gpr} for a recent discussion.
One of the most often discussed and most long-standing of these issues is the problem of 
`missing satellites' of the Milky Way \cite{Klypin:1999uc,Moore:1999nt}, as compared to 
the typical number expected in $\Lambda$CDM cosmology, which subsequently was 
complemented by an observed underabundance also of small galaxies in the field 
\cite{Zavala:2009ms,Papastergis:2011xe,Klypin:2014ira}. 

%Despite decades of intense efforts to reveal it, the identity of DM remains fundamentally unknown. 
According to the leading hypothesis, DM consists of a new type of elementary 
particles \cite{Bertone:2010zza}. 
The most often studied class of models postulates weakly interacting 
massive particles (WIMPs) to form the DM and connects the observed DM abundance in a theoretically 
compelling way to extensions of the standard model of particle physics
%indications that the standard model of particle physics may be incomplete
at energies beyond the electroweak scale. Standard WIMPs,
like the supersymmetric neutralino \cite{Jungman:1995df} or the first Kaluza-Klein 
excitation of the photon \cite{Hooper:2007qk}, are prototype examples of cold dark 
matter (CDM) as required by the $\Lambda$CDM paradigm. Null searches for such
WIMPs at the CERN LHC \cite{Cakir:2015gya,Aad:2015baa} or in direct detection
experiments located deep underground \cite{Aprile:2012nq,Akerib:2013tjd}, however, 
start to severely
limit this possibility. Furthermore, from a theoretical 
perspective WIMPs are by far not the only possible option for a good DM 
candidate \cite{Steffen:2008qp,Feng:2010gw,Baer:2014eja,Adhikari:2016bei}.
DM particles may instead have significantly stronger non-gravitational interactions, either 
within a yet to be explored dark sector or with ordinary standard model 
particles. This may visibly affect the distribution of the observed structure in the
universe \cite{Goldberg:1986nk,Gradwohl:1992ue,Bode:2000gq,Hannestad:2000gt,Boehm:2000gq,
Boehm:2001hm,Boehm:2004th,Cembranos:2005us,Hooper:2007tu,Feng:2009mn,
Kaplan:2009de,Ackerman:mha,Kaplan:2011yj,
vandenAarssen:2012ag,Aarssen:2012fx,Cline:2012is,CyrRacine:2012fz,Tulin:2012wi,
Bringmann:2013vra,Dasgupta:2013zpn,Tulin:2013teo,Cyr-Racine:2013fsa,Shoemaker:2013tda,
Cline:2013zca,Fan:2013yva,Chu:2014lja,Ko:2014bka,Foot:2014uba,Cherry:2014xra,Buckley:2014hja,
Boehm:2014vja,Buen-Abad:2015ova,Lesgourgues:2015wza,Foot:2016wvj}. 
Recently, a framework for an 
effective theory of 
structure formation (ETHOS) has been developed \cite{Cyr-Racine:2015ihg} that will 
eventually allow to directly map the
particle physics parameters in such models to cosmological observables at low redshift.
This is particularly relevant for those types of models that would evade any of the 
more traditional ways to search for DM at colliders, in direct or indirect detection 
experiments. In this case, detailed observations of the distribution of matter at small scales,
for example in terms of the power spectrum, may be the only way to test the DM particle 
hypothesis. 

One of the most prominent, potentially observable features of this type would be an 
exponential suppression of power in the spectrum of matter density fluctuations at sub-Mpc 
scales. The classical way to achieve this is through the free streaming of warm DM (WDM)
particles, where keV sterile neutrinos provide the prototype example for a well motivated 
DM  candidate of this type \cite{Adhikari:2016bei}.
Such a cutoff in the power spectrum is strongly constrained by
observations of the Lyman-$\alpha$ forest, typically translated into 
a lower bound on the WDM mass. Recent analyses report limits as stringent 
as $m_\mathrm{WDM}\gtrsim4.35$\,keV \cite{Viel:2013apy,Baur:2015jsy},
which however has been
argued to be overly restrictive when taking into account that the warm intergalactic medium 
could mimic such a cutoff \cite{Garzilli:2015iwa}. Completely independent bounds of very 
roughly $m_\mathrm{WDM}\gtrsim1$\,keV arise from the observed phase-space densities 
of Milky Way satellites and from subhalo number counts in $N$-body 
simulations \cite{Horiuchi:2013noa}
as well as from weak lensing observations \cite{Inoue:2014jka}. This range for the WDM mass, and 
hence the location of the cutoff, is interesting because already a value of
$m_\mathrm{WDM}\sim2$\,keV would provide a solution to the missing satellites 
problem \cite{Schneider:2013wwa,Maccio':2009rx}, with slightly 
larger values at least alleviating it.
Historically, this was indeed one of the prime 
motivations to focus on WDM \cite{Bode:2000gq}. A drawback of WDM models in
this mass range, however, is that they cannot address the other, at least as 
pressing, small-scale problems briefly mentioned in the beginning. In particular, the cuspy 
inner density profile of DM haloes expected in CDM cosmology is not affected in any 
significant way \cite{Maccio:2012qf,Shao:2012cg}, leaving the cusp-core 
problem \cite{deBlok:1997zlw,Oh:2010ea,Walker:2011zu} unexplained.

An alternative, and much less explored way of creating a cutoff in the power spectrum at  
these scales arises if {\it cold} DM is kept in local thermal equilibrium with a relativistic species until
the universe has cooled down to sub-keV temperatures \cite{Boehm:2000gq,Boehm:2001hm,
Chen:2001jz, Hooper:2007tu, Aarssen:2012fx,Boehm:2014vja}. In this case, DM thus
decouples kinetically much later than in the case of standard 
WIMPs \cite{Bringmann:2009vf}. The remaining viscous coupling between the two fluids
then typically leads to a characteristic `dark' oscillation pattern in the power spectrum, with a 
strong suppression at small scales \cite{Loeb:2005pm,Bertschinger:2006nq} as confirmed by
explicit numerical simulations \cite{Vogelsberger:2015gpr,Schewtschenko:2015rno}.
Interestingly, this is a possibility that arises rather naturally in self-interacting
DM models, allowing to address not only the missing satellites problem but at the same
time also \emph{all} other shortcomings of the $\Lambda$CDM paradigm 
\cite{Aarssen:2012fx}.
This observation has already triggered significant interest and led to a number of
specific model-building attempts \cite{Shoemaker:2013tda,Bringmann:2013vra,Dasgupta:2013zpn,
Ko:2014bka,Cherry:2014xra,Chu:2014lja,Bertoni:2014mva} as well as first fully self-consistent 
numerical simulations of structure formation for this class of models \cite{Vogelsberger:2015gpr}. 

Here, we take a much broader perspective and aim at \emph{classifying, in a systematic way, 
the minimal possibilities that can lead to late kinetic decoupling} with an observationally 
relevant cutoff in the power spectrum. Such a cutoff may or may not be related to a solution 
of the missing satellites problem, but would in any case provide a fascinating observational 
signature that helps to narrow down the identity of DM\@.
We use the language of simplified models to describe the main ingredients that are 
necessary for any model building in this direction, depending on the spin of the CDM 
particle and its relativistic scattering partner. This relativistic particle may either be 
some form of dark radiation (DR), the photon or one of the active neutrinos (though we will
see that the first option is favoured). We note that the existence of such a DR component is 
cosmologically very interesting in its own
right \cite{Brust:2013xpv,Archidiacono:2013fha}
%some theory papers - the 2 I've cited are mainly from the particle
%physics perspective; about the cosmologists' view I don't know
and can even be invoked to improve the consistency of 
different cosmological data sets
\cite{Wyman:2013lza,Hamann:2013iba,Battye:2013xqa,Gariazzo:2013gua,Battye:2014qga}.
As a result of our encompassing approach, 
we recover all previously identified  configurations with scalars, fermions and vectors that 
lead to late kinetic decoupling, and also find further solutions that open new avenues for 
future model building.
In our analysis, we fully include recent developments 
in the theoretical description of the decoupling process
(see the discussion in the Appendix for more details).

This article is organized as follows. We start by discussing in Section \ref{sec:dr} the 
generic requirements and limits for any DM model to feature sufficiently late kinetic 
decoupling such as to leave an observable imprint on the power spectrum or to alleviate 
the missing satellites problem. In Section \ref{sec:2pm} we restrict ourselves to
simplified models containing only the CDM particle and its interaction with a
relativistic particle, and provide a classification of all such models with the sought-after 
properties. We extend this classification in Section \ref{sec:3pm} by allowing for a further, 
independent virtual particle mediating the interaction.  
We present a summary of our results and conclude in Section 
\ref{sec:conc}. In two Appendices, we provide a concise review of
the kinetic decoupling of DM particles from a thermal bath (App.~\ref{app:kd}) and
list the elastic scattering matrix elements, as well as the ETHOS parameters, for all 
models relevant for our discussion 
(App.~\ref{app:msq}).

 %%%%%%%%%%%%%%%%%%%%%%%%%%
 \begin{figure}[t!]
 	\includegraphics[width=0.50\columnwidth] {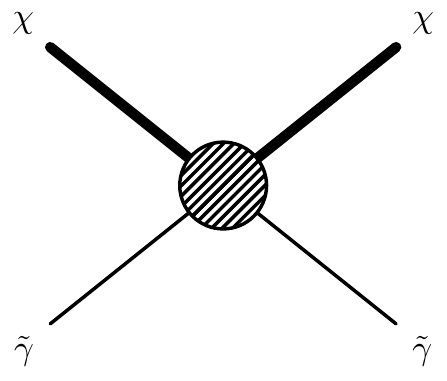}	
         \vspace*{-0.2cm}
 	\caption{Schematic illustration of the elastic scattering of a DM particle $\chi$ with 
	a (possibly dark) relativistic particle $\tilde\gamma$. Throughout this article,
	we use thick lines to denote heavy (non-relativistic) particles and thin lines
	to denotes light (relativistic) particles of any spin.
 	}
 	\label{fig:EffDiagram}
 \end{figure}		
 %%%%%%%%%%%%%%%%%%%%%%%%%%

%%%%%%%%%%%%%%%%%%%%%%%%%%%%%%%%%%%%%%%%%%%%
\section{Dark matter scattering with (dark) radiation}
\label{sec:dr}
%%%%%%%%%%%%%%%%%%%%%%%%%%%%%%%%%%%%%%%%%%%%

As motivated in the introduction, we are interested in scenarios where highly non-relativistic 
DM can be kept in local thermal equilibrium with a relativistic species until late times, via the
elastic scattering processes schematically shown in Fig.~\ref{fig:EffDiagram}. 
The kinetic decoupling of DM from this radiation component (see App.~\ref{app:kd} 
for details) then leads to a small-scale cutoff in the power spectrum of density 
fluctuations, corresponding to a minimal halo mass of \cite{Vogelsberger:2015gpr}
\be
\label{Mcut}
 M_\mathrm{cut, kd}=5\cdot10^{10}\left(\frac{T_\mathrm{kd}}{100\,\mathrm{eV}}\right)^{-3} 
 h^{-1}\, M_\odot\,,
\ee
where $T_\mathrm{kd}$ is the (photon) temperature at which decoupling occurs
and $h\simeq0.68$ \cite{Ade:2015xua} is the Hubble constant in units of 
100\,$\mathrm{km}\,\mathrm{s}^{-1}\mathrm{Mpc}^{-1}$
(note that this relation critically depends on how $T_\mathrm{kd}$ is defined,
see the discussion after Eq.~(\ref{xidef})).
This  
should be compared to the corresponding cutoff in the halo mass function \cite{Vogelsberger:2015gpr}
\be
 M_\mathrm{cut, WDM}=10^{11}\left(\frac{m_\mathrm{WDM}}{\mathrm{keV}}\right)^{-4} 
 h^{-1}\, M_\odot
\ee
that is expected for a standard {\it warm} DM candidate (which decouples at temperatures 
much higher than keV). In order for such a cutoff to be observable for \emph{cold} DM,
and to potentially address the missing satellites problem, we thus need kinetic
decoupling temperatures somewhat smaller than 1\,keV, i.e.~much smaller than 
the MeV to GeV temperatures one encounters for standard 
WIMPs \cite{Bringmann:2009vf}.

Let us stress again that we focus here on situations where the dominant suppression mechanism 
arises from acoustic oscillations of a CDM component \cite{Loeb:2005pm,Bertschinger:2006nq}, 
and this is the assumption under which Eq.~(\ref{Mcut}) is valid. Free streaming 
of the DM particles, which is the dominant effect for WDM,
in principle leads to an independent suppression of the power spectrum \cite{Green:2005fa}.
Following Ref.~\cite{Bringmann:2009vf}, we estimate that this effect is subdominant for DM 
masses above $100\,$keV, for $T_\mathrm{kd}\sim0.1$\,keV and for a DR temperature equal 
to that of the photons, while for significantly colder DR free streaming becomes important only 
for even smaller DM masses. For simplicity, we restrict ourselves to DM particles that are 
heavy enough to lie outside this intermediate regime between CDM and WDM.

Our goal is to systematically classify all (minimal) possibilities that could give rise to such 
a late kinetic decoupling of CDM. To this end, we choose to be completely agnostic about the 
nature of DM and the radiation component, so the latter could either be a form of 
dark radiation (e.g. sterile neutrinos) or be given by the standard cosmological photon or 
neutrino background. We simply assume that there is one DM species, denoted 
by $\chi$, and one radiation species scattering with $\chi$, denoted by $\tilde\gamma$. 
We allow arbitrary spins for both species, and scrutinize all relevant simplified model 
Lagrangians (which obviously could be embedded in more complete frameworks)
to see whether they allow for kinetic decoupling temperatures in the keV range or not. 

%%%%%%%%%%%%%%%%%%%%%%%%%%%%%%%%%%%%%%%%%%%%
\subsection{Generic requirements for late kinetic decoupling}
\label{sec:scatter_gen}

 %%%%%%%%%%%%%%%%%%%%%%%%%%	
\begin{figure}[t!]
\includegraphics[width=0.9\columnwidth]{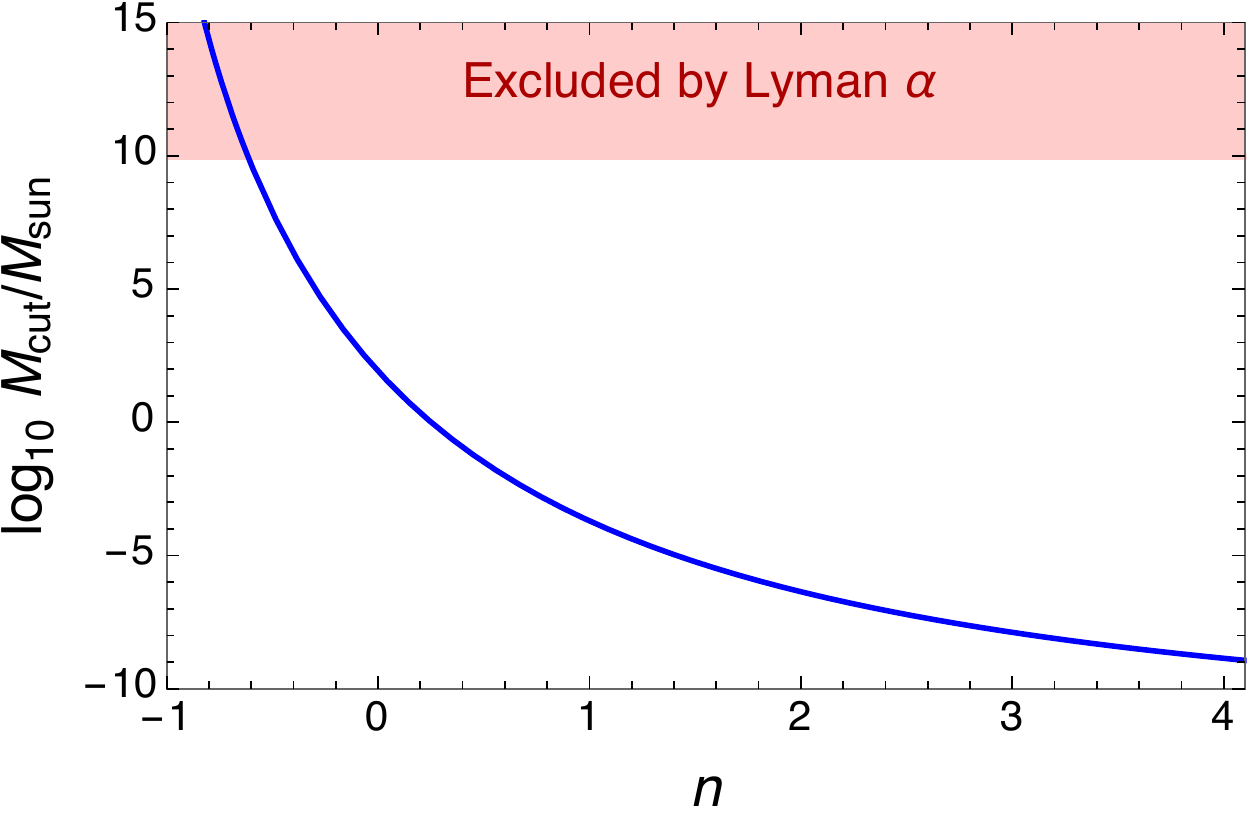}	
\vspace*{-0.3cm}
\caption{Cutoff in the halo mass function resulting from DM scattering with a radiation 
component $\tilde\gamma$. This assumes $M_\mathrm{cut}=M_n$ as introduced in 
Eqs.~(\ref{m2simp}, \ref{mcutsimp}), i.e.~an amplitude that scales with the energy 
$\omega$ of $\tilde\gamma$ as $\left|\mathcal{M}\right|^2\propto \omega^n$, a coupling 
strength roughly corresponding to the electroweak coupling and a DM mass of 100\,GeV. 
For comparison, we also indicate the value that is roughly excluded by Ly-$\alpha$ data,
%coresponding to $m_\mathrm{WDM}\sim2$\,keV,
with slightly smaller values allowing a potential solution to the missing satellites problem. 
}
\label{fig:Mcutsimp}
\end{figure}		
%%%%%%%%%%%%%%%%%%%%%%%%%%

Before we start this endeavor, let us illustrate the general challenges for model building 
that we should expect to encounter. Consider for 
simplicity the case where the scattering amplitude close to kinetic decoupling can be 
approximated by a power law in the energy $\omega$ of the relativistic scattering 
partner	
\be
\label{m2simp}
\left|\mathcal{M}\right|^2 \simeq c_n \eta_\chi (\omega/m_\chi)^n\,, 
 \ee
 where the matrix element squared here and in the following is understood to be {\it summed} 
 over the internal degrees of freedom (d.o.f.) $\eta$ of {\it both} initial and final states. For later 
 convenience, we have extracted the d.o.f.~of the initial DM particle, $\eta_\chi$, from the definition of $c_n$.
For $n>-1$, we can analytically solve the Boltzmann equation to determine 
$T_\mathrm{kd}$, 
see Eq.~(\ref{tkd_analytic}) in Appendix \ref{app:kd}, plug the result into Eq.~(\ref{Mcut}) and find
	\bea
\label{mcutsimp}
M_\mathrm{cut} \equiv M_\mathrm{cut, kd} &\simeq& 
M_n\, \xi^{3\frac{n+4}{n+2}}\left(\frac{c_n}{0.001}\right)^\frac{3}{n+2} \\
&& \times
\left(\frac{g_\mathrm{eff}}{3.36}\right)^{-\frac{3}{4+2n}}
\left(\frac{m_\chi}{100\,\mathrm{GeV}} \right)^{-3\frac{n+3}{n+2}}.\nonumber
\eea
Here, we have introduced 
\be 
\xi\equiv T_{\tilde\gamma}/T \,, 
\ee
$c_n\sim10^{-3}$ very roughly corresponds to the
case where the electroweak coupling mediates the scattering process, and $g_\mathrm{eff}$ is 
the usual effective number of relativistic degrees of freedom  around kinetic decoupling. 
$M_n$ is a numerical constant that is independent of the couplings or masses of the theory,
and plotted in Fig.~\ref{fig:Mcutsimp} as a function of $n$ and in units of $M_\odot$ (assuming 
a fermionic $\tilde\gamma$; for a bosonic  scattering partner,  $M_n$ would increase by an 
amount not visible at the resolution of the figure).
For reference, we also indicate the cutoff mass induced by a 2\,keV thermal WDM candidate;
as discussed in the introduction, this provides a rough distinction between what is ruled out
by Ly-$\alpha$ data and what would help to alleviate the missing
satellites problem.

Typical WIMP DM candidates are well described by the $n=2$ case, for which we have 
$M_2=4.4\cdot10^{-7}M_\odot$; for 100\,GeV neutralinos, for example, one finds roughly 
$10^{-7}M_\odot\lesssim M_\mathrm{cut}\lesssim 10^{-4} M_\odot$ 
\cite{Bringmann:2009vf}. Observable values of $M_\mathrm{cut}$, 
close to what is still allowed by Ly-$\alpha$ data, thus clearly require a significant deviation 
from the standard scenario. Looking at Eq.~(\ref{mcutsimp}), there appear only a handful 
of basic possibilities to increase 
$M_\mathrm{cut}$ in such a way. Let us briefly discuss them in turn.

\begin{itemize}
\item {\it Maximizing the radiation temperature}.
For DM scattering with photons or active neutrinos, we have by definition $\xi=1$ 
and $\xi=(4/11)^{1/3}=0.71$, respectively. If $\tilde\gamma$ constitutes a form of DR, on 
the other hand, $\xi$ is in principle a free parameter.\\
Observations of the cosmic microwave background (CMB), however, 
exclude the existence of an additional radiation component corresponding to the
contribution of one more massless neutrino (i.e.~$\Delta N_\mathrm{eff}=1$) by more than 
$3\sigma$  \cite{Ade:2015xua}.
For a fermionic (bosonic) $\tilde\gamma$, this translates to 
$\xi< 0.85 \, (0.82) /\eta_{\tilde\gamma}^{1/4}$,
inevitably implying a further {\it suppression} of $M_\mathrm{cut}$ with respect to what 
is shown in Fig.~\ref{fig:Mcutsimp}.\footnote{%
\label{noCMBbound}%
The CMB bound can in principle be evaded if $\tilde\gamma$ becomes non-relativistic right
after kinetic decoupling, i.e.~at $100\,\mathrm{eV}\gtrsim T\gtrsim 10$\,eV.
Assuming that there is no entropy production in the dark sector afterwards, however, this 
implies a warm (or even hot) DM density today of  
$\rho_{\tilde \gamma}^0 = \left({ \zeta(3)}/{\pi^2 } \right) N_{\tilde\gamma}
m_{\tilde \gamma}  \eta_{\tilde \gamma} \xi^3T_0^3$, where $N_{\tilde\gamma}=1$ 
($N_{\tilde\gamma}=3/4$) for a bosonic (fermionic) $\tilde\gamma$. Demanding that this
contribution make up at most a fraction $f$ of the total observed DM density 
translates to a bound on $\xi$ which turns out to be comparable to the CMB bound,
$\xi <0.8\, (f/N_{\tilde\gamma} \eta_{\tilde \gamma})^{1/3} (m_{\tilde \gamma}/10\,\mathrm{ eV})^{-1/3}$. 
%reference for a bound on f for sub-dominant WDM component would be nice...
Even if $\tilde\gamma$ is kept in thermal equilibrium with another, relativistic species this
conclusion does not change qualitatively; we discuss this case in Section 
\ref{sec:scatter_t}.
} 
Let us also mention that there exists an independent, weaker constraint on $\xi$ from 
big bang nucleosynthesis \cite{Nollett:2014lwa}. Interestingly, this constraint actually 
{\it favors} a small DR component (unlike the one from CMB observations).

Note, finally, that there often exists a \emph{lower} bound on the value of $\xi$ that can
be achieved in a given model-building framework. If $\tilde \gamma$ has been in thermal
equilibrium with photons down to some temperature $T_\mathrm{eq}$, for example, and there
was no additional entropy production in the visible sector afterwards,  
we have
$\xi%= \left(\frac{2+7/8\cdot2\cdot 3 \left(4/11\right)}{28 + 7/8\cdot 90}\right)^{1/3} 
\gtrsim 0.34\,\left[g_\mathrm{eff}(T_\mathrm{eq})/100\right]^{-1/3}\eta_{\tilde \gamma}^{1/3}$.

\item {\it Minimizing the energy dependence of the scattering matrix}.
It will obviously help if $n$ is as small as possible. The simplest way to
achieve a {\it constant} scattering amplitude ($n=0$), in particular, is a contact
interaction, i.e.~an (effective) 4-point vertex. We will study this option further in 
Sec.~\ref{sec:scatter_point}.\\
 In situations with propagators almost on shell (see the next 
point), the scattering rate can in extreme cases even {\it increase} with decreasing energy 
$\omega$ (corresponding to $n<0$). This happens in particular when the DR particle
appears in the $t$-channel, a scenario which we discuss in detail in Sections 
\ref{sec:scatter_t} and \ref{sec:mixed}.

%%%%%%%%%%%%%%%%%%%%%%%%%%	
 \begin{figure}[t!]
 	\includegraphics[width=\columnwidth]{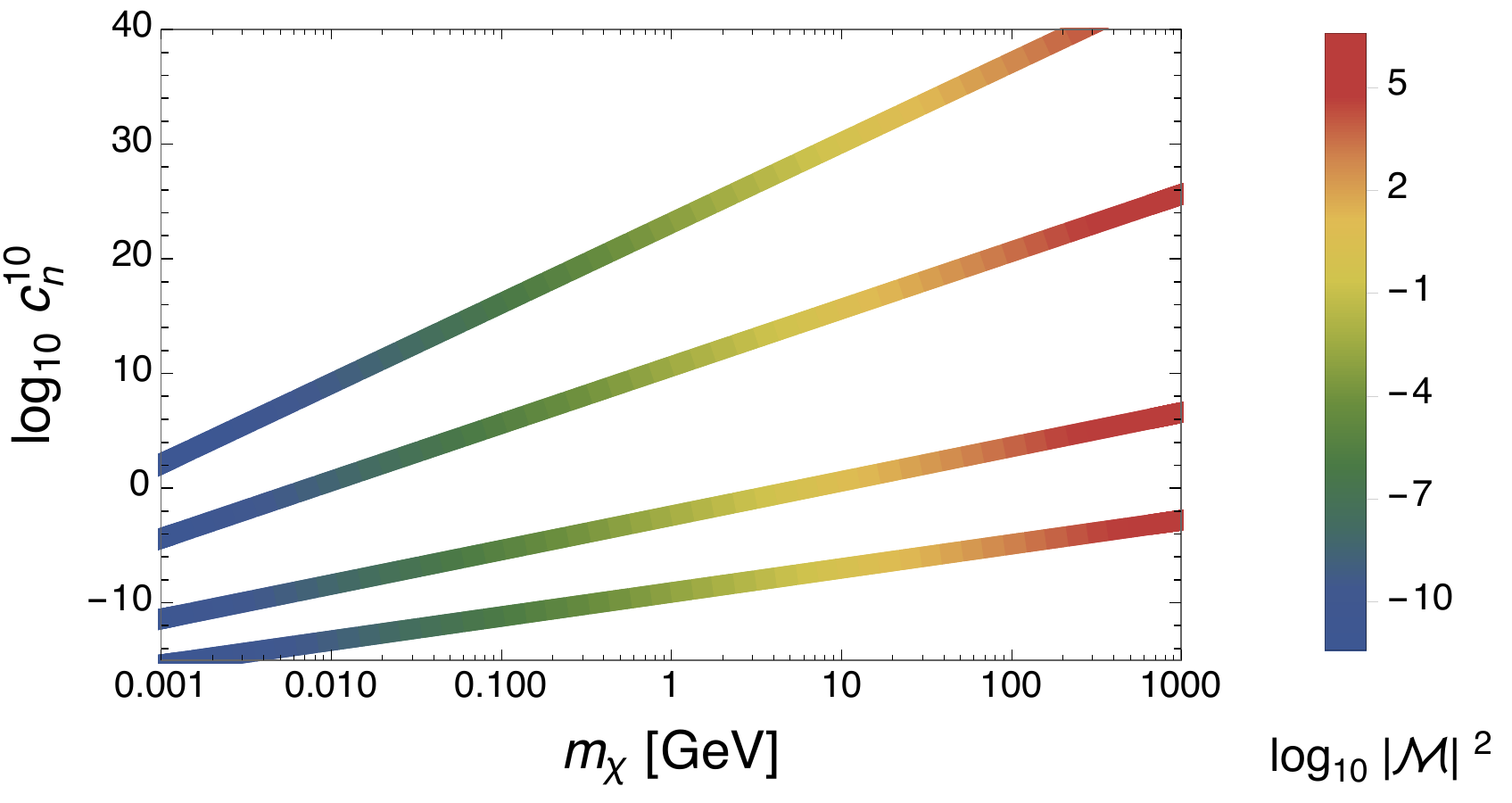}	
 	\caption{For a scattering amplitude 
	$\left|\mathcal{M}\right|^2 \simeq c_n (\omega/m_\chi)^n$ close to kinetic 
	decoupling, this figure shows the value of $c_n$ that results in  
	$M_\mathrm{cut}=10^{10}\,M_\odot$, as a function of the DM mass $m_\chi$. 
	From top to bottom, the lines correspond to $n=(4,2,0,-0.9)$, and we have 
	throughout fixed $\xi=T_{\tilde\gamma}/T=0.5$ and $g_\mathrm{eff}=3.36$.
	The color scale indicates the size of the matrix element for 
	 $\omega\to\langle\omega\rangle_{T_\mathrm{kd}}$.
	\label{fig:cn10}}
 \end{figure}		
 %%%%%%%%%%%%%%%%%%%%%%%%%%

\item {\it Increasing the effective coupling strength}. In Fig.~\ref{fig:cn10} we show 
which value of  $c_n$ in Eq.~(\ref{m2simp}) is needed to produce 
$M_\mathrm{cut}=10^{10}\,M_\odot$ (for $\xi=0.5$ and 
$g_\mathrm{eff}=3.36$). As can be seen, very large $c_n$, and hence
efficient enhancement mechanisms for the amplitude, are needed for $n>0$ and
$m_\chi\gtrsim1$\,GeV\@.
%which represents the typical WIMP case. 
For large DM masses, the required amplitude even becomes so large that unitarity
violation starts to become a possible concern; in critical cases, this needs to be 
checked on a model-by-model basis (see, e.g., Ref.~\cite{Kahlhoefer:2015bea}).
Perturbativity restricts couplings to satisfy $\alpha\lesssim1$, so the only option to 
achieve such large amplitudes is a virtual 
particle that is almost on shell in the particular kinematical situation
we are interested in ($\omega \ll m_\chi$ and hence $-t\ll m_\chi^2$). This can
be arranged both in the $s/u$-channel and in the $t$-channel. As these possibilities
typically correspond to unrelated interaction terms in the Lagrangian, we will
discuss them separately in Sections \ref{sec:2pm} and \ref{sec:3pm}.

\item {\it Decreasing the DM mass}.
From the discussion so far, a DM mass significantly smaller than our reference value,
$100\,\mathrm{keV}\lesssim m_\chi \ll 100$\,GeV, appears as the maybe most 
straightforward way to achieve a larger $M_\mathrm{cut}$.
Indeed, DM particles much lighter than typical  WIMP DM
candidates are by no means a problem {\it per se} -- though, as we will see, they might be 
disfavored in given model frameworks.
Recall that the lower bound here simply results from the range of validity of Eq.~(\ref{Mcut});
for lower DM masses free streaming effects would have to be taken into account 
(unless $\xi\ll1$), which is beyond the scope of this work. 
Note that \emph{if} the DM relic density is set by chemical decoupling from any of the 
standard model particles, the lower bound tightens to $m_\chi \gtrsim1$\,MeV \cite{Boehm:2013jpa}.
\end{itemize}
 
 %%%%%%%%%%%%%%%%%%%%%%%%%%%%%%%%%%%%%%%%%%%%
\subsection{Bounds from inherently related processes}
\label{sec:gen_annself}

Models with a large scattering 
rate $\chi\tilde\gamma\leftrightarrow\chi\tilde\gamma$ typically imply large 
annihilation rates, $\chi\chi\to\tilde\gamma\tilde\gamma$, and in some cases significant 
DM self-interaction rates, $\chi\chi\to\chi\chi$, as well. In the following, we will discuss
the {\it generic} constraints for model building that result from these processes (while we 
leave more model-specific considerations for later).

\subsubsection{Dark matter annihilation $\chi\chi\to\tilde\gamma\tilde\gamma$}
\label{sec:gen_ann}

So far, we have avoided to make any assumptions about how DM was produced
in the first place.
However, the fact that we consider interactions between $\chi$ and 
the thermally distributed $\tilde\gamma$ in order to achieve late kinetic decoupling
indeed strongly suggests that DM was thermally produced in the early universe.
For cold DM, the relic density then typically scales roughly as 
$\Omega_\chi h^2\propto m_\chi^2/{g'}^4$, with $g'$ being
the effective coupling to drive chemical decoupling. More concretely, if we assume 
$\mathcal{M}_{\chi\chi\to\tilde\gamma\tilde\gamma}\simeq {g'}^2$ at chemical freeze-out
and only take into account the leading, velocity-independent part of the cross section,
$\sigma v={g'}^4/32\pi m_\chi^2$,
the relic density is given by \cite{Gondolo:1990dk}
\be
\label{RDswave}
 \Omega_\chi h^2 \simeq 8.81\cdot10^{-5}\frac{x_f}{g_\mathrm{eff}^{1/2}(T_\mathrm{cd})}
 {g'}^{-4}\left(\frac{m_\chi}{100\,\mathrm{GeV}}\right)^2 ,
\ee
where $v$ is the relative velocity between the two annihilating DM particles 
(if the leading contribution to the cross section in the zero-velocity limit is instead given by
$\sigma v={g'}^4/32\pi m_\chi^2 v^2$, this expression must be multiplied by $x_f/3$).
Here, $x_f\equiv m_\chi/T_\mathrm{cd}$ and the effective number of degrees of freedom 
$g_\mathrm{eff}$ are evaluated at the temperature $T_\mathrm{cd}$ of {\it chemical} 
decoupling. A relic density in agreement
with the observed value of $\Omega_\chi h^2=0.1188 \pm 0.0010$ 
%including external data, see Table 4
\cite{Ade:2015xua} can
thus not only be obtained for the specific combination of weak-scale masses 
$m_\chi\sim100$\,GeV and couplings ${g'}^2\sim 0.04$, 
% assuming $\frac{x_f}{g_\mathrm{eff}^{1/2}(T_\mathrm{cd})}=2$, this is 0.0385
but also 
for any other combination of $m_\chi$ and $\alpha'$ that leaves the ratio of these
quantities constant
(this is sometimes referred to as \mbox{WIMP}less DM \cite{Feng:2008ya}). 

Let us now denote with $g$ the effective coupling for the scattering process, so we 
expect $c_n\sim g^4$ in Eq.~(\ref{m2simp}). Rotating the corresponding diagrams, 
schematically shown in Fig.~\ref{fig:EffDiagram}, we see that 
$\chi\chi\to\tilde\gamma\tilde\gamma$ must at least 
{\it contribute} to the total DM annihilation rate. Demanding that these annihilation
processes do not deplete the DM abundance below the observed value implies a 
rough {\it upper bound} of  
$g^2\lesssim{g'}^2\sim 0.04\, (m_\chi/100$\,GeV) for $\xi\sim1$ 
(note that this argument applies even if the initial DM abundance was
\emph{not} produced 
thermally). Using Eq.~(\ref{mcutsimp}), this in turn restricts the cutoff mass  
approximately to
\be
\label{McutmaxRD}
M_\mathrm{cut}\lesssim M_n
%\, \xi^{3\frac{n+4}{n+2}}
%\left(\frac{g_\mathrm{eff}}{3.36}\right)^{-\frac{3}{4+2n}}
\left(\frac{m_\chi}{100\,\mathrm{GeV}} \right)^{-3\frac{n+1}{n+2}}.
\ee
Even when taking into account the impact on the DM abundance, considering lighter
DM particles will thus in general help significantly to achieve a larger value of 
$M_\mathrm{cut}$.

Due to the different kinematics of scattering and annihilation processes, an intermediate
particle nearly on shell in the former case is not on shell in the latter. This means that 
the value of the matrix element can be much larger for scattering than for annihilation, 
even though the `same' diagrams are involved. As can easily be checked, the  
argument of the preceding paragraph still runs through, in exactly the same way, when 
taking into account that $c_n/g^4$ may in fact be much larger than unity because of this 
effect -- a possibility which we will make excessive use of. In this case, the right-hand 
side of Eq.~(\ref{McutmaxRD}) should be multiplied by a factor of 
$\left( {c_n/g^4} \right)^{3/(n+2)}$.

\subsubsection{Dark matter self-scattering $\chi\chi\to\chi\chi$}
\label{sec:gen_self}

Coupling DM to (dark) radiation inevitably implies that DM will be self-interacting, too.
If $\tilde\gamma$ is bosonic and there is a direct coupling $\chi$-$\chi$-$\tilde\gamma$, for 
example, this will induce a Yukawa-type potential with strength $\alpha_\chi=g_\chi^2/4\pi$
and range $1/m_{\tilde \gamma}$ between the DM particles, resulting in a characteristic 
velocity-dependent self-scattering cross section.\footnote{%
For pseudoscalar mediators, the situation is much more involved
\cite{Bellazzini:2013foa,Archidiacono:2014nda,Bedaque:2009ri,Dolan:2014ska}
and a full analysis beyond the scope of this work.
} 
We are typically interested in the classical limit 
($m_\chi v \gg m_{\tilde\gamma}$), where the velocity-weighted transfer cross section 
peaks at around $v_\mathrm{max}\sim0.1g_\chi\sqrt{m_{\tilde\gamma}/m_\chi}$, resulting in
 $\sigma_T (v_\mathrm{max})\sim 10/m_{\tilde\gamma}^2$;
for larger velocities $v\gg v_\mathrm{max}$, the transfer cross section then drops sharply as 
$\sigma_T\propto v^{-4}$ 
(see e.g.~Ref.~\cite{Feng:2009hw}). Observations of dwarf-scale systems result in upper 
bounds on the self-interaction rate of roughly 
$\sigma_T/m_\chi\lesssim 10\,\mathrm{cm}^2/\mathrm{g}\simeq
4.6\cdot 10^{4}/\mathrm{GeV}^3$ 
\cite{Wandelt:2000ad,Spergel:1999mh,Rocha:2012jg,Zavala:2012us,Kaplinghat:2015aga}, 
while constraints on cluster scales are up to two orders of magnitudes more stringent 
\cite{Markevitch:2003at,Randall:2007ph}.
Given that $v_\mathrm{max}$ is smaller than the typical velocities encountered in dwarf 
galaxies, however, the former limit constrains  the coupling $g_\chi$ much more severely in our 
case.

 %%%%%%%%%%%%%%%%%%%%%%%%%%	
 \begin{figure}[t!]
 	\includegraphics[width=1\columnwidth]{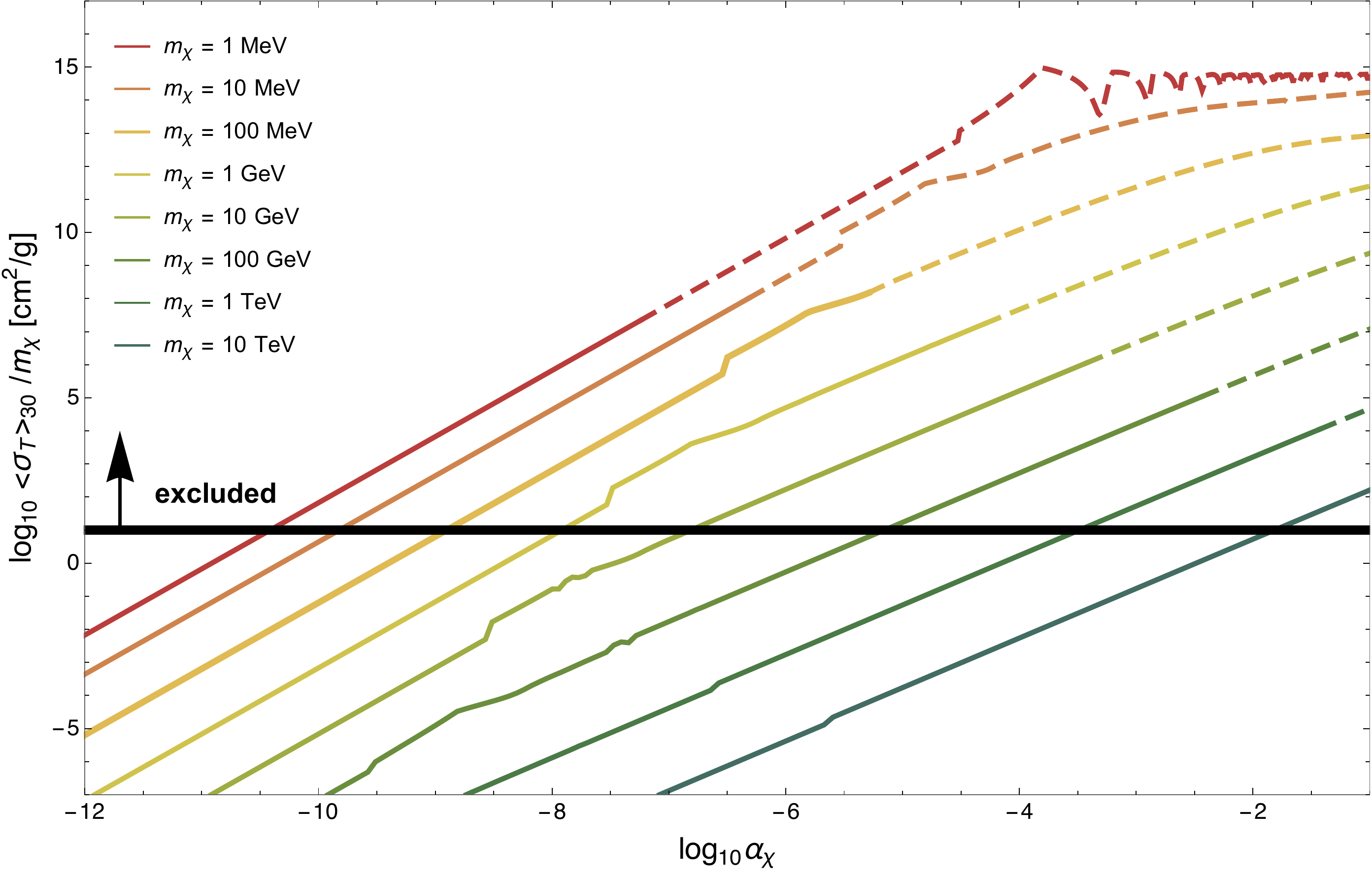}	
 	\caption{DM self-interaction in dwarf-scale systems, induced by a direct coupling
	between light bosonic $\tilde\gamma$ and DM, as a function of
$\alpha_\chi=g_\chi^2/4\pi$. 
	From top to bottom, the curves show the case for a DM mass increasing from 
	$m_\chi=1$\,MeV to $m_\chi=10$\,TeV\@. We have throughout assumed 
	 $m_{\tilde \gamma} =100$\,eV; lighter masses would give higher cross
	sections. The dashed parts of the curves indicate where the coupling strength $g_\chi$ 
	is so large that $\chi\chi\to\tilde\gamma\tilde\gamma$ would deplete the DM
	abundance below the observed value. Values above the horizontal thick line are
	excluded by dwarf galaxy observations.}
 	\label{fig:DR_yukawa}
 \end{figure}		
 %%%%%%%%%%%%%%%%%%%%%%%%%%	

Following Ref.~\cite{Cyr-Racine:2015ihg}, we define $\langle\sigma_T\rangle_{30}$
as the transfer cross section $\sigma_T$ averaged over a Maxwellian velocity distribution 
with a most probable velocity of $v_M=30$\,km/s, a value representative for dwarf galaxies. 
For $\sigma_T(v)$ we use the perturbative result \cite{Feng:2009hw} 
in the Born limit ($\alpha_\chi m_\chi\ll m_{\tilde\gamma}$), the parameterization
obtained in Ref.~\cite{Cyr-Racine:2015ihg} for the classical limit 
($m_\chi v \gg m_{\tilde\gamma}$), and an analytical result outside these two regimes
which results from approximating the  Yukawa potential by a Hulth\'en potential 
\cite{Tulin:2013teo}. 
In Fig.~\ref{fig:DR_yukawa}, we plot $\langle\sigma_T\rangle_{30}/m_\chi$
as a function of the coupling $g_\chi$ for various masses $m_\chi$. We choose
a reference value of $m_{\tilde\gamma}=100$\,eV, noting that the cross sections would 
become even larger for lighter $\tilde\gamma$ particles. In the figure, one can clearly identify 
the different regimes for $\sigma_T$ (as well as the imperfect matching conditions, which are 
an artifact of our parameterization and bear no physical significance). For
$\alpha_\chi\lesssim 10^{-7}(m_\chi/\mathrm{GeV})^{-1}$ we are thus in the Born regime (for the choice
of $m_{\tilde\gamma}$ adopted in this figure), while for larger coupling we are in the classical
regime (apart from very small DM masses, where the characteristic resonances from the
intermediate regime start to appear). Besides the bound on the self-interaction rate mentioned 
above, we also include in the figure the generic upper bound
on $g_\chi$ that results from DM annihilation processes $\chi\chi\to\tilde\gamma\tilde\gamma$,
see the discussion in the previous subsection.

%note also: if thermal production, we should not be too far away from where
%the dashed lines start!
As one can clearly see, the self-interaction bounds are typically much stronger than those from the 
relic density. 
This implies in particular that the annihilation process 
$\chi\chi\to\tilde\gamma\tilde\gamma$ cannot be responsible for setting the correct DM 
density in the first place: in this case one would need coupling strengths close to the 
transition between solid and dashed lines in Fig.~\ref{fig:DR_yukawa}. 
For DM masses below the TeV scale, the self-interaction limits become in any case so 
severe that extremely tiny couplings are needed to evade them. In the classical 
regime, in particular, we find a simple scaling 
\be
\label{s30fit}
\frac{\langle\sigma_T\rangle_{30}}{m_\chi}\sim 5.3
\left(\frac{\alpha_\chi}{10^{-5}}\right)^{1.5} 
\left(\frac{m_\chi}{100\,\mathrm{GeV}}\right)^{-2.5}
\left(\frac{m_{\tilde\gamma}}{\mathrm{keV}}\right)^{-0.5}
\frac{\mathrm{cm}^2}{\mathrm{g}}\,,
\ee
which is valid for DM masses of 
\be
m_\chi\gtrsim \frac{m_{\tilde\gamma}}{100\,\mathrm{eV}}\,\max{\left[1,\frac{1}{10^7\alpha_\chi}\right]}\,\mathrm{GeV}\,. 
\ee
For a direct coupling of 
DM to bosonic $\tilde\gamma$, the self-interaction bound thus 
makes it generically very hard to obtain as large cutoff masses as desired. Plugging the 
resulting constraint on $g_\chi$ into the expression for the cutoff mass, Eq.~(\ref{mcutsimp}),
we obtain the rough estimate of
\be
M_\mathrm{cut}\lesssim M_n\left(\frac{c_n}{10^6g_\chi^4} \right)^\frac{3}{n+2}
\left(\frac{m_\chi}{100\,\mathrm{GeV}}\right)^\frac{1-3n}{n+2},
\ee
where we have conservatively assumed $m_{\tilde\gamma}=1$\,keV and $M_n$ is
shown in Fig.~\ref{fig:Mcutsimp}.
This implies that, compared to the generic expectation of 
$c_n\sim g_\chi^4$, very large enhancements of the amplitude from almost on-shell virtual 
particles are necessary to achieve sufficiently large values of $M_\mathrm{cut}$.
Interestingly, it is also no longer favorable to consider small DM masses when taking
into account the self-interaction bound.

 %%%%%%%%%%%%%%%%%%%%%%%%%%
 \begin{figure}[t!]
 	\includegraphics[width=0.49\columnwidth]{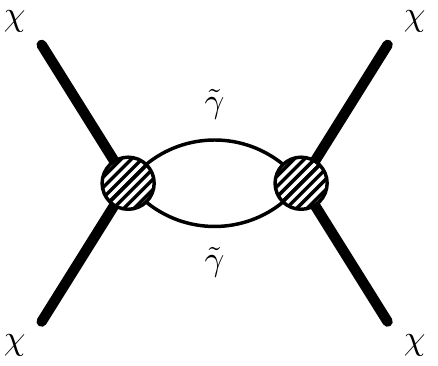}
 	\includegraphics[width=0.49\columnwidth,trim={0 0.4mm 0 0}]{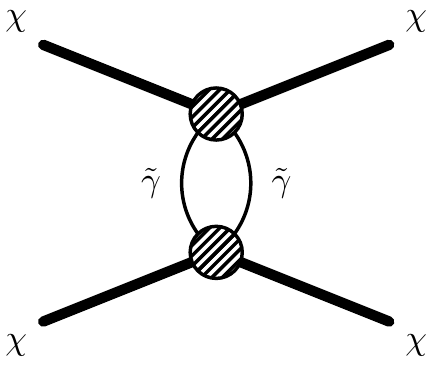}	
 	\caption{%
	{\it Minimal} effective DM self-interaction resulting from the
    $\chi$--$\tilde\gamma$
	interactions shown in Fig.~\ref{fig:EffDiagram}.
 	}
 	\label{fig:self_simp}
 \end{figure}		
 %%%%%%%%%%%%%%%%%%%%%%%%%%

Let us however stress that DM self-interactions do not only provide useful {\it constraints} on
the type of models that we want to consider. Rather, as briefly mentioned in the 
introduction, they have been invoked as {\it solutions} to a number of shortcomings of $\Lambda$CDM 
cosmology at small scales. In particular, it has been argued \cite{Feng:2009hw,Buckley:2009in,Loeb:2010gj,Vogelsberger:2012ku} that a velocity-dependent transfer
cross section $\sigma_T$ resulting from a Yukawa potential may successfully address both the 
{\it cusp-core} problem \cite{deBlok:1997zlw,Oh:2010ea,Walker:2011zu} and the 
{\it too-big-to-fail} problem \cite{BoylanKolchin:2011de,Papastergis:2014aba} that appear
at the scale of dwarf galaxies, without violating the stringent constraints at cluster scales. 
For this solution to work, self-interaction cross sections not more than one order of magnitude below the 
constraints on $\langle\sigma_T\rangle_{30}/m_\chi$ shown in Fig.~\ref{fig:DR_yukawa} are needed.
Another recent observation that has sparked significant interest in DM
self-interactions is  
the cluster Abell 3827, where one of the member galaxies falling towards the center of the cluster 
appears to be displaced from its own gravitational well \cite{Massey:2015dkw,Kahlhoefer:2015vua}.

\begin{figure*}[t!]
    \centering
    \begin{tabular}{|c|c||C{2.0cm}C{2.0cm}C{2.0cm}||C{2.0cm}C{2.0cm}C{2.0cm}||C{2.0cm}||}
    \hline
        \multicolumn{2}{|c||}{$\tilde\gamma$ \textbackslash \;$\chi$ } & \multicolumn{3}{c||}{\bf Scalar}    &  \multicolumn{3}{c||}{\bf Fermion} & \multirow{2}{*}{ \bf Vector }  \\[3ex]
        \hline
        & TOP &  LKD & TP &$\sigma_T $  & LKD & TP & $ \sigma_T $  &  \\[4ex]
        \hline
        \hline
        \multirow{7}{*}{ \bf Scalar} & $4p$ & $m_\chi \lesssim$ MeV & Yes & Constant &  
        \multicolumn{3}{c||}{ \cellcolor{Gray!40}  (only dim $ > 4$)   } &  \multirow{3}{*}{\cellcolor{Gray} }\\[4ex]
        & $t$  & $m_{\tilde\gamma}\sim1$\,keV \mbox{\tiny $m_\chi\gtrsim100\alpha_\chi^{3\!/\!5}$\,TeV}& 
        \cellcolor{Gray} ~ $\langle \sigma_T\rangle_{30}$ \quad{\tiny (for $m_\chi\gtrsim1$\,MeV)}& Yukawa &   
        $m_{\tilde\gamma}\sim1$\,keV \mbox{\tiny $m_\chi\gtrsim100\alpha_\chi^{3\!/\!5}$\,TeV}& 
        \cellcolor{Gray} ~ $\langle \sigma_T\rangle_{30}$ \quad{\tiny (for $m_\chi\gtrsim1$\,MeV)}& Yukawa &  
        \multirow{3}{*}{\cellcolor{Gray} $\langle \sigma_T\rangle_{30}$ } \\[6ex]
        & $s/u$  &  \multicolumn{3}{c||}{\cellcolor{Gray} $\langle \sigma_T\rangle_{30} $}  &  
        \multicolumn{3}{c||}{\cellcolor{Gray} $\langle \sigma_T\rangle_{30} $} &  \cellcolor{Gray}\\[4ex]
        \hline
        \hline
        &  &  \multicolumn{3}{c||}{ \cellcolor{Gray!40}} & 
        \multicolumn{3}{c||}{ \cellcolor{Gray!40}  } &   
        \cellcolor{Gray}  \\ %[1ex]
        \bf Fermion &  &  \multicolumn{3}{c||}{ \cellcolor{Gray!40} (only dim $ > 4$ due to $Z_2  $) } & 
        \multicolumn{3}{c||}{ \cellcolor{Gray!40} (only dim $ > 4$)  } &    \multicolumn{1}{c||}{ \cellcolor{Gray}  $Z_2  $}  \\[3ex]
        \hline
        \hline
        \multirow{6}{*}{\bf Vector} & $4p$ & \multicolumn{3}{c||}{\cellcolor{Gray!40} (only dim $ > 4$)} & \multicolumn{3}{c||}{\cellcolor{Gray!40} (only dim $ > 4$)} &
        \multirow{4}{*}{\cellcolor{Gray} $Z_2$ } \\[4ex]
        & $s/u$  &  \multicolumn{3}{c||}{$\langle \sigma_T\rangle_{30}  $ 
        \cellcolor{Gray}} &   \multicolumn{3}{c||}{$\langle \sigma_T\rangle_{30} $ \cellcolor{Gray}} & \multirow{1}{*}{\cellcolor{Gray}}\\[4ex]
        & $SU(N)$  &   $m_{\tilde\gamma}\sim1$\,keV \mbox{\tiny $m_\chi\gtrsim10\alpha_\chi^{3\!/\!5}$\,TeV}  &
        \cellcolor{Gray} ~ $\langle \sigma_T\rangle_{30}$ \quad{\tiny (for $m_\chi\gtrsim1$\,MeV)}& Yukawa &  
         $m_{\tilde\gamma}\sim1$\,keV \mbox{\tiny $m_\chi\gtrsim10\alpha_\chi^{3\!/\!5}$\,TeV} &
         \cellcolor{Gray} ~ $\langle \sigma_T\rangle_{30}$ \quad{\tiny (for $m_\chi\gtrsim1$\,MeV)}& Yukawa & 
        {\cellcolor{Gray!40}\mbox{\small (only broken} {{\scriptsize $SU(M)\to SU(N)$}{\small)}} \vspace*{-0.8cm}} \\[6ex]
        \hline
    \end{tabular}
    \caption{Overview of results for the 2-particle models we have considered, where the DM particle
    $\chi$ and the (dark) radiation particle $\tilde\gamma$ can be scalars, Dirac fermions and vectors, 
    respectively. 
    Here, (TOP) denotes the topology of the (dominant) DM-DR scattering amplitude. Late kinetic decoupling 
    (LKD) indicates whether in this type of models a small-scale cutoff as large as  
    $M_\mathrm{cut}\sim10^{10}\,M_\odot$ can be arranged.
    Thermal production (TP) indicates whether the observed DM density can be explained by thermal 
    production via $\chi\chi\to\tilde\gamma\tilde\gamma$, and $ \sigma_T$ indicates the type of the
    DM self-interaction rate (only for viable models).
    A white cell indicates that LKD is possible and that models of this type additionally satisfy the
    indicated property (e.g.~TP).  Dark gray indicates that the model is either ruled out, or that it is not possible
    to achieve LKD, for the reason stated. Here, $\langle\sigma_T\rangle_{30}$  indicates the DM 
    self-interaction strength at the scale of dwarf galaxy scales and $Z_2$ the assumed symmetry to stabilize
    DM. Operators with dimension (dim) larger than 4 map to the scalar/scalar 4-point case if they lead
    to an approximately constant scattering amplitude; otherwise they are too small to lead to LKD\@.
    Note that $t$-channel scattering with vector DR is only possible in a non-Abelian gauge theory and hence 
    covered in the $SU(N)$ part.\label{tab:Results2p}}
\end{figure*}

Light fermions $\tilde\gamma$, on the other hand, obviously do
not generate a Yukawa potential. A lower bound on the DM self-interaction rate can then
be obtained by considering the diagrams displayed in Fig.~\ref{fig:self_simp}, i.e.~by
forming a loop of the scattering process shown in Fig.~\ref{fig:EffDiagram}. 
Assuming again that the shaded blobs in Fig.~\ref{fig:self_simp} are
of the order of $g^2$, we roughly estimate
$|\mathcal{M}|^2 \sim g^8/16\pi^2$.
The transfer cross section will thus be of 
the order of $\sigma_T\sim 10^{-5} g^8/m_\chi^2$.
Even when using the stronger 
constraint $\sigma_T/m_\chi\lesssim 0.1\,\mathrm{cm}^2/\mathrm{g}$ from cluster scales 
(as $\sigma_T$ is to a good approximation velocity-independent in this case),
this only results in the essentially insignificant bound
$g^2\lesssim(m_\chi/\mathrm{MeV})^{3/4}$ -- which is always much weaker than
the relic density bound considered above for the range of DM masses
relevant for our 
discussion, $m_\chi\gtrsim0.1$\,MeV. We can in fact turn this argument around: if the 
effective couplings represented by the blobs in Fig.~\ref{fig:self_simp} were large enough 
to result in a significant self-interaction rate, this would imply a DM self-annihilation rate 
too large to be consistent with the  DM abundance observed today.
Stronger bounds from self-interactions can arise in specific models, however, as we will see in the following.

%%%%%%%%%%%%%%%%%%%%%%%%%%%%%%%%%%%%%%%%%%%%
\section{2-particle models}
\label{sec:2pm}
\vspace{-2cm}
We first consider models with the minimal possible particle content, i.e.~we assume that 
there is no additional particle mediating the scattering process 
$\chi\tilde\gamma\to\chi\tilde\gamma$. This leaves three basic topologies that we will 
study in more detail in the following: {\it A)} contact interactions, {\it B)} $s/u$-channel 
mediated scattering processes and {\it C)} dominantly $t$-channel mediated scattering 
processes. 
For each of these cases, we will discuss all possible spin combinations that potentially 
could lead to late kinetic decoupling, i.e.~we allow in principle 
both $\chi$ and $\tilde\gamma$ to be a scalar, (Dirac) fermion or vector particle, respectively
(for the sake of brevity, we will however not explicitly consider pseudoscalars and axial 
vectors in our analysis).
For simplicity, and to avoid unphysical results, we will assume that a vector boson is 
always associated with a gauge symmetry -- which may however be spontaneously broken
to allow for $m_{\tilde\gamma}\neq0$. We further require that DM is stabilized by a $Z_2$
symmetry, i.e.~we do not allow for vertices with an odd number of $\chi$ particles.
For a quick overview of our results for this type of models, we refer the reader to 
Fig.~\ref{tab:Results2p}.

%%%%%%%%%%%%%%%%%%%%%%%%%%%%%%%%%%%%%%%%%%%%
\subsection{Point-like interactions}
\label{sec:scatter_point}

%%%%%%%%%%%%%%%%%%%%%%%%%%
\begin{figure}[t!]
	\includegraphics[width=0.50\columnwidth]{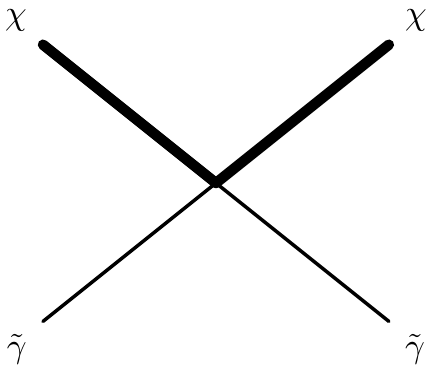}	
	\caption{Diagram illustrating a {\it point-like interaction} of a DM
particle $\chi$ with 
	a (possibly dark) relativistic particle $\tilde\gamma$. Because we
focus on 
	unsuppressed 
	interactions, only dimension-4 operators are considered, which
restricts the analysis
	of this topology to bosonic particles.
	}
	\label{fig:4pDiagram}
\end{figure}		
%%%%%%%%%%%%%%%%%%%%%%%%%%

Let us start with the case of a {\it single}, point-like contact interaction, as
depicted in 
Fig.~\ref{fig:4pDiagram}. The simplest possibility to obtain this is
with a dimension-4 
operator. Due to gauge invariance, the 
only such operator that we need to study separately is in fact the case
of a 
`portal interaction' between a scalar $\chi$ and a scalar
$\tilde\gamma$, leading to a constant scattering amplitude. This is because
a 4-point coupling involving (broken or unbroken) 
gauge fields would imply the existence of further
3-point couplings that unavoidably lead to additional diagrams of the
form studied in the
subsequent Sections \ref{sec:scatter_su} and \ref{sec:scatter_t}. 

Higher dimensional operators that lead to an (almost) constant
scattering rate will have the 
same phenomenology, albeit with a suppressed amplitude, and hence do not
have to be 
studied separately. Alternatively, a higher-dimensional operator containing derivatives or fermionic DR could
add an energy dependence to the scattering rate. As any such operator is irrelevant in the 
language of effective field theory, i.e.~suppressed at low energies,
it will necessarily yield $n>0$ in Eq.~(\ref{m2simp}).  Given that $c_n$ is suppressed by a large
mass scale,  Fig.~\ref{fig:cn10} then tells us that this possibility will not succeed
in producing 
sufficiently large values of $M_\mathrm{cut}$.

The only point-like interaction we have to consider in more detail at
this point is thus a portal interaction 
of the form $\mathcal{L}\supset \frac{\lambda}4 \chi^2\tilde\gamma^2$.
This implies 
$|\mathcal{M}|^2=\lambda^2=\langle|\mathcal{M}|^2\rangle_t$
and, cf.~Eqs.~(\ref{m2simp}, \ref{mcutsimp}),
% this assumes geff=3.36
\be
\label{mcut_SSSS}
M_\mathrm{cut}^{4S}\simeq8.4\times 10^{10}\,
\xi^6\lambda^3\left(\frac{m_\chi}{10\,\mathrm{GeV}}\right)^{-9/2}\,
M_\odot\,,
\ee
seemingly implying that a cutoff in the desired range can be obtained
for any DM mass
smaller than a few GeV.

As stressed in Section \ref{sec:gen_ann}, however, an additional upper
bound on $\lambda$
results from the requirement that the DM annihilation rate 
$\chi\chi\to\tilde\gamma\tilde\gamma$ should not become so large that it
would deplete the
initial DM abundance (thermally produced or not) below the currently
observed value.
In Eq.~(\ref{RDswave}) we should thus simply replace
$g'\to\sqrt{\lambda}$, and require
that the resulting value for $\Omega_\chi$ is not smaller than the
observed one.
This leads to\footnote{%
\label{RDfoot}%
We note that $x_f$ depends logarithmically on the DM mass, and we used here 
the approximation given in Kolb\,\&\,Turner \cite{Kolb:1990vq}.
Furthermore, we took into account the  impact of $T_{\tilde\gamma}\neq T$ 
during freeze-out by assuming $x_f\propto \xi$ in  
Eq.~(\ref{RDswave}). We checked this latter assumption explicitly by solving the 
full Boltzmann equation provided in Ref.~\cite{Gondolo:1990dk}, finding that 
the actual scaling is more accurately given by $x_f\propto \xi^r$, with $1.1\lesssim r\lesssim1.2$
(where $r$ is larger for smaller values of $m_\chi$ and/or $\xi$).
Note that we assume that $\xi$ remains
constant between chemical and kinetic decoupling.
%confirmed numerically!
% "proof":
%Assuming that the cross sections $\langle \sigma v\rangle'$ (for
%$\xi\neq1$) and 
% $\langle \sigma v\rangle$ (for $\xi=1$) are adjusted such as to give
% the same yield,
% we get the relation $$ \frac{n_\chi(\xi T_{cd}')}{s(T_{cd}')} =
% \frac{n_\chi( T_{cd})}{s(T_{cd})}. 
% $$ Combining this with the equations $$ H(T_{cd}) = n_\chi(T_{cd})
% \langle \sigma v\rangle 
% $$ and $$  H(T_{cd}') = n_\chi(\xi T_{cd}') \langle \sigma v\rangle'.
% $$ This results in the 
% following scaling of $\langle \sigma v\rangle'$ 
%	$$ \langle \sigma v\rangle' = \frac{H(T_{cd}')}{n_\chi( T_{cd})}
%	\frac{s(T_{cd})}{s(T_{cd}')} = \langle \sigma v\rangle
%	\sqrt{\frac{g_*(T_{cd}')}{g_*(T_{cd})}}\frac{g_{*S}(T_{cd})}{g_{*S}(T_{cd'})}
%	\frac{T_{cd}}{T_{cd}'}.$$ 
%If $T_{cd} \ll m_\chi$ then a good approximation is $ T_{cd}' \sim \xi
%T_{cd}$. 
}
\be
\label{mcut_SSSS_RD}
M_\mathrm{cut}^{4S}\lesssim3\times 10^{10}\,
\xi^{15/2}\left(\frac{m_\chi}{\mathrm{MeV}}\right)^{-3/2}\,
M_\odot\,,
\ee
where the maximal value for $\lambda$, and hence $M_\mathrm{cut}$, is
achieved if DM is 
actually produced thermally (and this process is dominated by the same
portal
coupling between DM and DR).
Given that  $\xi\gtrsim1$ is strongly constrained by CMB observations
(see also footnote \ref{noCMBbound}),  DM in this simplest scenario must thus be lighter than 
about $1\,\mathrm{MeV}$ in order to produce a cutoff in an observationally interesting 
range. As discussed, free streaming effects start to 
further increase the cutoff mass for $m_\chi\lesssim 0.1\,\mathrm{MeV}$
(or even lighter DM masses if $\xi\ll1$).
The resulting additional suppression of structure  implies that the 
same value of $M_\mathrm{cut}$ can be achieved for smaller values of $\xi$,
which allows to satisfy the strong CMB constraints on this quantity by an even
larger margin (while $m_\chi\ll0.1$\,MeV would simply result in the standard WDM case). 
We leave a full exploration of this interesting regime for future work.

For this mass range, the DM annihilation bound becomes 
$\lambda\lesssim7\cdot10^{-7}\xi^{0.5}\, m_\chi/\mathrm{MeV}$. 
Even though it is parametrically suppressed by a factor of 
$\lambda^4$, however, the
induced DM self-coupling (see Fig.~\ref{fig:self_simp}) for this model is
actually log-divergent. To be able to remove this divergence by renormalization, we thus must
add an interaction term \mbox{$\Delta\mathcal{L}=(\lambda'/4!)\chi^4$}.
Its finite part can thus be tuned to any desired value, independent of the above
discussion, leading to a velocity-independent DM self-interaction cross section.
Let us conclude the discussion of this case by remarking that the required
small value of $\lambda$ might most naturally be realized by a $\text{dim}>4$ 
operator (as long as it leads to an approximately constant scattering rate,
with $n=0$). See Refs.~\cite{Hooper:2007tu,Shoemaker:2013tda} for  
examples of large 
cutoff masses resulting from such effective operators (and sub-GeV DM
masses).

%%%%%%%%%%%%%%%%%%%%%%%%%%%%%%%%%%%%%%%%%%%%
\subsection{Scattering exclusively via $s/u$-channel}
\label{sec:scatter_su}

\begin{figure}[t!]
	\includegraphics[width=0.49\columnwidth]{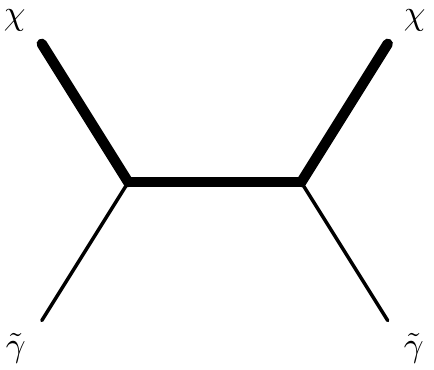}	
	\includegraphics[width=0.49\columnwidth,trim={0 0.8mm 0
0}]{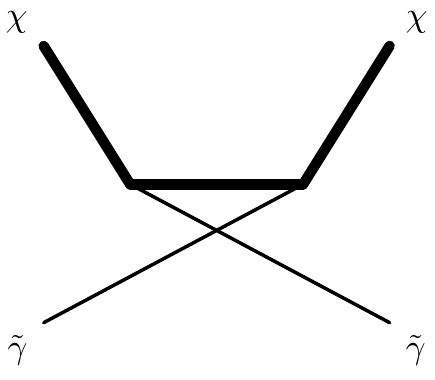}
	\caption{As Fig.~\ref{fig:4pDiagram}, but in the presence of a 
	$\chi$-$\chi$-$\tilde\gamma$ coupling, which leads to a resonance in the
$s$- (left)
	and $u$-channel (right). 
	}
	\label{fig:suDiagram}
\end{figure}

Let us next consider situations where the scattering proceeds
exclusively via the $s$- (and 
hence also $u$-) channel. The requirement to stabilize DM via a $Z_2$
symmetry then 
implies, as illustrated in Fig.~\ref{fig:suDiagram}, that the virtual
particle must be $\chi$
(recall that by assumption no further particle beyond $\chi$ and
$\tilde\gamma$ can be 
involved in our simplified 2-particle model). This  means that
$\tilde\gamma$ must
be bosonic, and we fully need to take into account the stringent
constraints on DM 
self-interactions discussed in Section \ref{sec:gen_self}.
Here, we only consider the situations that arise when $\tilde\gamma$ is a scalar or an
Abelian gauge boson, and $\chi$ is either a scalar or a fermion. 
Otherwise -- i.e.~for non-Abelian DR or  vector DM -- there are necessarily both 
4-point, $s/u$- and $t$-channel diagrams involved in the scattering process;
we defer the treatment of these cases to Section \ref{sec:mixed}.
We calculate all relevant matrix elements in Appendix \ref{app:msq}, and
list the results in 
Table \ref{tab:Msq_2p}.

%%%%%%%%%%%%%%%%%%%%%%%%%%	
%%%%%%%%%%%%%%%%%%%%%%%%%%

While both diagrams in Fig.~\ref{fig:suDiagram} individually contain a
resonance, those
leading contributions cancel exactly in the $t\to0$ limit in all cases.
The result is an effective 
scattering 
amplitude that is to a very good approximation independent of the energy 
$\omega$ of the relativistic scattering partners. We thus obtain the
{\it same} result as in
the contact interaction case, Eq.~(\ref{mcut_SSSS}), with the
understanding that we should
replace $\lambda^2$ by the corresponding expression for
$\langle|\mathcal{M}|^2\rangle/\eta_\chi$
stated in Table \ref{tab:Msq_2p}, where $\eta_\chi$ denotes the number of
internal degrees of freedom of the DM particle. The essential difference, however, is
that now we have a 
three-point
coupling giving rise to a strong Yukawa potential between the DM
particles. We can thus 
combine the result for the cutoff mass, Eq.~(\ref{mcutsimp}), with the
constraint 
$\langle \sigma_T\rangle_{30}/ m_\chi\lesssim10\,\mathrm{cm}^2/\mathrm{g}$ on
the
transfer cross section, where $\langle \sigma_T\rangle_{30}$ is supplied
in 
Eq.~(\ref{s30fit}). This results in
\be
\label{sucutmax}
M_\mathrm{cut}^{s/u}\lesssim 2\cdot10^{-7}\xi^6 r^\frac32
\left(\frac{m_{\tilde\gamma}}{\mathrm{keV}}\right)
\left(\frac{m_\chi}{100\,\mathrm{GeV}}\right)^\frac12 M_\odot\,.
\ee
Here, we have introduced $r\equiv\langle|\mathcal{M}|^2\rangle/(\eta_\chi g_\chi^4)$, where
$g_\chi$ is the dimensionless coupling constant that enters the Yukawa potential -- for 
the scalar/scalar (fermion/vector) case, e.g., we have 
$r=1/2$ ($r=16/3$).
%since we are summing over spins and particle/antiparticle i.e. $\eta_\chi = 4$.

Equation~(\ref{sucutmax}) clearly demonstrates that the strong constraints on DM 
self-interactions make it impossible to achieve late kinetic decoupling
if the scattering is 
only mediated through $s$- and $u$-channel diagrams. We note that we
arrived at this 
conclusion completely independently of the DM production mechanism.
We \emph{have} assumed in this argument, however, that the DM self-scattering
takes place in the classical regime. For very light, (sub-)MeV DM  (e.g.~scalar DM 
scattering with hidden $U(1)$ vectors \cite{Feng:2009mn}) it may thus be possible 
to achieve large cutoff values and evade the self-scattering constraints.

%%%%%%%%%%%%%%%%%%%%%%%%%%%%%%%%%%%%%%%%%%%%
\subsection{Scattering dominantly via $t$-channel}
\label{sec:scatter_t}

 %%%%%%%%%%%%%%%%%%%%%%%%%%	
 \begin{figure}[t!]
 	\includegraphics[width=0.50\columnwidth]{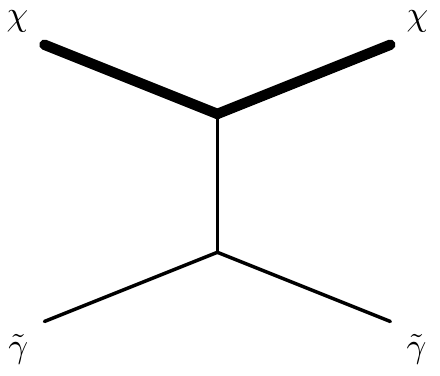}	
 	\caption{As Fig.~\ref{fig:suDiagram}, but in the presence of an additional 
	3-$\tilde\gamma$ coupling, which leads to a resonance in the $t$-channel
	{\it in addition} to the $s/u$ resonances shown in Fig.~\ref{fig:suDiagram}.
 	}
 	\label{fig:tDiagram}
 \end{figure}		
 %%%%%%%%%%%%%%%%%%%%%%%%%%	

Due to the $Z_2$ symmetry for the
$\chi$ particles, any scattering diagram involving a $t$-channel exchange is of the form 
displayed in Fig.~\ref{fig:tDiagram}. Just as for the $s/u$ case, this topology thus
only allows scalar or non-Abelian $\tilde\gamma$. Here, we only consider the former case,
deferring a dedicated discussion of the latter case to the next subsection. 
Such models have two independent coupling constants for the $\chi$-$\chi$-$\tilde{\gamma}$ and 
$\tilde{\gamma}$-$\tilde{\gamma}$-$\tilde{\gamma}$ vertices. The presence of the former induces
$s$- and $u$-channel diagrams of the type discussed above. Here, we will thus require that 
those couplings are small enough to satisfy the self-interaction constraints of 
Fig.~\ref{fig:DR_yukawa}, cf.~Eq.~(\ref{s30fit}), and that the $t$-channel diagram
dominates the scattering process. We note that this is indeed the generic situation, 
even for a $\tilde{\gamma}$-$\tilde{\gamma}$-$\tilde{\gamma}$ coupling much smaller than the
$\chi$-$\chi$-$\tilde{\gamma}$ coupling, because of the strong kinematic enhancement of 
the $t$-channel diagram.  We calculate the two relevant 
matrix elements, i.e.~those for scalar and fermionic DM, in Appendix \ref{app:msq} and list the 
results in Table \ref{tab:Msq_2p}.

As expected from the familiar Coulomb case, the scattering amplitude 
from the $t$-channel exchange of a massless particle diverges, and has to be regulated
by introducing a non-vanishing DR mass term. In fact, such a mass term can be argued 
to arise from requiring the potential to be bounded from below: in our simplified model, 
this can only be achieved by adding a 4-point interaction term 
$(\lambda/4!)\tilde\gamma^4$ to the scalar potential 
$V(\tilde\gamma)= (m_0^2/2)\tilde\gamma^2+(\mu_{\tilde\gamma}/3!)\tilde\gamma^3$;  
considering the global minimum of this potential then leads to the conclusion that 
$m_{\tilde\gamma}\sim \max(m_0,\mu_{\tilde\gamma}/\sqrt{\lambda})\gtrsim \mu_{\tilde\gamma}$, 
largely independent of the value of $m_0$.
In general, Debye screening will furthermore generate a thermal mass of the 
order of $m_{\tilde\gamma}^\mathrm{Debye}\sim \sqrt{\lambda} T_{\tilde\gamma}$. 
At temperatures $T_{\tilde\gamma}\gg \mu_{\tilde\gamma}$, the combination of this effect 
and the requirement of vacuum stability thus even lead to 
$m_{\tilde\gamma}\gg \mu_{\tilde\gamma}$ -- essentially independent of the size of 
$\lambda$.

In the case of scalar DM, we have another dimensionful constant $\mu_\chi$ which 
denotes the $\chi$-$\chi$-$\tilde\gamma$ coupling. Perturbativity and the absence of 
a global minimum in the scalar potential with $\chi \neq 0$, which would break the 
DM-stabilizing $Z_2$ symmetry, restrict $\mu_\chi$ to be sufficiently smaller than $m_\chi$.
However, generically we still expect $\mu_\chi \gg m_{\tilde\gamma}$. Lastly, we would like
to mention that for scalar DM and DR one will generally also have a portal interaction term 
in the Lagrangian as discussed in Section \ref{sec:scatter_point}. Due to the strong 
kinematic enhancement of the $t$-channel diagram, however, this term will not
have any significant effect unless $\mu_\chi\ll m_{\tilde\gamma}$.

In the limit where DR is highly relativistic, we find for fermionic DM an average scattering
amplitude of 
\be
\label{tchannel_generic}
\langle |\mathcal{M}|^2 \rangle_t = \frac{g_\chi^2\mu_{\tilde\gamma}^2m_\chi^2}{\omega^4}\ln{\frac{4\omega^2}{m_{\tilde \gamma }^2}}.
\ee
For scalar DM we find the same expression after replacing the dimensionless  DM-DM-DR 
coupling $g_\chi$ with the corresponding dimensionful coupling 
$\mu_\chi$ as $g_\chi\to \mu_\chi/(\sqrt{8}m_\chi)$.
We remind the reader that for such an energy dependence, the general analytic solution
referred to in Eq.~(\ref{mcutsimp}) is no longer valid. We can still immediately see that
in this case the momentum transfer rate $\gamma$, cf.~Eqs.~(\ref{fT}, \ref{maverage}), 
would fall with temperature  {\it less} rapidly than the Hubble rate. In such a situation, 
DM and DR would initially {\it not} be in local thermal equilibrium. Once they enter it, 
however, they would not leave it anymore -- leading to a depletion of structure on large 
scales that is unacceptable from an observational point of view (unless the couplings are 
chosen so small that thermal equilibrium would only be reached late during matter 
domination).

In view of this rather unexpected behavior, let us lift our general assumption of 
ultra-relativistic DR and investigate which effect an increased DR mass 
$m_{\tilde\gamma}$ would have on the cosmological behavior of this class of models. 
We will assume that $\tilde\gamma$ still follows a thermal distribution,\footnote{%
In contrast to the effectively massless case, this requires thermal 
equilibrium of $\tilde\gamma$ with at least one further relativistic species $\varphi$ 
of temperature $T_{\tilde\gamma}$.  Since we consider here by construction a situation in which 
$\tilde\gamma$ is non-relativistic around kinetic decoupling of the DM particles, the 
CMB bound on the energy density of additional degrees of freedom (i.e.~on $\xi$) thus 
becomes independent of $\eta_{\tilde\gamma}$ and only depends on $\eta_\varphi$.
}
so we should expect that at some point the Boltzmann suppression of the $\tilde\gamma$
number density will dominate over the $T_{\tilde\gamma}^{-4}$ scaling from 
Eq.~(\ref{tchannel_generic}), leading to a suppression of the momentum transfer rate
and hence kinetic decoupling relatively shortly after the DR has become non-relativistic.

To investigate this in more detail, we solve the full Boltzmann Eq.~(\ref{tchidt}) numerically, 
noting that the solution close to kinetic decoupling, and \emph{for a given value of 
$T_\mathrm{kd}$}, only depends on two 
parameter ratios, $m_{\tilde\gamma}/\xi$ and
$g_\chi^2\mu_{\tilde\gamma}^2/m_\chi$. 
We find that the former quantity is essentially fixed by the 
requirement to obtain a cutoff mass of $M_\mathrm{cut}=10^{10}M_\odot$, varying only from 
 $m_{\tilde\gamma}/\xi=0.53$\,keV for 
 $g_\chi^2\mu_{\tilde\gamma}^2/m_\chi=10^{-20}$\,keV to 
 $m_{\tilde\gamma}/\xi=2.0$\,keV for $g_\chi^2\mu_{\tilde\gamma}^2/m_\chi=10^{-10}$\,keV.  
 For very small couplings leading to $g_\chi ^2\mu_{\tilde\gamma}^2/m_\chi\lll10^{-20}$\,keV, 
 on the other hand, it is no longer
 possible to achieve a cutoff mass of $M_\mathrm{cut}=10^{10}M_\odot$ because $\chi$ 
 and $\tilde\gamma$
 would not reach local thermal equilibrium early enough in the first place.

 \begin{figure}[t!]
     \includegraphics[width=\columnwidth]{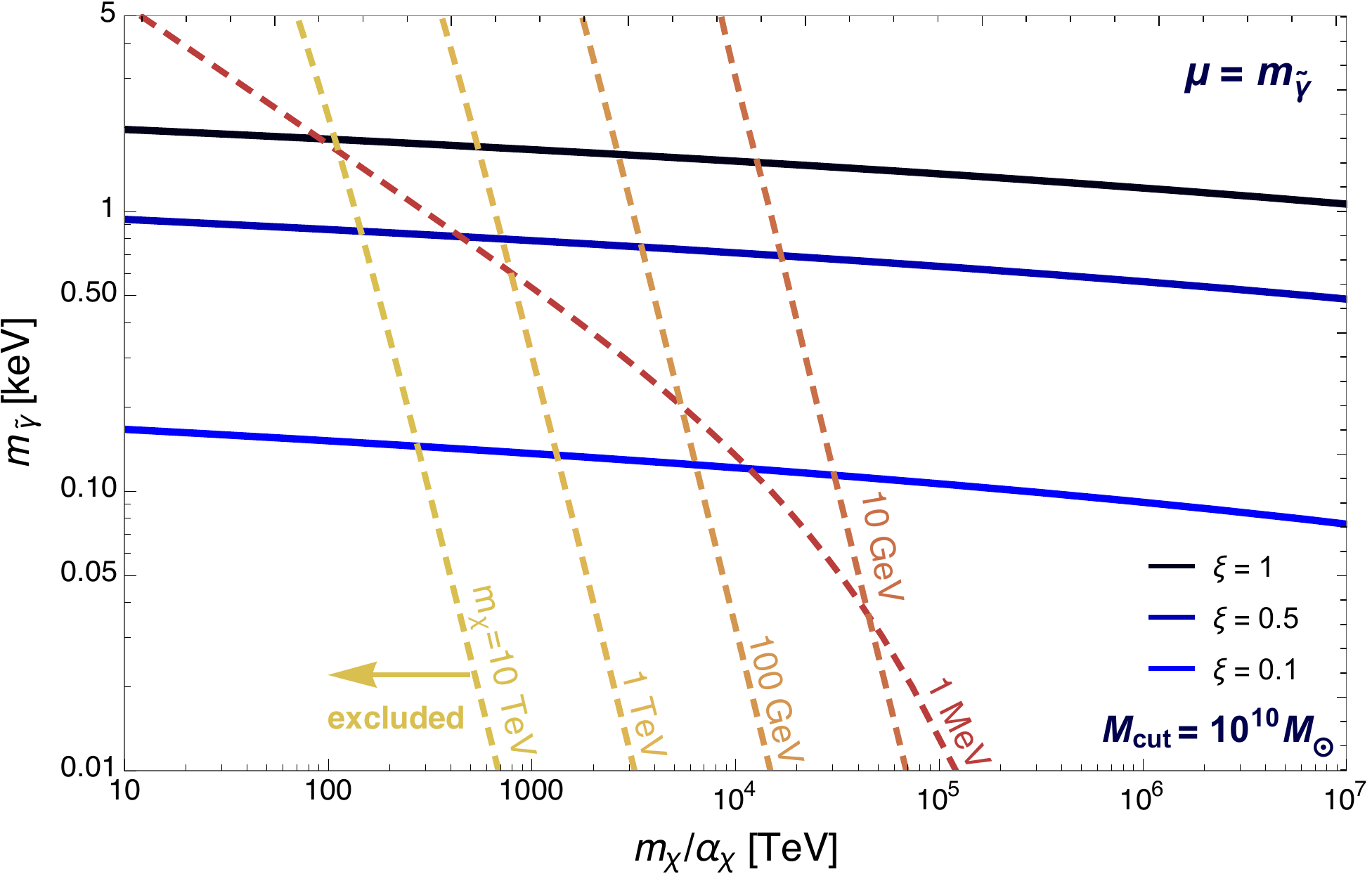}    
     \caption{Fermionic DM of mass $m_\chi$ scattering with scalar dark radiation 
     of mass $m_{\tilde\gamma}$, with $\alpha_\chi\equiv g_\chi^2/(4\pi)$ and the cubic
     DR self-coupling fixed to $\mu=m_{\tilde\gamma}$. Solid lines show the 
     parameter combinations that lead to $M_\mathrm{cut}\simeq10^{10}M_\odot$, for 
     a DR to photon temperature ratio of $\xi=1,0.5, 0.1$ (from top to bottom).
     Dashed lines give the constraints that result from DM self-scattering, for 
     a given DM mass; everything left to the respective dashed curve is excluded.
     \label{fig:2bt}
     }
\end{figure}

In Fig.~\ref{fig:2bt}, we show the DR mass that is required in this
scenario to obtain a 
cutoff mass of $M_\mathrm{cut}\simeq10^{10}M_\odot$, as a function of $m_\chi/\alpha_\chi$ 
and for several values of $\xi$ (solid lines). We also show, for various values of $m_\chi$, 
the constraints that arise from conservatively requiring that the DM self-interaction does not 
become too strong, namely $\langle\sigma_T\rangle_{30}<30\,\mathrm{cm}^2/g$; everything
to the left of the dashed lines is thus excluded. For $m_\chi\gtrsim1$\,GeV, these constraints
scale as expected for the classical regime, cf.~Eq.~(\ref{s30fit}), 
i.e.~$(m_\chi/\alpha_\chi)_\mathrm{min}\propto m_\chi^{-2/3}$. Decreasing the DM mass
below about 1\,GeV, the limits do not tighten significantly anymore. As shown
exemplarily for  $m_\chi=1$\,MeV, they feature instead a much stronger dependence on 
$m_{\tilde\gamma}$ in this regime. This implies, as expected, that for DM masses even
closer to the DR mass of order keV (required for sufficiently late kinetic decoupling)  those 
limits start to become \emph{less} stringent again. In the plot, we have fixed 
$\mu=m_{\tilde\gamma}$. Smaller values will shift the solid lines to the left, by a
factor of $m_{\tilde\gamma}^2/\mu^2$. As long as $\mu$ is still large enough to bring 
DM and DR into local thermal equilibrium, this has hardly any effect on the allowed range 
of parameters in this model.

Models with fermionic DM that couple to a keV-scale scalar with a cubic self-coupling thus
allow to have both large cutoff masses and DM self-interaction strengths relevant at the 
scale of dwarf galaxies, for a broad range of DM masses. 
For $m_\chi\gtrsim1$\,GeV, this is very roughly achieved for 
a coupling strength of $\alpha_\chi\sim 10^{-6}(m_\chi/10\,\mathrm{GeV})^{5/3}$,
while smaller DM masses require a stronger coupling than expected from this simple
scaling law. Interestingly, for DM masses smaller than around 1\,MeV, the constraints
on the DM self-interaction rate are no longer stronger than those from the DM
annihilation rate (as it would be the case for $m_{\tilde\gamma}\ll1$\,keV). 
This implies that in this setup one may in fact have thermally produced DM, with
both $M_\mathrm{cut}$ and $\langle\sigma_T\rangle_{30}$ in a range that is
interesting from the point of view of $\Lambda$CDM small-scale problems.
Let us finally stress that the above discussion applies in full analogy to the case of 
scalar DM, with the already mentioned replacement $g_\chi\to \mu_\chi/(\sqrt{8}m_\chi)$.

%%%%%%%%%%%%%%%%%%%%%%%%%%%%%%%%%%%%%%%%%%%%
\subsection{DM-DR interactions through all channels}
\label{sec:mixed}	
Lastly, we consider those cases where treating $s/u$- and  $t$-channel (as well as 4-point) diagrams 
separately is no longer possible because of gauge invariance. We first note that a vector DM 
particle is generally not allowed if fermions exist that are charged under the same gauge group, 
because the assumed  $Z_2$ symmetry would be incompatible with covariant derivatives.
In a dark sector with a minimal field content without fermions, on the other hand, the spontaneous breaking 
of a $U(1)$ symmetry necessarily leads to a massive vector that obeys a $Z_2$ symmetry and hence
constitutes a very natural DM candidate, which has been discussed e.g.~in the context 
of Higgs portal models \cite{Lebedev:2011iq}. In a similar fashion, breaking a non-Abelian group leads to 
{\it two} independent $Z_2$ symmetries and hence two different DM particles \cite{Gross:2015cwa} 
-- a situation which we
will not study further because at this point we are only interested in scenarios with a single DM particle.

As shown in Appendix \ref{app:msq}, the scattering amplitude for Abelian vector DM and scalar DR
is independent of the DR energy in the limit that we are considering, and thus leads to the same
phenomenology as discussed in Section \ref{sec:scatter_su} for interactions that proceed exclusively 
through $s/u$-channel exchange. This implies in particular that late kinetic 
decoupling cannot be achieved for Abelian vector DM because the required coupling strength is ruled out by the resulting 
strong DM self-interaction.

In the context of the 2-particle models that we consider here, the only case that we have left out
from our discussion so far is non-Abelian DR\@. The DM particle can then be either
a scalar or a fermion, which leads to identical results for the scattering rates (up to a 
constant factor of order unity, see Table \ref{tab:Msq_2p}, and a sub-dominant contribution from the 4-point 
coupling in the scalar case).\footnote{%
It was only recently pointed out \cite{Ko:2016fcd} that it is also possible to break a non-Abelian group 
{\it partially} in such a way
that the gauge bosons of a residual non-Abelian subgroup would constitute DR, and DM would consist of vector particles
stabilized by a $Z_2$ symmetry. The phenomenology of such a setup depends on the exact
breaking pattern, and contains anyway more than one DM particle for the concrete situation considered in \cite{Ko:2016fcd}. Hence,
we do not further consider this possibility among the minimal scenarios we focus on here.
}
We note that the case of fermionic DM scattering with non-Abelian DR has been 
previously studied in Ref.~\cite{Buen-Abad:2015ova}, where it was also pointed out
that the necessarily small gauge couplings imply that confinement is irrelevant.  DR can 
hence  be described as a perfect fluid just like in all the other model types we study here.

Also in this case, a DR mass has to be introduced in order to regularize the scattering 
amplitude. Such a mass arises inevitably from screening effects in the thermal plasma
and can be estimated as $m_{\tilde\gamma}^\mathrm{Debye}\sim g_\chi T_{\tilde\gamma}$
\cite{Arnold:1995bh}. On top of this thermal mass, there can of course also be a 
temperature-independent mass if the gauge symmetry is spontaneously broken.
In the limit where the DR is still ultra-relativistic, the squared scattering amplitude is in any 
case of the same form as Eq.~(\ref{tchannel_generic}), but with the leading $\omega^{-4}$ 
dependence replaced by a $\omega^{-2}$ dependence. 
Such a dependence implies that the momentum transfer rate scales
as $\gamma\propto T^2_{\tilde\gamma}$, i.e.~(for constant $\xi$)
in the same way as 
the Hubble rate during radiation domination. During matter domination, on the other hand, 
$\gamma$ will quickly fall behind $H\propto T^{3/2}$.

If the leading contribution to the DR mass is thermal, this can result in a very interesting 
phenomenology, where all density perturbation modes that enter the 
horizon before matter-radiation equality are suppressed in a smooth way (while those that
enter after equality are essentially unaffected). Tuning the $SU(N)$ coupling strength to
$\alpha_\chi\sim10^{-9}$ (for $N=2$ and $m_\chi\sim\text{TeV}$), in particular, would help to 
alleviate a certain tension 
in the normalization of the power spectrum of density fluctuations as inferred from
different types of observations \cite{Buen-Abad:2015ova,Lesgourgues:2015wza}. Measurements
of the CMB
\cite{Ade:2015xua}, in particular, predict a value of the observable $\sigma_8$ that is about 
$2\sigma$ larger than what is obtained from large scale structure data 
\cite{Beutler:2014yhv,Battye:2014qga}. Adopting the above parameter values, we 
find that the resulting DM self-interaction becomes 
$\langle\sigma_T\rangle_{30}/m_\chi\lesssim 1\,\mathrm{cm}^2/\mathrm{g}$ for 
$m_{\tilde\gamma}\gtrsim 10^{-8}$\,eV,
%\sim g\xi T_\mathrm{CMB}^0$ today. 
thus evading the observational constraints on this quantity (see the discussion in 
Sec.~\ref{sec:gen_self}).
We note that the necessarily small value of $\alpha_\chi$ implies that the process 
$\chi\chi\to\tilde\gamma\tilde\gamma$ cannot be responsible for the thermal production of 
DM in this scenario.

Let us instead entertain the possibility, as in the preceding Section, that the non-Abelian
gauge bosons also have a \emph{constant} mass term which starts to dominate around keV
temperatures. Requiring again that $\tilde\gamma$ is somehow kept in chemical 
equilibrium even after it becomes non-relativistic, this would lead to the characteristic 
exponential cutoff in the power spectrum that is the main focus of this article.
We thus solve the full Boltzmann Eq.~(\ref{tchidt}) numerically, requiring again that 
$M_\mathrm{cut}=10^{10}M_\odot$. Similar to the $t$-channel case discussed above,
this fixes the ratio $m_{\tilde\gamma}/\xi$ as a function of the ratio $\alpha_N^2 m_{\tilde\gamma}^2/m_\chi$,
where we have defined $\alpha_N\equiv g_\chi^2 \sqrt{(N^2-1)}/(4\pi)$. Again,
the allowed value of $m_{\tilde\gamma}$ shows very little variation, from 
$m_{\tilde\gamma}=2.2$\,keV for $\alpha_N^2m_{\tilde\gamma}^2/m_\chi=10^{-10}$\,keV to 
$m_{\tilde\gamma}=0.18$\,keV for $\alpha_N^2m_{\tilde\gamma}^2/m_\chi=10^{-25}$\,keV\@.
For $\alpha_N^2m_{\tilde\gamma}^2/m_\chi\lll10^{-25}$\,keV, $\chi$ and $\tilde\gamma$
 are not in local thermal equilibrium at high temperatures.
 
 \begin{figure}[t!]
     \includegraphics[width=\columnwidth]{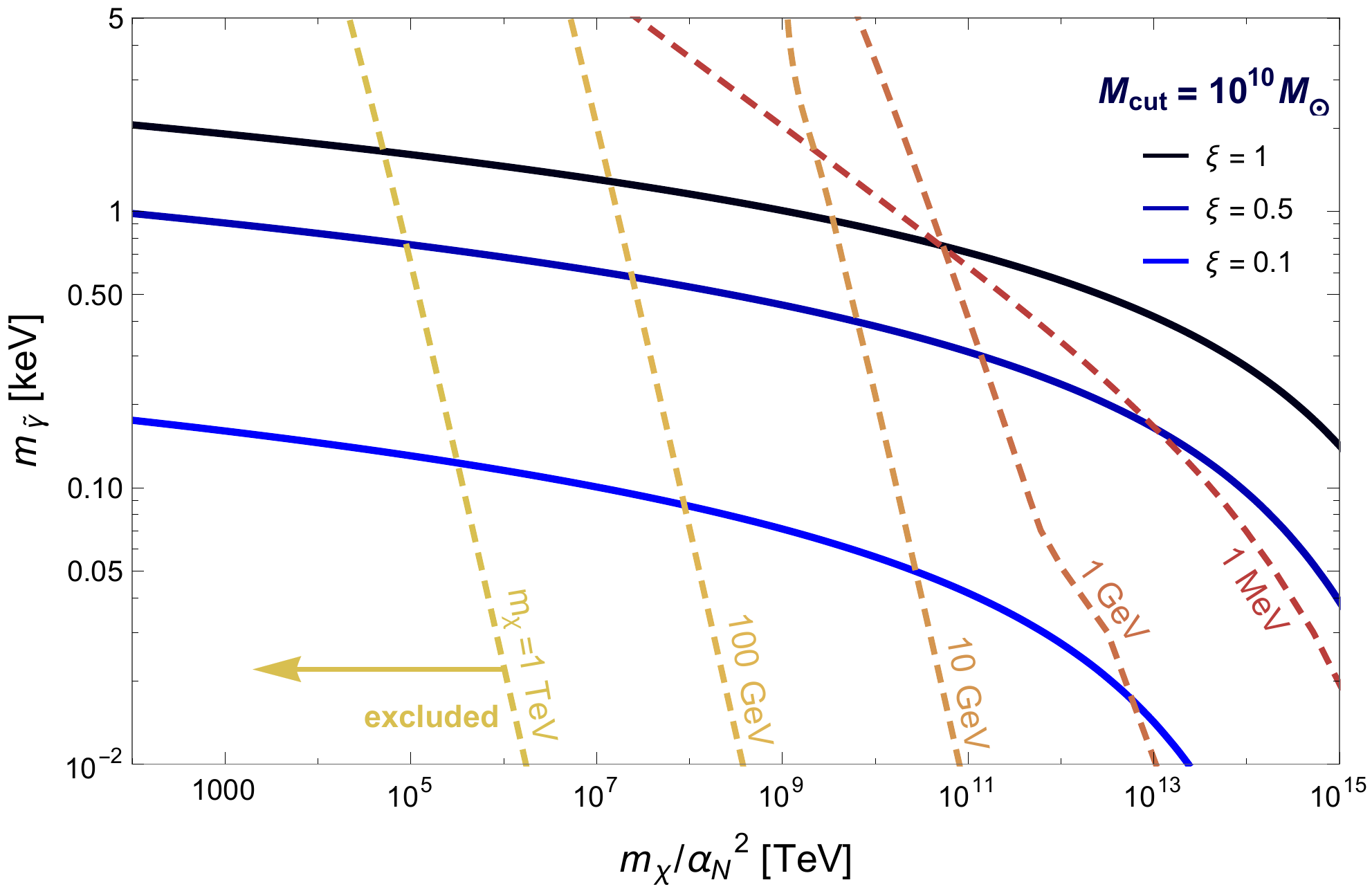}    
     \caption{Fermionic DM of mass $m_\chi$ scattering with non-Abelian DR of mass
     $m_{\tilde\gamma}$. For DR consisting of $SU(N)$ gauge bosons, we define 
     $\alpha_N\equiv g_\chi^2 \sqrt{(N^2-1)}/(4\pi)$. Solid lines show the 
     parameter combinations that lead to $M_\mathrm{cut}=10^{10}M_\odot$, for 
     a DR to photon temperature ratio of $\xi=1,0.5, 0.1$ (from top to bottom).
     Dashed lines show the constraints that result from DM self-scattering, 
     for $N=2$ and 
     a given DM mass; everything left to the respective curve is excluded.
     Larger values of $N$ result in weaker constraints. 
     \label{fig:2b_SUN}
     }
\end{figure}

In analogy to the case of scalar DR in the $t$-channel discussed in the previous subsection,
we plot in Fig.~\ref{fig:2b_SUN} the value of of the dark gluon mass $m_{\tilde\gamma}$ that is 
needed for a  cutoff mass $M_\mathrm{cut}\simeq10^{10}M_\odot$, as a function of 
$m_\chi/\alpha_N^2$ and for several values of $\xi$ (solid lines).
We also show, as dashed lines, the constraints connected to DM self-interactions; here,
we simply re-scaled the available parameterizations for $\sigma_T$ by the difference
between $SU(N)$ mediators and $U(1)$ mediators expected at tree level.\footnote{%
Concretely, we find $|\mathcal{M}|^2_{SU(N)}/|\mathcal{M}|^2_{U(1)} =
(N^2 -1)/4\equiv{\alpha_\chi'}^2/\alpha_\chi^2$, after {\it summing} over 
all colors, and then use
$\sigma_{T,SU(N)}(\alpha_\chi) = \sigma_{T,U(1)}(\alpha_\chi') / N^2$ to account for the color average in 
the initial state. We stress that this prescription is only an approximation to the full higher-order 
 $\sigma_T$, in the non-perturbative regime, but note that it reproduces the exact result in the Born regime.
}
For $m_\chi\gtrsim10$\,GeV, these constraints
scale as expected for the classical regime, cf.~Eq.~(\ref{s30fit}), 
i.e.~$(m_\chi/\alpha_\chi^2)_\mathrm{min}\propto m_\chi^{-7/3}$.
For smaller DM masses, the limits weaken with respect to this scaling for the largest DR
masses shown in the figure. Also in this case, this implies that there is a small region
in parameter space where relatively light \emph{thermal} DM, with $m_\chi\lesssim1$\,MeV, 
can produce an observable cutoff in the power spectrum \emph{and} feature an 
observationally relevant, but not yet excluded self-interaction rate.

%%%%%%%%%%%%%%%%%%%%%%%%%%%%%%%%%%%%%%%%%%%%
%%%%%%%%%%%%%%%%%%%%%%%%%%%%%%%%%%%%%%%%%%%%
%\newpage
\section{3-particle models}
\label{sec:3pm}

We now extend our discussion to simplified models where the scattering between $\chi$
and $\tilde\gamma$ is mediated by a {\it different} particle. As before, we will 
require that 
DM is stabilized by a $Z_2$ symmetry; this time, however, we will allow for further, heavier
particles to carry the same parity (which corresponds to the standard situation
in typical scenarios with WIMP DM candidates, like supersymmetry or universal extra 
dimensions). This restricts the logical possibilities to the {\it same topologies} as considered 
in the previous section, i.e.~scattering exclusively via a heavy particle $\chi'$ in the 
$s/u$-channel (as depicted in Fig.~\ref{fig:suDiagram}) or scattering via a light particle 
$\tilde\gamma'$ in the  $t$-channel (as depicted in Fig.~\ref{fig:tDiagram}).
For simplicity, we also do not explicitly study the possibility of vector DM (see Section
\ref{sec:mixed} for some general considerations concerning this option), and
restrict the discussion to couplings described by dimension-4 
operators (though we comment in Appendix \ref{app:msq3} on some opportunities that 
arise when lifting this assumption).

The fact that $\chi$ and $\chi'$ (as well as $\tilde\gamma$ and $\tilde\gamma'$) may
differ in both spin and mass opens several new avenues for model building and the 
phenomenology of these models. Most strikingly, more combinations of particle spins
are now possible (including fermionic $\tilde\gamma$) and the scattering can proceed
{\it exclusively} through the $t$-channel. The new mass scale, furthermore, can help
to avoid bounds on the self-interaction of DM, and qualitatively change the resonance 
structure of the $s/u$-channel diagrams.
As before, we will discuss the two fundamental topologies separately, focussing on those 
models and aspects that result in a qualitative difference to the 2-particle models.

%%%%%%%%%%%%%%%%%%%%%%%%%%%%%%%%%%%%%%%%%%%%
\subsection{Scattering via $s/u$-channel}
\label{sec:3pscatter_su}

Let us first consider models with a mediator particle $\chi'$ that is slightly heavier than 
$\chi$ and shares the same $Z_2$ parity. Defining $\Delta m \equiv m_{\chi'} - m_\chi$, we 
restrict our analysis to masses for which we have 
$ m_\chi \gg \Delta m \gg \omega \gg m_{\tilde \gamma}$. Larger values of 
$\Delta m$ would simply result in suppressed scattering rates; very small values, on the
other hand, would typically involve serious fine-tuning in concrete models (and, furthermore,
 in many  cases just lead to situations that are fully analogous to the $s/u$-channel 
 2-particle models discussed in the previous Section).

A complete list of the relevant models, along with results for the scattering matrix elements,
is given in Tab.~\ref{tab:Msq_3p}. Note that this time there appear no vector particles in this 
classification. This is because non-Abelian gauge bosons are not compatible with the imposed 
$Z_2$ symmetry, for the topology considered here, while Abelian 
gauge bosons only couple to a pair of \emph{identical} particles (appearing e.g.~in the situation 
studied in Sec.~\ref{sec:2pm}).
To leading order, the squared amplitudes are all of the
 form
\be
\label{3bsuMsq}
\langle |\mathcal{M}|^2 \rangle_t = \frac{r \eta_\chi g_\chi^4}{\delta^{2}} \left(\frac{\omega}{m_\chi}\right)^n,
\ee
where $n=0$ for scalar DR, and $n=2$ if $\tilde\gamma$ is a fermion. 
Here, $g_\chi$ denotes the $\chi$-$\chi'$-$\tilde\gamma$ coupling (divided by $m_\chi$
in the one case it is dimensionful, namely when all particles are scalars), $\delta$ is 
given by $\delta \equiv \Delta m /m_\chi$ and $r$ is a 
model-dependent constant with $1\leq r\leq16$.
Defining $M_{10}\equiv M_\mathrm{cut}/10^{10} M_\odot$, we can use the analytic 
expression (\ref{mcutsimp}) for the cutoff mass and find 
in terms of the parameters introduced above\footnote{%
The leading number refers to a bosonic $\tilde\gamma$, the one in parentheses to a 
fermionic $\tilde\gamma$. 
}
\bea
 M_{10}^{n=0} &\simeq& 8.4 \,(6.8)\, \xi^{6} \left(\frac{r g_\chi^4 }{\delta^2}\right)^\frac32 \left(\frac{m_\chi}{10\,\mathrm{GeV}}\right)^{-\frac92}, \label{mcutmax3b0}\\
 M_{10}^{n=2} &\simeq&  7.9 \,(7.7)\, \xi^\frac92 \left(\frac{r g_\chi^4 }{\delta^2}\right)^\frac34 \left(\frac{m_\chi}{10\,\mathrm{MeV}}\right)^{-\frac{15}4}. \label{mcutmax3b2}
 %\\
% M_{10}^{n=4} &\simeq&  1.4 (1.4)\, \xi^{4} \left(\frac{r g^4 }{\delta^2}\right)^\frac12\left(\frac{m_\chi}{1\,\mathrm{MeV}}\right)^{-\frac72}  \label{mcutmax3b4}
\eea

The main constraint on this type of models typically results from the requirement that
the pair-annihilation rate of $\chi$ should not be  so large that it would deplete the number 
density of $\chi$ below the cosmological abundance of DM\@. Following the discussion in
Sec.~\ref{sec:gen_ann}, we thus have to demand 
that $r^{1/4}g_\chi$ is smaller than the value of $g'$ in Eq.~(\ref{RDswave}) that is needed
for $\Omega_\chi h^2\simeq 0.119$.\footnote{%
Here, we do not include the $\delta$-dependence in the comparison of the effective 
coupling constants because this derives
from an on-shell enhancement that is absent for the annihilation process.
Note also that in models where $\chi'$ is close in mass to
$\chi$, coannihilations \cite{Edsjo:1997bg} become important. This will 
{\it increase} the effective annihilation rate during freeze-out, hence leading to a 
{\it stronger} constraint on $g$. The actual limits on $M_\mathrm{cut}$ are thus 
slightly more stringent than stated in Eqs.~(\ref{mcutmax3b0RD}, \ref{mcutmax3b2RD}) 
-- apart from models with $p$-wave rather than $s$-wave annihilation, where the additional 
factor of $x_f/3$ in Eq.~(\ref{RDswave}) has the opposite effect.
}
Using furthermore $x_f\propto \xi$, this leads to the following upper bounds on the cutoff 
mass:
\bea
 M_{10}^{n=0} &\lesssim& 0.9\,\xi^{\frac{15}2} \left(\frac{\delta}{0.01}\right)^{-3} \left(\frac{m_\chi}{10\,\mathrm{GeV}}\right)^{-\frac32}, \label{mcutmax3b0RD}\\
 M_{10}^{n=2} &\lesssim& 4\, \xi^\frac{21}4 \left(\frac{\delta}{0.01}\right)^{-\frac32} \left(\frac{m_\chi}{100\,\mathrm{keV}}\right)^{-\frac94}. \label{mcutmax3b2RD}
 %\\
% M_{10}^{n=4} &\lesssim& 8\, \xi^\frac92 \left(\frac{\delta}{0.01}\right)^{-1} \left(\frac{m_\chi}{10\,\mathrm{keV}}\right)^{-\frac52}
\eea
We stress that while the actual bounds are model-dependent, because the DM annihilation rate may
be dominated by other processes than $\chi\chi\to\tilde\gamma\tilde\gamma$, the above expressions
provide very useful order-of-magnitude estimates that allow to classify in which models
cutoff masses $M_\mathrm{cut}\sim\mathcal{O}\left(10^{10}M_\odot \right)$ can in 
principle be achieved.

For cases where the scattering rate is almost constant ($n=0$),  such large cutoffs
can relatively easily be obtained for DM masses up to around 10 GeV (or even 
larger DM masses if one is willing to accept a fine tuning between $m_\chi$ and $m_{\chi'}$ 
beyond the percent level). We note that Eq.~(\ref{mcutmax3b0}) reproduces as expected 
the result for a scalar four-point coupling that we earlier derived in Eq.~(\ref{mcut_SSSS}), 
after replacing $r \eta_\chi g_\chi^4/\delta^2\to\lambda^2$. The different conclusions about the 
maximal 
mass scale of the DM particles in these cases ($\sim$$10$\,GeV vs.~$\sim$$1$\,MeV) arise 
thus exclusively due to the on-shell enhancement resulting from $\delta\ll1$. For an example
similar to this type of model, where a fermionic DM particle interacts with a fermionic mediator and 
{\it pseudo}scalar DR particles, see Ref.~\cite{Chu:2014lja} (but note that in this case the model 
contains a further scalar $t$-channel mediator, as discussed in Section IV B).

For cases with $n=2$, i.e.~for a fermionic $\tilde\gamma$, viable models in the above sense 
are restricted to a small range of sub-MeV DM masses (for an example of such a model, 
where fermionic DM couples to neutrinos via a scalar, see Ref.~\cite{Bertoni:2014mva}). 
Similar to the situation discussed in Section~\ref{sec:scatter_point}, the mass range of
interest extends to $m_\chi\lesssim100$\,keV, where free streaming effects
have to be taken into account. In this regime, $\xi$ can hence be chosen small
enough to avoid any tension with CMB data and yet suppress the power spectrum
as desired.

It is very interesting to 
note that \emph{all} models discussed in this Section are in principle viable, if only for a
relatively small range of DM masses and mediator particles that are highly degenerate
in mass with the DM particles. None of these models, on the other hand, gives  naturally 
rise to large DM self-interaction rates. Similar to the case of the simple scalar
4-point interaction discussed above, those would have to be added by hand.

%%%%%%%%%%%%%%%%%%%%%%%%%%%%%%%%%%%%%%%%%%%%
\subsection{Scattering via $t$-channel}
\label{sec:3pscatter_t}

In this Section,  we consider models where we add a light bosonic particle 
$\tilde \gamma'$ to mediate the interaction between $\chi$ and  $\tilde \gamma$
via a $t$-channel diagram. 
For simplicity we only consider models with the following hierarchy of energy scales: 
$m_\chi \gg m_{\tilde\gamma'} \gg \omega \gg m_{\tilde \gamma}$. This ensures that
we are sufficiently far away from the situation discussed in the 2-particle case, while 
retaining the possibility of large scattering rate enhancements through an 
almost on-shell mediator particle $\tilde\gamma'$.

We provide a complete list of the relevant models, as well as results for the scattering 
matrix elements, in Tab.~\ref{tab:Msq_3p}. For a dimensionful
$\tilde\gamma$-$\tilde\gamma$-$\tilde\gamma'$ coupling $\mu_{\tilde\gamma}$, the scattering amplitudes
are always constant, to leading order, and given by
\be
\label{3btn0}
\langle |\mathcal{M}|^2 \rangle_t = r \eta_\chi g_\chi^2 \left(\frac{\mu_{\tilde\gamma}}{m_{\tilde \gamma'}}\right)^2 \left(\frac{m_\chi}{m_{\tilde \gamma'}}\right)^2.
\ee
Otherwise, they take the form
\be
\label{3btn2}
\langle |\mathcal{M}|^2 \rangle_t = r \eta_\chi g_\chi^2 g_{\tilde\gamma}^2 \left(\frac{m_\chi}{m_{\tilde \gamma'}}\right)^4 \left(\frac{\omega}{m_\chi}\right)^2.
\ee
Here, $g_\chi$ denotes the $\chi$-$\chi$-$\tilde\gamma'$ coupling (divided by $m_\chi$ in 
cases where it is dimensionful), and
$g_{\tilde\gamma}$ denotes the $\tilde\gamma$-$\tilde\gamma$-$\tilde\gamma'$ coupling;
$r$ is a 
model-dependent constant in the range $1\leq r\leq128/3$.
 In all these cases, the form of the amplitude allows us to  use the analytic expression 
(\ref{mcutsimp}) for the cutoff mass $M_{\textrm{cut}}$. 
For the constant amplitude, Eq.~(\ref{3btn0}), this leads to 
\bea
 M_{10}^{n=0} &\simeq& 8.4 \,(6.8)\, \xi^{6} r^\frac32 \left(\frac{g_\chi\mu_{\tilde\gamma}}{m_{\tilde\gamma'}}\right)^3 \left(\frac{m_{\tilde\gamma'}}{\mathrm{GeV}}\right)^{-3}  \left(\frac{m_\chi}{\mathrm{TeV}}\right)^{-\frac32},\nonumber\\
\label{mcut3btn0}
\eea
while for the $n=2$ case, Eq.~(\ref{3btn2}), the resulting cutoff mass becomes 
\bea
 M_{10}^{n=2} &\simeq& 1.4 \,(1.4)\, \xi^\frac92 \left(r g_\chi^2 g_{\tilde\gamma}^2\right)^\frac34 \left(\frac{m_{\tilde\gamma'}}{\mathrm{MeV}}\right)^{-3} \left(\frac{m_\chi}{\mathrm{TeV}}\right)^{-\frac34}.\nonumber\\
\label{mcut3btn2}
\eea

 %%%%%%%%%%%%%%%%%%%%%%%%%%    
 \begin{figure}[t!]
     \includegraphics[width=\columnwidth]{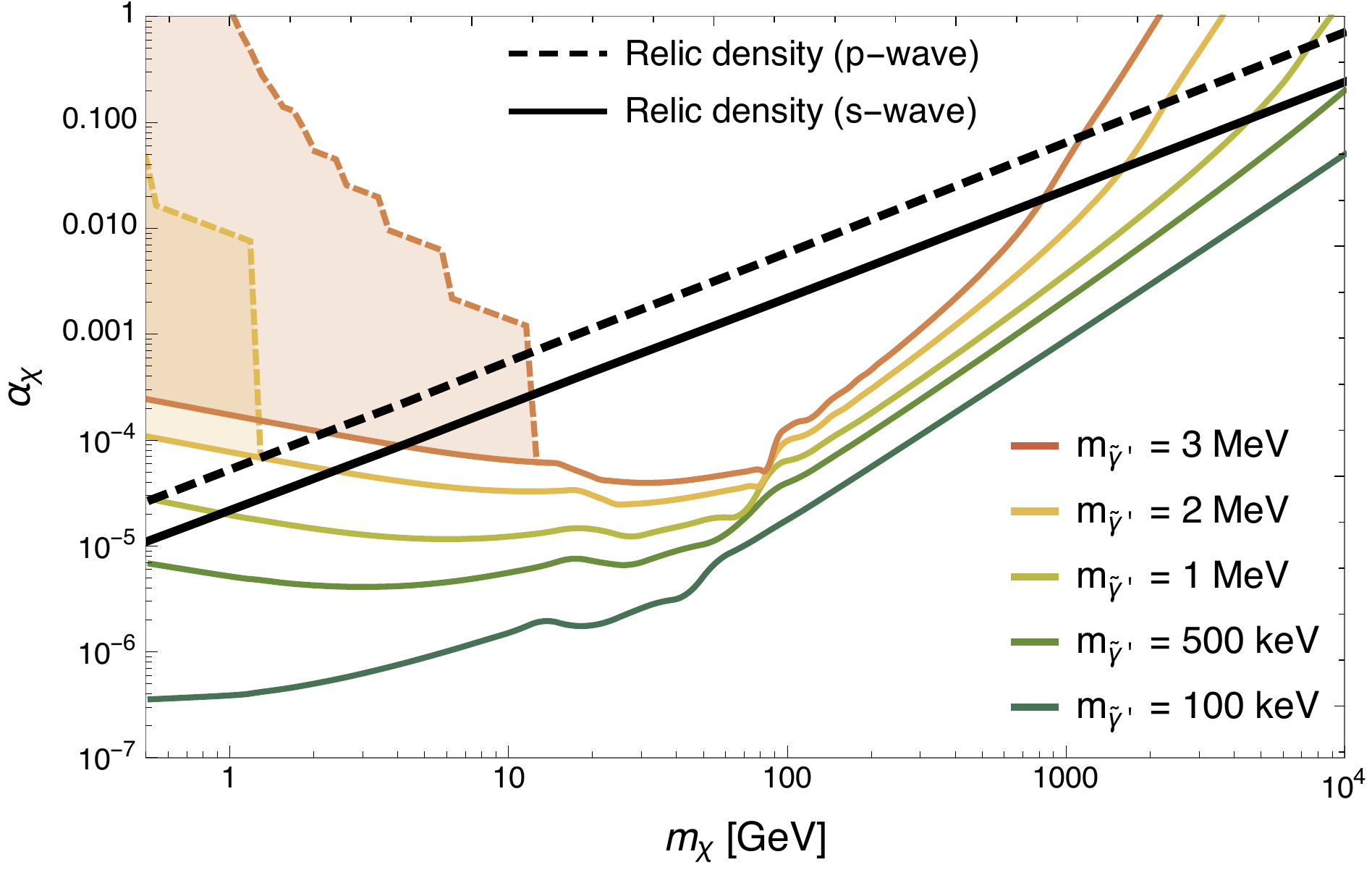}    
     \caption{Combinations of $\alpha_\chi$ and $m_\chi$ that lead to 
     $\langle\sigma_T\rangle_{30}/m_\chi=1\,\mathrm{cm}^2/\mathrm{g}$, for various mediator masses 
     $m_{\tilde \gamma'}$. The shaded areas show the range of parameters where the appearance of
     resonances in $\sigma_T$ allows multiple solutions to this condition.
     For comparison, we also show the value of $\alpha_\chi$ that results
     in the correct relic density when considering only the process $\chi\chi\to\tilde\gamma'\tilde\gamma'$,
     assuming that $\xi=0.5$ (see text for further details).
     \label{fig:tRDvsSIDM}
     }
 \end{figure}        
 %%%%%%%%%%%%%%%%%%%%%%%%%%    

An important phenomenological difference of these models, as compared to the 3-particle models in 
the $s/u$-channel, is that the light mediator $\tilde\gamma'$ will mediate a significant velocity-dependent 
DM self-interaction. Because we now have the freedom to choose $m_{\tilde\gamma'}\gg m_{\tilde\gamma}$,
the DM self-interaction rate can be be much more easily arranged to be in an observationally relevant 
range (e.g.~such as to mitigate the $\Lambda$CDM small-scale problems). In fact, this can be done while 
at the same time allowing for thermally produced DM\@. In Fig.~\ref{fig:tRDvsSIDM} we illustrate this point 
by plotting
the value of $\alpha_\chi=g_\chi^2/4\pi$ as a function of $m_\chi$ that is required to obtain 
$\langle \sigma_T\rangle_{30}/m_\chi=1\,\mathrm{cm}^2/\mathrm{g}$, for various values of 
$m_{\tilde\gamma'}$.  For large DM masses, $m_\chi\gtrsim100$\,GeV in the plot, we are in the classical 
regime for $\sigma_T$; for DM masses below about $10$\,GeV and small values of $\alpha_\chi$, we are instead
in the Born regime. The small jumps that are visible in between are not physical but result from 
the fact that the parameterizations that we adopt here do not connect the various regimes described in 
Section \ref{sec:gen_self} in a perfectly smooth way. Once the ratio of mediator to DM mass becomes
large enough, strong resonances develop in $\sigma_T$. As indicated by shaded areas in 
Fig.~\ref{fig:tRDvsSIDM}, this allows multiple solutions to 
$\langle \sigma_T\rangle_{30}/m_\chi=1\,\mathrm{cm}^2/\mathrm{g}$. Here,  the steps in the upper 
envelopes of these shaded areas reflect the number of resonances where this condition can be met.

In the same figure we show, for comparison, the value of $\alpha_\chi$ that follows
from Eq.~(\ref{RDswave}) when assuming that the process $\chi\chi\to\tilde\gamma'\tilde\gamma'$ 
proceeds with a rate of $\sigma v=\pi\alpha_\chi^2/2 m_\chi^2$ and is fully responsible for setting the correct 
relic density (labelled `$s$-wave'). We also show the case of $\sigma v=\pi\alpha_\chi^2 v^2/2 m_\chi^2$
(labelled  `$p$-wave'). We note that a more accurate treatment would depend on the concrete model. 
For a fermionic DM particle $\chi$ annihilating to a vector $\tilde\gamma'$ ($s$-wave) or scalar 
$\tilde\gamma'$ 
($p$-wave), for example, the actual annihilation rate at lowest order is larger by a factor of $2$ and $3/2$, 
respectively (this corresponds to models \ref{L3_T_FSS} to \ref{L3_T_FVF} in Appendix \ref{app:msq}).
Accordingly, the $s$-wave ($p$-wave) line would move downwards by about 30\% (20\%).
In general, the annihilation rate also receives an enhancement due to the Sommerfeld effect. 
Compared to what is shown in the figure, this will result in a slightly smaller value of $\alpha_\chi$ 
that is necessary to achieve the correct relic density; even in the vicinity of resonances, however, this is 
only an $\mathcal{O}(1)$ effect \cite{Feng:2010zp} which again would hardly be visible at the resolution 
given here.

Even though the details are somewhat model-dependent, Fig.~\ref{fig:tRDvsSIDM} clearly illustrates that for 
mediator masses
$m_{\tilde\gamma'}\gtrsim1$\,MeV and DM masses in the TeV range it is in general possible to
accommodate thermal DM production and a DM self-interaction rate that is sufficiently large to visibly 
affect the inner structure of subhalos at the scale of dwarf galaxies.  Let us now investigate the 
consequences for the cutoff in the power spectrum in this regime, assuming again that the above $s$-wave 
annihilation cross section is responsible for setting the relic density. This fixes $g_\chi$ in 
Eqs.~(\ref{mcut3btn0}, \ref{mcut3btn2}), which thus become
\bea
 \!\!M_{10}^{n=0} &\simeq& \xi^\frac{27}{4} r^\frac32 
 \!\left(\frac{\mu_{\tilde\gamma}}{m_{\tilde\gamma'}}\right)^3 \left(\frac{m_{\tilde\gamma'}}{\mathrm{GeV}}\right)^{-3} \!\left(\frac{x_f}{g_\mathrm{eff}(T_\mathrm{cd})} \right)^\frac34\!\!,\label{mcutmax3b0tRD}\\
 \!\!M_{10}^{n=2} &\simeq& \frac12  \xi^\frac{39}{8} r^\frac34 
g_{\tilde\gamma}^\frac32 \left(\frac{m_{\tilde\gamma'}}{\mathrm{MeV}}\right)^{-3}
 \left(\frac{x_f}{g_\mathrm{eff}(T_\mathrm{cd})} \right)^\frac38. \label{mcutmax3b2tRD}
\eea
Note that now there is only a very weak dependence of the cutoff on the DM mass, through $x_f$ and 
$g_\mathrm{eff}$ (as well as the Sommerfeld effect, see \cite{Aarssen:2012fx,Bringmann:2013vra} for 
examples). For MeV mediators and couplings $\alpha_{\tilde\gamma}\sim\alpha_\chi\sim10^{-2}$, thermally
produced DM can thus lead to $M_{10}\sim1$ in the $n=2$ case, implying in particular that for 
$m_\chi\sim\text{TeV}$ a simultaneous solution of {\it all} small scale problems of $\Lambda$CDM is possible.
A similar phenomenology is obtained for $n=0$, i.e.~for models with dimensionful
$\tilde\gamma$-$\tilde\gamma$-$\tilde\gamma'$ couplings, if one adopts 
$\mu_{\tilde\gamma}\sim10^{-3}\,m_{\tilde\gamma}$.
Indeed, the $n=2$ case corresponds exactly to the situation first described in
Ref.~\cite{Aarssen:2012fx}, and followed up by several concrete examples for model-building
with vector mediators and fermionic  DM and fermionic DR 
\cite{Bringmann:2013vra,Dasgupta:2013zpn,Ko:2014bka,Cherry:2014xra} as well as scalar DR 
\cite{Chu:2014lja}. 
As one can see from this discussion, however, there exists a rather large variety of models  
that fall into this class, including the possibility of scalar mediators. The possibilities for future 
model building that we have pointed out here thus go clearly beyond the specific 
settings considered so far.

The very large values of the tree-level scattering amplitude we need for TeV-scale DM
particles, see also Fig.~\ref{fig:cn10}, may lead to worries about the reliability of the 
calculation, since higher-order corrections could be important.
Therefore, we calculated the full one-loop correction arising from the exchange of one 
additional scalar mediator for fermionic DM and scalar DR using 
\textsf{LoopTools} \cite{Hahn:1998yk}. It 
turned out that this correction can safely be neglected.
This can be traced back to the fact that the DR particle in the loop is highly virtual since the 
change of 4-momentum upon entering the loop is of order 
$\sqrt{|t|} \sim T_\text{kd} \gg m_{\tilde\gamma}$.
The kinematical situation is thus different from the scattering or annihilation of non-relativistic 
particles, which remain nearly on-shell when exchanging one or more light mediators and thus 
experience Sommerfeld enhancement \cite{Hisano:2004ds,Iengo:2009ni,ArkaniHamed:2008qn}.

To conclude this section, let us point out that there is yet another class of thermally produced 
DM models, visible in the {\it low-mass} part of  Fig.~\ref{fig:tRDvsSIDM}, where the self-
interaction rate is at the right level to
potentially address the cusp-core or the too-big-to-fail problem. In contrast
to the class of solutions discussed in the previous paragraph, here the transfer cross section $\sigma_T$ 
is either in the Born or in the resonant regime. Given that the cutoff is almost independent of 
$m_\chi$ for all thermally produced models considered in this section, however, the same 
conditions on $m_{\tilde\gamma}$ and $g_{\tilde\gamma}$ (or $\mu_{\tilde\gamma}$) as just 
discussed above will lead to $M_{10}\sim1$ -- though of course the different values of $\alpha_\chi$
and $m_\chi$ will lead to different requirements for concrete model building.
We have thus identified a whole new class of GeV DM models
that could potentially address all $\Lambda$CDM small-scale problems simultaneously. We leave
a more detailed investigation of the expected rich phenomenology as an interesting
direction for future work.

%%%%%%%%%%%%%%%%%%%%%%%%%%%%%%%%%%%%%%%%%%%%
\section{Conclusions}
\label{sec:conc}
%%%%%%%%%%%%%%%%%%%%%%%%%%%%%%%%%%%%%%%%%%%%

If cold DM is kept in local thermal equilibrium with a relativistic species (`dark radiation', DR) until the 
universe has cooled down to temperatures below $\sim1$\,keV, this results in a characteristic 
suppression of the power spectrum of matter density fluctuations for scales below what corresponds 
roughly to the size of the smallest dwarf galaxies. 
Such a cutoff may help to alleviate the problem of missing satellites in the cosmological 
concordance model. More importantly, it provides quite in general a fascinating way of  probing
new particle physics in the dark sector by using astrophysical observables connected to the 
distribution of  cosmological structure. This type of probe is thus highly complementary to
traditional attempts to identify the particle nature of DM.

In this article, we have provided a systematic classification of the minimal model-building options
that allow for such a scenario. The simplest solution turns out to be a contact interaction 
between a DM particle with $m_\chi\lesssim1$\,MeV and a relativistic DR particle, either in the 
form of a 4-point `portal' interaction 
between two scalars or via a suppressed, higher-dimensional operator (Section \ref{sec:scatter_point}).
Scenarios where DM couples via a 3-point coupling to DR, on the other hand, are severely constrained 
by observational
bounds on the strength of DM self-interactions, leaving no room for a sufficiently late kinetic 
decoupling (Section \ref{sec:scatter_su}). This problem may be circumvented by allowing for a mediator 
particle that is slightly
heavier than DM (Section \ref{sec:3pscatter_su}) or lighter than DM but
significantly heavier than DR  (Section \ref{sec:3pscatter_t}). In the first class of models, DM
cannot be too heavy (typically $m_\chi\lesssim\mathcal{O}(10\,\mathrm{GeV})$); in the second 
class, the possibility to get an observable cutoff for thermally produced DM turns out to be 
almost independent of the DM mass.
In Appendix \ref{app:msq}, we provide the corresponding ETHOS 
\cite{Cyr-Racine:2015ihg,Vogelsberger:2015gpr} parameters of our simplified particle physics 
models; similar
parameters will result in almost identical results when performing full numerical simulations of 
structure formation for such DM candidates.

Within the classes of models considered, we do not find examples where photons could play the
role of the `dark' radiation component to achieve sufficiently late kinetic decoupling. As discussed 
in Appendix \ref{app:msq3},
this may change to some extent if higher-dimensional operators are included in the discussion of 
possible interactions between the mediator and the radiation component. The fact that the left-handed
leptons of the standard model are contained in $SU(2)$ doublets makes it furthermore challenging to 
construct models where late kinetic decoupling can be achieved with (active) neutrinos,
see again Appendix \ref{app:msq3} for a discussion.

The main phenomenological difference to WDM scenarios, which lead to a similar cutoff in the 
power spectrum, is that  in particular the class of models featuring $t$-channel mediators 
much lighter than DM (and much heavier than DR) gives naturally rise to relatively large DM 
self-interaction rates. For a few concrete models with TeV scale DM particles and MeV scale mediators, 
it has been noticed before that this fact can 
be used to simultaneously alleviate \emph{all} small-scale problems of $\Lambda$CDM 
cosmology for thermally produced DM\@. We have not only demonstrated that the models studied 
so far fall into a much broader class of viable solutions with this property, but also identified 
a new class of GeV scale DM models with similar properties (Section \ref{sec:3pscatter_t}). This 
opens promising and largely unexplored model-building avenues.

We have furthermore shown that the cubic self-interaction of a scalar 
DR particle makes the DM-DR interaction increasingly efficient for small energies 
(Section \ref{sec:scatter_t}). In this case, DM and DR would inevitably be in local thermal equilibrium
at late times, though not necessarily at early times. If the dark `radiation' particles are instead 
massive, with $m_{\tilde\gamma}\sim1$\,keV, they will however decouple around the same time as in the other
cases discussed here. For (sub-)MeV DM, such a scenario would in fact also allow for thermally
produced DM with self-interaction rates in the observationally relevant rate (similar to the case of 
non-Abelian dark `radiation', see Section \ref{sec:mixed}). We leave a more detailed
investigation of this interesting observation for future work.

%%%%%%%%%%%%%%%%%%%%%%%%%%%%%%%%%%%%%%%%%%%
\vfill
\section*{Acknowledgements}
%%%%%%%%%%%%%%%%%%%%%%%%%%%%%%%%%%%%%%%%%%%

We would like to thank Tobias Binder, Thomas Hahn, Jasper Hasenkamp, Andrzej Hryczuk, J\"org J\"ackel,
Felix Kahlhoefer, Oleg Lebedev and Kai Schmidt-Hoberg for very fruitful discussions.

\bigskip
{\bf Note added.} While finalizing this manuscript, two studies appeared on the arXiv
that independently identified some of the new scenarios for very late kinetic decoupling 
that we have described and classified here. In particular, T.~Binder \emph{et al.} 
\cite{Binder:2016pnr} pointed out that the fermion/scalar/fermion combination shown as
entry \ref{L3_T_FSF} in Table \ref{tab:Msq_3p} provides such a new solution, while
Y.~Tang \cite{Tang:2016mot} identified interacting scalar DR and fermionic DM as a further 
possibility (see the fermion/scalar case in Table \ref{tab:Msq_2p}, and the discussion in 
Section \ref{sec:scatter_t}).

\newpage
\appendix

%%%%%%%%%%%%%%%%%%%%%%%%%%%%%%%%%%%%%%%%%%%%
\section{Kinetic decoupling}
\label{app:kd}
%%%%%%%%%%%%%%%%%%%%%%%%%%%%%%%%%%%%%%%%%%%%
The kinetic decoupling of DM particles from a thermal bath can be described from
first principles by solving the underlying Boltzmann equation 
\cite{Bertschinger:2006nq,Bringmann:2006mu} -- just as the standard way of calculating 
the relic density of thermally produced DM particles \cite{Gondolo:1990dk} is based on 
solving the Boltzmann equation during an earlier epoch of chemical decoupling.
The original formalism \cite{Bringmann:2006mu} was later extended to non-relativistic 
scattering partners \cite{Bringmann:2009vf} that may have a temperature differing from 
that of photons \cite{Aarssen:2012fx, Bringmann:2013vra}, situations in which the DM
number density or the  effective number of relativistic degrees of freedom can change 
during  or after decoupling \cite{Bringmann:2009vf,vandenAarssen:2012ag} and, most 
recently, to the case where the scattering amplitude is not Taylor expandable around small
momentum transfer \cite{KasaharaPHD,Gondolo:2012vh}. Here, we provide a brief 
summary taking into account these more recent developments.
% (see also Ref.~\cite{Bringmann:2006mu} for a pedagogic review).

Consider a particle $\tilde\gamma$ with a thermal distribution $g^\pm$ of temperature 
$T_{\tilde\gamma}$, and a non-relativistic DM particle $\chi$ that can interact with 
$\tilde\gamma$. The Boltzmann equation that governs the evolution of the DM 
phase-space distribution $f$ in an expanding Friedmann-Robertson-Walker universe is 
then given by $L[f] = C[f]$, with the Liouville operator\footnote{%
All momenta that appear in these expressions are physical, as opposed to co-moving,
and the time-dependence of $f$ is understood to arise, to leading order, exclusively 
from the expansion of the universe via $\mathbf{p}\propto 1/a$ (where $a$ is the 
scalefactor). \\
Note further that we are throughout using conventions for the normalization of quantum
fields and their interactions that are consistent with those of Peskin \& Schroeder \cite{Peskin:1995ev}.
}
\be
 L[f]=E\left(\partial_t - H \mathbf{p}\cdot\partial_\mathbf{p} \right) f(\mathbf{p})\,,
\ee
and a collision term 
\bea
  \label{Cfull}
   C[f]&=&\frac{1}{2\eta_\chi}\int\frac{d^3k}{(2\pi)^32\omega}
   \int\frac{d^3\tilde k}{(2\pi)^32\tilde \omega}\int\frac{d^3\tilde p}{(2\pi)^32\tilde E} \\
   && \times (2\pi)^4\delta^{(4)}(\tilde p+\tilde k-p-k)
   \left|\mathcal{M}\right|^2_{\chi\tilde\gamma\leftrightarrow\chi\tilde\gamma} \nonumber\\
  &&  \left[(1\!\mp\!g^\pm)(\omega)\, g^\pm(\tilde\omega)
  f(\mathbf{\tilde p})-(1\!\mp\!g^\pm)(\tilde\omega)\, g^\pm(\omega)f(\mathbf{p})\right].\nonumber
\eea
Here, $H=\dot a/a$ is the Hubble parameter, and $(E,\mathbf{p})$ and 
$(\omega,\mathbf{k})$ are the 4-momenta of the incoming 
particles $\chi$ and $\tilde\gamma$, respectively (outgoing momenta are denoted with a 
tilde). The scattering amplitude $|\mathcal{M}|^2$ is {\it summed} over all internal degrees 
of freedom, and the phase-space densities are normalized such that, e.g., the number 
density of the particle $\chi$ is given by $n_\chi=\eta_\chi\int d^3p\, f(\mathbf{p})/(2\pi)^3$. 

Even when the DM particle is no longer in local thermal equilibrium, one can now {\it 
define} a parameter
\be
 T_\chi\equiv\frac{\eta_\chi}{3m_\chi n_\chi}\int\frac{d^3p}{(2\pi)^3}\mathbf{p}^2f(\mathbf{p})\,.
\ee
Introducing further the dimensionless parameters
\bea
  \label{xdef}
  x&\equiv& m_\chi/T\,, \\
    \label{ydef}
 y&\equiv&{m_\chi T_\chi}{s^{-2/3}}\,,
%   \label{xidef}
% \xi&\equiv& T_{\tilde\gamma}/T\,,
\eea
the 2nd moment of the full Boltzmann equation, keeping only leading terms in 
$\mathbf{p}^2/m_\chi^2$, reduces to\footnote{%
Note that this assumes a constant comoving DM particle number density  
during and after kinetic decoupling. Otherwise an additional term must be added to this 
equation that couples the evolution of $T_\chi$ to that of $n_\chi$ 
\cite{vandenAarssen:2012ag}. This can be relevant, e.g., in the presence of resonances or
a strong Sommerfeld enhancement of the DM annihilation rate.
}
\be
\label{dydx}
\frac{d\log y}{d\log x}=\left(1-\frac{1}{3}\frac{d \log g_{*S}}{d\log x}\right)
\frac{\gamma(T_{\tilde\gamma})}{H(T)}\left(\frac{y_\mathrm{eq}}{y}-1\right)\,.
\ee
Here, $s$ is the total entropy density, $g_{*S}$ are the effective entropy degrees of 
freedom of {\it all} relativistic particles in the universe and $y_\mathrm{eq}$ is given by 
Eq.~(\ref{ydef}) with $T_\chi\to T_{\tilde\gamma}$. The momentum transfer rate 
$\gamma$, finally, is given by
\bea
\label{fT}
 \gamma (T_{\tilde\gamma})&=&\frac{1}{48\pi^3\eta_\chi T_{\tilde\gamma}m_\chi^3}\\
 &&\times
 \int d\omega\,k^4
 \left(1\mp g^\pm\right)g^\pm(\omega)
  \mathop{\hspace{-12ex}\left|\mathcal{M}\right|^2_{t=0}}_{\hspace{4ex}s=m_\chi^2+2m_\chi\omega+m_\mathrm{\tilde\gamma}^2}\,, \nonumber
\eea
where $k\equiv\left| \mathbf{k}\right|$.
The above expression only holds if $|\mathcal{M}|^2$ is Taylor expandable around $t=0$
(in the sense that $|\mathcal{M}|^2=|\mathcal{M}|^2_{t=0}\left[1+\mathcal{O}(\omega^2/m_\chi^2)\right]$,  
taking into account that $t$ is of the same order as $\omega^2$). 
While this is typically a good assumption, it fails for example if the denominator 
is suppressed by $\omega$ or $t$ (because the propagator is almost on shell).
In such situations, we have to make the replacement \cite{KasaharaPHD} 
\be
\label{maverage}
 \mathop{\hspace{-12ex}\left|\mathcal{M}\right|^2_{t=0}}_{\hspace{4ex}s=m_\chi^2+2m_\chi\omega+m_\mathrm{\tilde\gamma}^2}
\!\!\! \longrightarrow
\left<\left|\mathcal{M}\right|^2\right>_t 
\!\!\equiv \frac{1}{8k^4}\int_{-4k^2}^0
\!\!\!\! dt(-t)\left|\mathcal{M}\right|^2.
\ee
This allows us to re-write $\gamma (T_{\tilde\gamma})$ in terms of the total 
transfer cross section, $\sigma_T\equiv\int d\Omega\,(1-\cos\theta)d\sigma/d\Omega$,
as
\be
\label{gammast}
\gamma (T_{\tilde\gamma})=\frac{1}{3\pi^2 \eta_\chi m_\chi}
 \int d\omega\, g^\pm(\omega)\,\partial_\omega\!\left(k^4\sigma_T\right)\,,
\ee
where we have used that 
$g^\pm(1\mp g^\pm)(\omega)=-T_{\tilde\gamma}\partial_\omega g^\pm(\omega)$.

The solution to Eq.~(\ref{dydx}) before and after DM leaves local thermal equilibrium with 
the $\tilde\gamma$ particles is of the form
\be
\label{Tchisimp}
T_\chi(T)=\left\{
\begin{array}{cc}
T_{\tilde\gamma} & \mathrm{for}~T\gtrsim T_\mathrm{kd} \\
 \mathcal{C}/a^2& \mathrm{for}~T\lesssim T_\mathrm{kd}
\end{array}
\right.
\ee
with a constant $\mathcal{C}$ that is uniquely determined by the solution of the 
differential equation. Given that the transition between the two regimes typically 
happens rather fast, it is natural to {\it define} the kinetic decoupling temperature 
as the point where the two asymptotics meet (see Fig.~1 in 
Ref.~\cite{Bringmann:2009vf}). This is equivalent to re-writing $\mathcal{C}$, and hence 
Eq.~(\ref{Tchisimp}), as
\be
\label{Tchicorr}
\,T_\chi(T)=\left\{
\begin{array}{cc}
T_{\tilde\gamma}(T) & \mathrm{for}~T\gtrsim T_\mathrm{kd} \\
 \!\xi T_\mathrm{kd} \left[a(T_\mathrm{kd})/a(T)\right]^2& \mathrm{for}~T\lesssim T_\mathrm{kd}
\end{array}
\right. 
\ee
where we have introduced
%comment on $T$-dependence of this quantity...
\be
\label{xidef}
\xi\equiv T_{\tilde\gamma}/T\,.
\ee
%In practice, one solves Eq.~(\ref{dydx}) numerically starting from the boundary condition 
%$y=y_\mathrm{eq} T$ at early time and follows the solution to small temperatures until % $XXX$ converges to a constant value (equal to $T_\mathrm{kd}$).
Other definitions of the kinetic decoupling temperature exist in the literature (see 
e.g.~\cite{Profumo:2006bv,Shoemaker:2013tda,Cyr-Racine:2013fsa, Cherry:2014xra,Visinelli:2015eka}), 
for example requiring 
that $\gamma=H$ at the time of kinetic decoupling, 
which all are related to the definition advocated here by multiplying the small temperature 
regime of Eq.~(\ref{Tchicorr}) with a constant different from unity.

The most prominent observable connected to kinetic
decoupling is that of a cutoff in the power spectrum of matter density perturbations.
% NON-LINEAR!? -> CHECK!
For very late kinetic decoupling, the dominant mechanism of suppressing the growth
of DM perturbations are dark acoustic oscillations 
\cite{Loeb:2005pm,Bertschinger:2006nq} (unless DM is very light, in which case 
free-streaming effects \cite{Green:2005fa} start to dominate).
As recently 
confirmed numerically  \cite{Vogelsberger:2015gpr}, the resulting minimal halo mass
is then given by Eq.~(\ref{Mcut}), which is in rather good agreement with earlier analytic 
estimates. Note that 
$T_\mathrm{kd}$ in this expression is calculated by using the definition given by 
Eq.~(\ref{Tchicorr}); for an alternative definition, the expected magnitude of 
$M_\mathrm{cut}$ has to be correspondingly re-scaled.

Let us conclude this Section by making explicit how the above general analysis 
simplifies for the purpose of the specific application we are interested in for most of this article: 
DM scattering
with a highly relativistic species, resulting in kinetic decoupling in the keV range. The 
latter implies that we are still well in the radiation dominated era, 
$H^2=(4\pi^3G/45)g_\mathrm{eff}T^4$, with a constant  
number of effective relativistic degrees of freedom
$g_\mathrm{eff}=3.36$.\footnote{%
If $\tilde\gamma$ constitutes some form of {\it dark} radiation, this would in principle
contribute additional degrees of freedom on top of those from the standard model 
neutrinos and photons taken into account here. Such an additional contribution is
observationally strongly constrained \cite{Ade:2015xua}, and would anyway change the 
prediction for $T_\mathrm{kd}$ only by a factor $\lesssim \frac14 \Delta g_\mathrm{eff}/g_\mathrm{eff}$, cf.~Eq.~(\ref{tkd_analytic}).
}
Eq.~(\ref{dydx}) then becomes
\be
\label{tchidt}
\frac{dT_\chi}{dT}-2\frac{T_\chi}{T}=\left({T_\chi} -\xi T\right)
\frac{f(T_{\tilde\gamma})}{T^3}\,,
\ee
where
\bea
\label{fDR}
 f(T_{\tilde\gamma})&=&  \frac{3\sqrt{5 /\pi}}{2\pi g_\mathrm{eff}^{1/2}} {M_\mathrm{Pl}}\,   \gamma (T_{\tilde\gamma})
 \\
 &=&\frac{\sqrt{5 /\pi}}{2(2\pi)^4}\frac{M_\mathrm{Pl}}{g_\mathrm{eff}^{1/2}\eta_\chi m_\chi^3}
 \int d\omega\, g^\pm\partial_\omega
 \left( \omega^4 \left<\left|\mathcal{M}\right|^2\right>_t \right).\nonumber
\eea
As in Eq.~(\ref{maverage}), we can evaluate the amplitude at $t=0$ instead of taking the
average {\it if} the Taylor series around this point locally provides a good approximation. 
In general, the above two equations need to be solved numerically to determine 
$T_\mathrm{kd}$ according to Eq.~(\ref{Tchicorr}), as implemented in 
\ds~\cite{DSweb,Bringmann:2009vf}. In many cases of practical interest, the amplitude is 
furthermore well approximated by a power law for small energies,\footnote{\label{foot:average}%
Note that this introduces the coefficient $c_n$ with the correct prescription of summing and
averaging the amplitude squared over initial and final states, in the sense that this is how it
enters in the momentum transfer rate, see Eqs.~(\ref{fT}) and (\ref{fDR}).
}
\be
\label{powerlaw}
\frac1{\eta_\chi} %\sum_{\mathrm{all~states}}
\left<\left|\mathcal{M}\right|^2\right>_t = c_n\frac{\omega^n}{m_\chi^n}
 +\mathcal{O}\left( \frac{\omega^{n+1}}{m_\chi^{n+1}} \right).
\ee
In this case, Eqs.~(\ref{tchidt},\ref{fDR}) can be solved analytically even for non-integer 
$n>-1$  
\cite{Bringmann:2006mu}, and the kinetic decoupling temperature as defined in 
Eq.~(\ref{Tchicorr}) is given by
\be
\label{tkd_analytic}
\frac{T_\mathrm{kd}}{m_\chi} = 
\left(\frac{\xi T^2}{m_\chi T_\chi}\right)_{T\lesssim T_\mathrm{kd}}
=\left[\left(\frac{a}{n+2}\right)^{1/(n+2)}\Gamma\left(\frac{n+1}{n+2}\right)\right]^{-1},
\ee
with
\be
\label{eq:a}
a= \sqrt{\frac{5}{2(2\pi)^9g_\mathrm{eff}}}
(n+4)!\,\zeta(n+4)
\, \xi^{n+4}c_n\frac{M_\mathrm{Pl}}{m_\chi}
\ee
for a bosonic $\tilde\gamma$. If $\tilde\gamma$ is a fermion, the above expression has to 
be multiplied by a factor of $1-2^{-(n+3)}$.

%%%%%%%%%%%%%%%%%%%%%%%%%%%%%%%%%%%%%%%%%%%%
\section{Scattering matrix elements}
\label{app:msq}
%%%%%%%%%%%%%%%%%%%%%%%%%%%%%%%%%%%%%%%%%%%%

In this Appendix, we provide the Lagrangians and scattering amplitudes for all 
models included in our analysis. 
While we use the full expressions to calculate the kinetic decoupling temperature, we
state here only the leading terms for $\left|\mathcal{M}\right|^2$ in the limit 
$m_\chi \gg \omega \gg m_{\tilde \gamma}$. In each case, we also check explicitly 
whether keeping only these leading order terms provides a good estimate for the 
calculation of $T_\mathrm{kd}$, and whether simply evaluating $\left|\mathcal{M}\right|^2$ 
for $t=0$ leads to a reliable estimate of $T_\mathrm{kd}$ or whether 
one has instead to use the $t$-averaging prescription given in Eq.~(\ref{maverage}).
See Appendix \ref{app:kd} for more details about how to calculate $T_\mathrm{kd}$.

%%%%%%%%%%%%%%%%%%%%%%%%%%%%%%%%%%%%%%%%%%%%
\subsection{2-particle models}
Let us first consider those simplified models that only contain the (cold) DM particle 
$\chi$ and the (relativistic) scattering partner $\tilde\gamma$. As motivated in Section 
\ref{sec:2pm}, we are then interested in the following interaction terms 
(to indicate the spin of the involved particles, we denote scalars 
always with $\phi$, vectors with $V$ and fermions with $\psi$).

\begin{enumerate}
\item {\it Scalar four-point interaction}
\be
\label{L_SSSS}
\Delta \mathcal{L} = \frac{\lambda}{4} \phi_\chi^2 \phi_{\tilde \gamma}^2.
\ee

\item {\it DM -- DR interactions through $s/u$-channel}
\begin{itemize}
\item{Scalar -- Scalar}
\be
\label{L_SU_SS}
\Delta \mathcal{L} = \frac{\mu_\chi}{2} \phi_\chi^2 \phi_{\tilde \gamma}.
\ee
\item{Fermion -- Scalar}
\be
\label{L_SU_FS}
\Delta \mathcal{L} = g_\chi \bar\psi_\chi \psi_\chi \phi_{\tilde \gamma}.
\ee
\item{Scalar -- $U(1)$ Vector}
\bea
\label{L_SU_SV}
\Delta \mathcal{L} &=& ig_\chi \left[(\partial_\mu\phi^\dagger_{\chi})\phi_{\chi} - \phi^\dagger_{\chi}(\partial_\mu\phi_{\chi})\right] V^\mu_{\tilde \gamma} \nonumber\\
 &&{}
 - g^2_\chi (V^\mu_{\tilde \gamma})^2 |\phi_\chi|^2.
\eea
\item{Fermion -- $U(1)$ Vector}
\be
\label{L_SU_FV}
\Delta \mathcal{L} = g_\chi \bar\psi_\chi \slashed{V}_{\tilde \gamma} \psi_\chi.
\ee
\end{itemize}

\item {\it DM -- DR interactions through $t$-channel}
\begin{itemize}
\item{Scalar -- Scalar}
\be
\label{L_T_SS}
 \Delta \mathcal{L} =  \frac{\mu_\chi}{2} \phi_\chi^2 \phi_{\tilde \gamma} +  \frac{\mu_{\tilde \gamma}}{6}  \phi_{\tilde \gamma}^3.
\ee

\item{Fermion -- Scalar}
\be
\label{L_T_FS}
\Delta \mathcal{L} = g_\chi \bar\psi_\chi \psi_\chi \phi_{\tilde \gamma} +  \frac{\mu_{\tilde \gamma}}{6}  \phi_{\tilde \gamma}^3.
\ee
\end{itemize}

\item \emph{DM -- DR interactions through \emph{all} channels}
\begin{itemize}
\item{Vector -- Scalar}
\bea
\label{L_V_S}
 \Delta \mathcal{L} &=& 
 %- \frac{1}{4} \left( F_{\mu\nu}\right)^2   +\frac{1}{2} (\partial_\mu \phi_{\tilde{\gamma}})^2 \nonumber\\
 % & & + \frac{1}{2} m^2_\chi V_{\chi\mu} V_{\chi}^\mu -\frac{1}{2} m_{{\tilde{\gamma}}}^2 \phi_{\tilde{\gamma}}^2 \nonumber\\
 g_\chi m_\chi V_{\chi \mu} V_{\chi}^\mu \phi_{\tilde{\gamma}} + \frac{1}{2}g_\chi^2  V_{\chi \mu} V_{\chi}^\mu \phi_{\tilde{\gamma}}^2  \nonumber\\
&& - \frac{1}{2}g_\chi \frac{m_{\tilde{\gamma}}^2}{m_\chi} \phi_{\tilde{\gamma}}^3\,.
%-\frac{1}{8} g^2_\chi \frac{m^2_{\tilde{\gamma}}}{m_\chi^2}\phi_{\tilde{\gamma}}^4 
%+ \frac{m^2_{\tilde{\gamma}} m^2_\chi}{8g^4_\chi} .
\eea
(resulting from a spontaneously broken $U(1)$ symmetry;  the imaginary part of the original scalar 
field $\Phi$ thus gives the mass to $V_\chi^\mu$).

\item{Scalar -- $SU(N)$ Vector}
\bea
\label{L_SUN_SV}
 \Delta \mathcal{L} &=& - \frac{1}{2} \text{Tr}\left[ \left( F_{\mu\nu}^a\right)^2 \right]  - g^2_\chi V^{a\mu}_{\tilde \gamma}V^{b}_{ \tilde \gamma\mu} \Phi^\dagger_{\chi} t^a t^b \Phi_{\chi} \nonumber\\
  &&{} + ig_\chi \left[(\partial_\mu\Phi^\dagger_{\chi})t^a\Phi_{\chi} - \Phi^\dagger_{\chi}t^a(\partial_\mu\Phi_{\chi})\right] V^{a\mu}_{\tilde \gamma}. \nonumber\\
\eea

\item{Fermion -- $SU(N)$ Vector}
\bea
\label{L_SUN_FV}
\Delta \mathcal{L} &=& - \frac{1}{2} \text{Tr}\left[ \left( F_{\mu\nu}^a\right)^2 \right] + g_\chi \bar\Psi_\chi \slashed{V}^{a}_{\tilde\gamma}t^a \Psi_\chi.
\eea
\end{itemize}

\end{enumerate}

\begin{table*}[t!]
\renewcommand{\arraystretch}{2.2}
    \centering
    \begin{tabular}{c|c|c|c|c|c|c}
        DM / DR & $\Delta\mathcal{L}$ & $\langle |\mathcal{M}|^2 \rangle_t$ & $|\mathcal{M}|^2_{t=0}$ & $\dfrac{T_\mathrm{kd}\left(\langle |\mathcal{M}|^2 \rangle_t\right)}{T_\mathrm{kd}\left(|\mathcal{M}|^2_{t=0}\right)}$ & $|\mathcal{M}|^2\propto\left(\dfrac{\omega}{m_\chi}\right)^n$ & ETHOS parameters\\[1.5ex]
        \hline
        \hline
         \multicolumn{7}{c}{\it 4-point (contact interaction only)} \\
        \hline
    scalar/scalar  & \ref{L_SSSS} & $\lambda^2$ & $\lambda^2$& 1.0 &  \ding{51}
    & $\{a_2, \alpha_{l\geq2}=1 \}$ \\
        \hline

         \multicolumn{7}{c}{\it $s/u$-channel} \\
        \hline
        scalar/scalar & \ref{L_SU_SS} & $\dfrac{\mu^4_\chi}{2m_\chi^4}$ & $\mathcal{O}(m_{\tilde \gamma }^4)$&--- & \ding{51}/\ding{55}
        & $\{a_2, \alpha_2=\frac35, \alpha_{l\geq3}=\frac23 \}$ \\[1.2ex]
        \hline
        fermion/scalar & \ref{L_SU_FS} &  $\dfrac{16g_\chi^4}{3}$ & $16g_\chi^4$& 1.7 & \ding{51}
        & $\{a_2, \alpha_2=\frac35, \alpha_{l\geq3}=1 \}$ \\[1ex]
        \hline
        scalar/vector & \ref{L_SU_SV}  & $\dfrac{32 g^4_\chi}{3}$ &$16 g^4_\chi$& 1.2 & \ding{51}
        & $\{a_{ 2}, \alpha_2=\frac{9}{10}, \alpha_{l\geq3}=1 \}$\\[1ex]
        \hline
        fermion/vector & \ref{L_SU_FV}  & $\dfrac{64g^4_\chi}{3}$ & $32 g^4_\chi$& 1.2 & \ding{51}
        & $\{a_{ 2}, \alpha_2=\frac{9}{10}, \alpha_{l\geq3}=1 \}$ \\[1ex]
        \hline

         \multicolumn{7}{c}{\it $t$-channel} \\
        \hline
        scalar/scalar & \ref{L_T_SS}  & $\dfrac{\mu_\chi^2\mu_{\tilde \gamma}^2}{8\omega^4} \ln\dfrac{4\omega^2}{m_{\tilde \gamma }^2}$ &
       $\dfrac{\mu_\chi^2 \mu_{\tilde \gamma}^2}{m_{\tilde\gamma}^4}$ &
       --- & \ding{55} 
       & ($m_{\tilde\gamma}\to0$ undefined) \\[1.2ex]
        \hline
             fermion/scalar&  \ref{L_T_FS}  & $\dfrac{2g^2_\chi \mu_{\tilde \gamma}^2m_\chi^2\ln{\left(4\omega^2/m_{\tilde \gamma }^2\right)}}{\omega^4}$&
        $\dfrac{16g^2_\chi \mu_{\tilde \gamma}^2m_\chi^2}{m_{\tilde\gamma}^4}$ &
       --- & \ding{55} 
       & ($m_{\tilde\gamma}\to0$ undefined) \\[1.2ex]
        \hline
         \multicolumn{7}{c}{\it DM -- DR interactions through \emph{all} channels} \\
        \hline
                           vector/scalar& \ref{L_V_S}  & $4g^4_\chi $&
        $48 g^4_\chi $ & 3.5 &  \ding{51}
    & $\{a_2,\alpha_{2}=\frac{3}{5}, \alpha_{l\geq3}=1 \}$\\
        \hline
                   scalar/vector ($SU(N)$)& \ref{L_SUN_SV}  & $\dfrac{9g^4_\chi C_FC_A^2 m_\chi^2\ln{\frac{4\omega^2}{m_{\tilde \gamma }^2}}}{\omega^2}$&
        $\dfrac{72g^4_\chi C_FC_A^2m_\chi^2\omega^2}{m_{\tilde\gamma}^4}$ &
       --- & \ding{55} 
       & ($m_{\tilde\gamma}\to0$ undefined) \\[1.2ex]
        \hline
fermion/vector ($SU(N)$) & \ref{L_SUN_FV}  & $\dfrac{18g^4_\chi C_FC_A^2 m_\chi^2\ln{\frac{4\omega^2}{m_{\tilde \gamma }^2}}}{\omega^2}$ &
         $\dfrac{144g^4_\chi C_FC_A^2m_\chi^2\omega^2}{m_{\tilde\gamma}^4}$ &
       --- & \ding{55} 
       & ($m_{\tilde\gamma}\to0$ undefined)\\[1.2ex]
        \hline
    \end{tabular}
\caption{Full list of relevant 2-particle models. For the scattering matrix elements, only the
leading terms in $\omega/m_\chi$ are given, assuming
$m_\chi \gg \omega \gg m_{\tilde \gamma}$ (for the $t$-channel results, we have further
assumed that $\mu_{\tilde\gamma}$ is sufficiently large that the $t$-channel amplitude always dominates over
the $s/u$-channel amplitudes). 
In the fifth column, we state the ratio of the
kinetic decoupling temperature resulting from the $t$-averaging
prescription to that from the
$t\to0$ prescription. The next-to-last column indicates whether keeping only the leading order result
for the amplitude (after averaging or setting $t=0$) provides a good estimate for
$T_\mathrm{kd}$. In this case the analytical solution, 
Eq.~(\ref{tkd_analytic}), can be used;
otherwise, Eq.~(\ref{tchidt}) must be solved numerically.
The last column, finally, states the full set of ETHOS parameters \cite{Cyr-Racine:2015ihg} 
that describe the respective model, as defined in 
Eqs.~(\ref{an},\ref{alphan}).
\label{tab:Msq_2p}}
\end{table*}

We list the squared amplitudes for all those models, with and without averaging over $t$, in 
Table \ref{tab:Msq_2p}. In all cases, the amplitudes squared are summed (not averaged) 
over all external spins, polarization states and `colors', as well as over particles and antiparticles.
We also provide the ratio of kinetic 
decoupling temperatures that results when using the $t$-averaging prescription and 
the simplified $t=0$ prescription, respectively. To obtain this ratio, we calculated the kinetic
decoupling temperature by solving the full process Eq.~(\ref{tchidt}) numerically. In 
the next-to-last column of Table \ref{tab:Msq_2p}, we indicate whether expanding the 
amplitude as a power law in the energy of the relativistic scattering partner provides an
accurate estimate of the correct decoupling temperature (i.e.~whether the analytic result 
given in Eq.~(\ref{tkd_analytic}) agrees with the full numerical result at the percent level).

We are studying here situations with intrinsically large kinematic enhancements, i.e.~where
the presence of small quantites in propagators -- namely $t$, $\omega$ and
$m_{\tilde\gamma}$ -- can have significant effects on the amplitude. It should therefore not be
a surprise that we identify cases where the simple $t=0$ prescription
breaks down completely. 
One of those examples is the case of scalar/scalar scattering via the $s/u$ 
channel, where the amplitude evaluated at $t=0$ is proportional to $m_{\tilde\gamma}^4/m_\chi^4$
while the averaged amplitude is not suppressed by the small DR mass. A similar issue appears,
certainly not unexpected, in all cases where $\tilde\gamma$ appears in the $t$-channel. 
Apart from that, we confirmed that if the squared amplitude  takes the form of a power law in 
the energy \emph{close to kinetic decoupling}, as well as at slightly higher temperatures, the 
analytic solution 
(\ref{tkd_analytic}) for the kinetic  decoupling temperature provides a very reliable estimate for 
the full numerical result. For DR in the $t$-channel, however, the amplitude close to kinetic
decoupling is not of the form given in Eq.~(\ref{powerlaw}) and, consequently,  the analytic 
solution cannot be expected to apply. 

As mentioned in the introduction, finally, ETHOS \cite{Cyr-Racine:2015ihg} provides an efficient 
way of classifying the impact
of DM models on structure formation by means of a handful of phenomenological parameters -- 
in the sense that every DM model with similar ETHOS parameters leads to almost identical 
results in full numerical simulations. An important input here is the DR opacity to DM scattering, 
$\dot\kappa_{\tilde\gamma-\chi}$, which for relativistic DR
typically can be parameterized as
\bea
\frac{\dot\kappa_{\tilde\gamma-\chi}}{\Omega_\chi h^2}
&\equiv&\frac{-n_\chi}{16\pi m_\chi^2(1+z)\Omega_\chi h^2}\frac{\int d\omega\, \omega^3 g^\pm(\omega)\left[ A_0 - A_1\right]}{\int d\omega\, \omega^3 g^\pm(\omega)}\nonumber\\
&\simeq&-\sum_m a_m \left(\frac{1+z}{1+z_{\rm kd}}\right)^m\,,\label{an}
\eea
where $z$ denotes the cosmological redshift and
\be
A_l(\omega) \equiv \frac{1}{2}\int_{-1}^1 d{\cos\theta}\, P_l(\cos\theta) \frac{|\mathcal{M}|^2}{\eta_\chi\eta_{\tilde\gamma}}\Bigg{|}_{\begin{subarray}{l} t=2\omega^2(\cos\theta-1) \\ s=m_\chi^2+2\omega m_\chi \end{subarray}}.
\ee
In the above expression, $\theta$ is the angle between the incoming and outgoing DR 
particle, and $P_l$ denotes the $l$th Legendre polynomial. This means that $A_0-A_1$ is, 
up to a 
constant, essentially just the transfer cross section 
$\sigma_T$ for DM-DR scattering, and $\dot\kappa_{\tilde\gamma-\chi}$ thus closely related 
to the momentum transfer rate $\gamma$, see Eq.~(\ref{gammast}).
The only other relevant parameters for the models studied here are a set of
angular coefficients $\alpha_l $ defined by
\be
\label{alphan}
\alpha_l \equiv \frac{\int d\omega\, \omega^3 g^\pm(\omega)\left[A_0(\omega) -A_l(\omega)\right]}{\int d\omega\, \omega^3 g^\pm(\omega)\left[A_0(\omega) -A_1(\omega)\right]}\,, \quad l\geq2.
\ee

In Table \ref{tab:Msq_2p}, we provide for each model the full set of non-vanishing parameters 
$\{a_n,\alpha_l\}$ in the
limit considered here, namely $T_{\tilde\gamma}\ll m_\chi$. 
We also indicate those cases where the above expressions do not apply because 
the limit 
$m_{\tilde\gamma}\to0$ cannot be taken, a situation for which the ETHOS parameterization 
has not been worked out yet.
For models with the same set of parameters, the effect of the cutoff in the primordial power
spectrum on non-linear structure formation will  be identical. We note that the value of 
$\alpha_l$ has a rather limited impact in this respect, as it leaves shape and location of the 
{\it first} peak in the linear power spectrum mostly unaffected \cite{Cyr-Racine:2015ihg}.

%%%%%%%%%%%%%%%%%%%%%%%%%%%%%%%%%%%%%%%%%%%%
\subsection{3-particle models}
\label{app:msq3}

\begin{table*}[t!]
			    \centering
			    \renewcommand{\arraystretch}{2.2}
    \begin{tabular}{c|c|c|c|c|c|c}
        DM / mediator / DR & $\Delta\mathcal{L}$ & $\langle |\mathcal{M}|^2 \rangle_t$ & $|\mathcal{M}|^2_{t=0}$ & $\dfrac{T_\mathrm{kd}\left(\langle |\mathcal{M}|^2 \rangle_t\right)}{T_\mathrm{kd}\left(|\mathcal{M}|^2_{t=0}\right)}$ & $|\mathcal{M}|^2\propto\left(\dfrac{\omega}{m_\chi}\right)^n$ & ETHOS parameters\\[1.5ex]
        \hline
        \hline
		         \multicolumn{6}{c}{\it s/u-channel} \\
		         \hline
        scalar/scalar/scalar & \ref{L3_SU_SSS} & $\dfrac{\mu_\chi^4}{m_\chi^2 \Delta m^2}$ & $\dfrac{\mu_\chi^4}{m_\chi^2 \Delta m^2}$& 1.0&\ding{51}& $\{a_{ 2}, \alpha_{l\geq2}=1 \}$\\[1.2ex]
        \hline
       scalar/fermion/fermion & \ref{L3_SU_SFF} & $\dfrac{32g_\chi^4\omega^2}{3\Delta m^2}$ & $\dfrac{16g_\chi^4\omega^4}{\Delta m^4}$& --- &\ding{51}&$\{a_{ 4}, \alpha_{l\geq2}= \frac{3}{4} \}$ \\[1ex]
        \hline
     fermion/scalar/fermion & \ref{L3_SU_FSF} &  $\dfrac{88g_\chi^4\omega^2}{ 3\Delta m^2}$ & $\dfrac{40g_\chi^4\omega^2}{\Delta m^2}$&1.1&\ding{51}&$\{a_{ 4}, \alpha_{l\geq2}= \frac{12}{11} \}$ \\[1ex]    
        \hline
        fermion/fermion/scalar & \ref{L3_SU_FFS} & $\dfrac{64g_\chi^4 m_\chi^2}{\Delta m^2}$ & $\dfrac{64g_\chi^4 m_\chi^2}{\Delta m^2}$&1.0&\ding{51}&$\{a_{ 2}, \alpha_{l\geq2}=1 \}$\\[1ex]
        \hline

         \multicolumn{6}{c}{\it t-channel} \\
        \hline
				scalar/scalar/scalar &\ref{L3_T_SSS} &$\dfrac{\mu_\chi ^2\mu_{\tilde \gamma}^2}{m_{\tilde\gamma'}^4}$ & $\dfrac{\mu_\chi ^2\mu_{\tilde \gamma}^2}{m_{\tilde\gamma'}^4}$ & 1.0 &\ding{51}&$\{a_{ 2}, \alpha_{l\geq2}=1 \}$\\[1.2ex]
		\hline
						scalar/scalar/fermion &\ref{L3_T_SSF} &$\dfrac{32\mu_\chi ^2g_{\tilde \gamma}^2 \omega^2}{3m_{\tilde\gamma'}^4}$ & $\mathcal{O}(m_{\tilde \gamma}^2)$ & --- &\ding{51}/\ding{55}& $\{a_{ 4}, \alpha_{l\geq2}= \frac{3}{4} \}$\\[1.2ex]
		\hline
						scalar/scalar/vector &\ref{L3_T_SSV} &$\dfrac{4\mu_\chi^2\mu_{\tilde \gamma}^2}{m_{\tilde\gamma'}^4}$ & $\dfrac{4\mu_\chi^2\mu_{\tilde \gamma}^2}{m_{\tilde\gamma'}^4}$ & 1.0 &\ding{51}&$\{a_{ 2}, \alpha_{l\geq2}=1 \}$\\[1.2ex]
		\hline
		 scalar/vector/scalar &\ref{L3_T_SVS} &$\dfrac{64g_\chi ^2g_{\tilde \gamma}^2\omega^2 m_\chi^2}{m_{\tilde\gamma'}^4}$ & $\dfrac{64g_\chi^2g_{\tilde \gamma}^2\omega^2 m_\chi^2}{m_{\tilde\gamma'}^4}$& 1.0 &\ding{51}&$\{a_{ 4}, \alpha_{l\geq2}=1 \}$\\[1.2ex]
		\hline
		 scalar/vector/fermion &\ref{L3_T_SVF} &$\dfrac{128g_\chi^2g_{\tilde \gamma}^2\omega^2 m_\chi^2}{3m_{\tilde\gamma'}^4}$ & $\dfrac{128g_\chi^2g_{\tilde \gamma}^2\omega^2 m_\chi^2}{m_{\tilde\gamma'}^4}$&1.3 &\ding{51}&$\{a_{ 4}, \alpha_{l\geq2}= \frac{3}{2} \}$\\[1.2ex]
		\hline

		      fermion/scalar/scalar & \ref{L3_T_FSS}  &$\dfrac{16g_\chi^2\mu_{\tilde \gamma}^2 m_\chi^2}{m_{\tilde\gamma'}^4}$ & $\dfrac{16g_\chi^2\mu_{\tilde \gamma}^2 m_\chi^2}{m_{\tilde\gamma'}^4}$& 1.0 &\ding{51}&$\{a_{ 2}, \alpha_{l\geq2}= 1 \}$\\[1.2ex]
		        \hline
		    		fermion/scalar/fermion & \ref{L3_T_FSF}  &$\dfrac{512g_\chi^2g_{\tilde \gamma}^2\omega^2 m_\chi^2}{3m_{\tilde\gamma'}^4}$ & $\mathcal{O}(m_{\tilde \gamma }^2)$&---&\ding{51}/\ding{55}&$\{a_{ 4}, \alpha_{l\geq2}= \frac{3}{4} \}$\\[1.2ex]
		\hline
        fermion/scalar/vector &  \ref{L3_T_FSV} & $\dfrac{32g_\chi^2\mu_{\tilde \gamma}^2 m_\chi^2}{m_{\tilde\gamma'}^4}$ & $\dfrac{32g_\chi ^2\mu_{\tilde \gamma}^2 m_\chi^2}{m_{\tilde\gamma'}^4}$ & 1.0 &\ding{51}&$\{a_{ 2}, \alpha_{l\geq2}= 1 \}$ \\[1.2ex]
        \hline
		 fermion/vector/scalar & \ref{L3_T_FVS} &$\dfrac{128 g_\chi^2g_{\tilde \gamma}^2\omega^2 m_\chi^2}{m_{\tilde\gamma'}^4}$ &$\dfrac{128 g_\chi^2g_{\tilde \gamma}^2\omega^2 m_\chi^2}{m_{\tilde\gamma'}^4}$& 1.0 &\ding{51}&$\{a_{ 4}, \alpha_{l\geq2}= 1 \}$ \\[1.2ex]
		\hline
		fermion/vector/fermion & \ref{L3_T_FVF} &$\dfrac{256g_\chi^2g_{\tilde \gamma}^2\omega^2 m_\chi^2}{3m_{\tilde\gamma'}^4}$ & $\dfrac{256g_\chi^2g_{\tilde \gamma}^2\omega^2 m_\chi^2}{m_{\tilde\gamma'}^4}$&1.3 &\ding{51}&$\{a_{ 4}, \alpha_{l\geq2}= \frac{3}{2} \}$  \\[1.2ex]
		        \hline
    \end{tabular}	
    \caption{Full list of relevant 3-particle models -- including DM particles $\chi$, (dark) radiation particles 
    $\tilde\gamma$, and mediator particles $\chi'$ or $\tilde\gamma'$. 
    For $s/u$-channel and $t$-channel processes, we have assumed
     $m_\chi \gg \Delta m\equiv m_{\chi'}-m_\chi  \gg \omega \gg m_{\tilde \gamma}$ 
     and $m_\chi \gg m_{\tilde\gamma'}  \gg \omega \gg m_{\tilde \gamma}$, respectively; 
     in all cases, we state only the leading terms for the squared amplitude.
     The last three columns are defined as in Table \ref{tab:Msq_2p}.
     % NB : $\mu_{\tilde\gamma}$ are not independent of $m_{\tilde\gamma'}$ in any of these cases!
     % The phenomenlogy of $m_{\tilde\gamma'}\to0$ needs thus to be studied in each case separately
\label{tab:Msq_3p} }
\end{table*}

We now consider simplified models where the scattering between non-relativistic DM 
particles $\chi$ and relativistic particles $\tilde\gamma$ is mediated by a \emph{different}
particle -- either a `DM-like' particle $\chi'$ which is slightly heavier than $\chi$ (leading
to $s/u$-channel exchange) or a `DR-like' particle $\tilde\gamma'$ which is much heavier 
than $\tilde\gamma$ (leading to $t$-channel exchange).
As motivated in Section 
\ref{sec:3pm}, we are then interested in the following interaction terms of dimension 4
(to indicate the spin of $\chi$ and $\tilde\gamma$, we denote again scalars 
always with $\phi$, vectors with $V$ and fermions with $\psi$).

%\newpage
\begin{enumerate}

\item {\it DM -- DR interactions through a mediator $\chi'$ in the $s/u$-channels}
\begin{itemize}
\item{Scalar -- Scalar -- Scalar}
\bea
\label{L3_SU_SSS}
\Delta \mathcal{L}& =&   \mu_\chi \phi_\chi \phi_{\chi'} \phi_{\tilde \gamma} .
\eea

\item{Scalar -- Fermion -- Fermion}
\bea
\label{L3_SU_SFF}
\Delta \mathcal{L}& =&  g_\chi \phi_\chi \bar \psi_{\chi'} \psi_{\tilde \gamma} +\mathrm{h.c.}
\eea

\item{Fermion -- Scalar -- Fermion}
\bea
\label{L3_SU_FSF}
\Delta \mathcal{L}& =&  g_\chi \phi_{\chi'} \bar \psi_{\tilde \gamma} \psi_\chi +\mathrm{h.c.}
\eea

\item{Fermion -- Fermion -- Scalar}
\bea
\label{L3_SU_FFS}
\Delta \mathcal{L}& =&   g_\chi \phi_{\tilde \gamma} \bar \psi_{\chi'} \psi_\chi +\mathrm{h.c.}
\eea

\end{itemize}

\item {\it DM -- DR interactions through a mediator $\tilde\gamma'$ in the $t$-channel}
\begin{itemize}
\item{Scalar -- Scalar -- Scalar}
\bea
\label{L3_T_SSS}
\Delta \mathcal{L}& =&   \frac{\mu_\chi}{2} \phi_{\tilde\gamma'}  \phi_\chi^2  +  \frac{\mu_{\tilde \gamma}}{2} \phi_{\tilde\gamma'} \phi_{\tilde \gamma}^2 .
\eea

\item{Scalar -- Scalar -- Fermion}
\bea
\label{L3_T_SSF}
\Delta \mathcal{L}& =&   \frac{\mu_\chi}{2} \phi_{\tilde\gamma'}  \phi_\chi^2 +g_{\tilde \gamma} \bar\psi_{\tilde \gamma} \psi_{\tilde \gamma} \phi_{{\tilde\gamma'}} .
\eea

\item{Scalar -- Scalar -- Vector}
\bea
\label{L3_T_SSV}
\Delta \mathcal{L}& =&  \frac{\mu_\chi}{2}\phi_{\tilde\gamma'}  \phi_\chi^2  + \mu_{\tilde \gamma} \phi_{\tilde\gamma'} V_{\mu{\tilde \gamma}} V^\mu_{\tilde \gamma}
\eea
(assuming the gauge symmetry to be broken spontaneously by
$\langle\phi_{\tilde\gamma'}\rangle \propto \mu_{\tilde\gamma}$).
%NB: $\phi$ is complex in the unbroken theory, but the imaginary part is eaten by the vector
% -- so $\phi$ in (\ref{L3_T_SSV}) is real.

\item{Scalar -- Vector -- Scalar}
\bea
\label{L3_T_SVS}
\Delta \mathcal{L} &=& ig_\chi \left[(\partial_\mu\phi^\dagger_{\chi})\phi_{\chi} - \phi^\dagger_{\chi}(\partial_\mu\phi_{\chi})\right] V^\mu_{{\tilde\gamma'}} \nonumber\\
 && {}
 + ig_{\tilde \gamma} \left[(\partial_\mu\phi^\dagger_{\tilde \gamma})\phi_{\tilde \gamma} - \phi^\dagger_{\tilde \gamma}(\partial_\mu\phi_{\tilde \gamma})\right] V^\mu_{{\tilde\gamma'}}.
\eea

\item{Scalar -- Vector -- Fermion}
\bea
\label{L3_T_SVF}
\Delta \mathcal{L}& =&  ig_\chi \left[(\partial_\mu\phi^\dagger_{\chi})\phi_{\chi} - \phi^\dagger_{\chi}(\partial_\mu\phi_{\chi})\right] V^\mu_{{\tilde\gamma'}} \nonumber\\
&&{} +g_{\tilde \gamma} \bar\psi_{\tilde \gamma} \slashed{V}_{{\tilde\gamma'}} \psi_{\tilde \gamma}.
\eea

\item{Fermion -- Scalar -- Scalar}
\bea
\label{L3_T_FSS}
\Delta \mathcal{L}& =&   g_\chi \bar\psi_\chi \psi_\chi \phi_{{\tilde\gamma'}} +  \frac{\mu_{\tilde \gamma}}{2} \phi_{\tilde\gamma'} \phi_{\tilde \gamma}^2 .
\eea

\item{Fermion -- Scalar -- Fermion}
\bea
\label{L3_T_FSF}
\Delta \mathcal{L}& =&   g_\chi \bar\psi_\chi \psi_\chi \phi_{{\tilde\gamma'}} +g_{\tilde \gamma} \bar\psi_{\tilde \gamma} \psi_{\tilde \gamma} \phi_{{\tilde\gamma'}}  .
\eea

\item{Fermion -- Scalar -- Vector}
\bea
\label{L3_T_FSV}
\Delta \mathcal{L}& =&   g_\chi \bar\psi_\chi \psi_\chi \phi_{{\tilde\gamma'}} + \mu_{\tilde \gamma} \phi_{\tilde\gamma'} V_{\mu{\tilde \gamma}} V^\mu_{\tilde \gamma}.
\eea
(assuming again the gauge symmetry to be broken spontaneously by
$\langle\phi_{\tilde\gamma'}\rangle \propto \mu_{\tilde\gamma}$)

\item{Fermion -- Vector -- Scalar}
\bea
\label{L3_T_FVS}
\Delta \mathcal{L}& =&  g_\chi \bar\psi_{\chi} \slashed{V}_{{\tilde\gamma'}} \psi_{\chi}\nonumber\\
&&{} + ig_{\tilde \gamma} \left[(\partial_\mu\phi^\dagger_{\tilde \gamma})\phi_{\tilde \gamma} - \phi^\dagger_{\tilde \gamma}(\partial_\mu\phi_{\tilde \gamma})\right] V^\mu_{{\tilde\gamma'}} .
\eea

\item{Fermion -- Vector -- Fermion}
\bea
\label{L3_T_FVF}
\Delta \mathcal{L}& =&  g_\chi \bar\psi_{\chi} \slashed{V}_{{\tilde\gamma'}} \psi_{\chi}+g_{\tilde \gamma} \bar\psi_{\tilde \gamma} \slashed{V}_{{\tilde\gamma'}} \psi_{\tilde \gamma} .
\eea

\end{itemize}

\end{enumerate}

We note that the dimensionful coupling $\mu_{\tilde\gamma}$ cannot be chosen completely 
independently of the mediator mass $m_{\tilde\gamma'}$; in Eqs.~(\ref{L3_T_FSV}, 
\ref{L3_T_SSV}), e.g., it is the vacuum expectation value of the same field that gives rise
to both quantities.
We list the resulting squared amplitudes in Table \ref{tab:Msq_3p}, following the same
format as in Table \ref{tab:Msq_2p} for the 2-particle models. Note that we focus on 
parameter choices where we can expect qualitative differences with respect to the 
2-particle models discussed above. Sufficiently close to 
kinetic decoupling, we therefore require 
$m_\chi \gg \Delta m\equiv m_{\chi'}-m_\chi  \gg \omega \gg m_{\tilde \gamma}$ 
(for $s/u$-channel mediated processes) and 
$m_\chi \gg m_{\tilde\gamma'} \gg \omega \gg m_{\tilde \gamma}$ (for $t$-channel mediated 
processes), respectively.

Also in this case, we identify situations where the simple $t=0$ prescription leads to a
qualitatively wrong result for the inferred decoupling temperature. In the $s/u$-channel, 
this happens for the combination of scalar DM and fermionic $\chi'$ and $\tilde\gamma$,
where $\left|\mathcal{M}\right|_{t=0}^2\propto \omega^4$ while 
$\langle\left|\mathcal{M}\right|^2\rangle_t\propto\omega^2$.
In the $t$-channel the `critical' combinations concern fermionic $\tilde\gamma$ and scalar
$\tilde\gamma'$: unlike suggested by the result for $\left|\mathcal{M}\right|_{t=0}^2$, those
combinations do not lead to an insignificant scattering rate -- but rather to a scattering rate
that is (up to a constant factor) the same as in the case of exchanging a \emph{vector} particle
$\tilde\gamma'$. As long as one uses the correct description for calculating the matrix
element, on the other hand, the analytic solution 
(\ref{tkd_analytic}) for the kinetic  decoupling temperature always provides a very reliable estimate for 
the full numerical result. 

For convenience, we provide the ETHOS parameters also for all 3-particle models. 
For the case of $\{a_m,\alpha_l\}=\{a_4,\alpha_{l\geq2}=3/2\}$ -- which appears e.g.~for 
fermion-fermion scattering via vector exchange, see Eq.~(\ref{L3_T_FVF}), as studied in 
Ref.~\cite{Bringmann:2013vra,Dasgupta:2013zpn,Ko:2014bka,Cherry:2014xra}  --  detailed 
numerical simulations have already been performed \cite{Vogelsberger:2015gpr}. Those
simulations  included the effect of DM self-interactions, for which we do not explicitly list the 
relevant ETHOS parameters here.
We note, finally, that a recent computation for the models (\ref{L3_T_FSF}) and (\ref{L3_T_FVF}) 
resulted in identical linear power spectra for these two cases when neglecting the impact
of perturbations in the DR fluid \cite{Binder:2016pnr}. Including this effect, which is encoded
in the parameters $\alpha_l$, we thus expect (small) differences between the power spectra 
generated by DM-DR scattering mediated by scalar and vector mediators, respectively
(see the discussion in Ref.~\cite{Cyr-Racine:2015ihg}).

Let us, finally, point out that none of the models studied in this section (nor in the previous section
where we considered two-particle models) allows for the possibility that $\tilde\gamma$ is the SM
photon. For the Lagrangians in Eqs.~(\ref{L_SU_SV}, \ref{L_SU_FV}), for example, this is excluded 
because it would lead to too large DM self-interactions (as discussed in Section \ref{sec:scatter_su});
the Lagrangians (\ref{L3_T_SSV},\ref{L3_T_FSV}), on the other hand, are not compatible with an 
unbroken $U(1)$ gauge symmetry. We note that this conclusion may change when including 
higher-dimensional operators in the discussion, an interesting candidate being e.g.~a scalar $\tilde\gamma'$ 
in the $t$-channel that couples via $\tilde\gamma'F_{\mu\nu}F^{\mu\nu}$ to photons. As the effective
coupling $g_{\tilde\gamma'}$ of such higher-dimensional operators would necessarily be suppressed, 
however, we expect from Fig.~\ref{fig:tRDvsSIDM} that such solutions would require even smaller 
mediator masses $m_{\tilde\gamma'}$, and hence even lighter DM.

Similarly, it appears to be challenging for any concrete model building to identify $\tilde\gamma$ 
with the SM neutrino. The main reason is that due to $SU(2)$ gauge invariance any new state
coupling to the neutrino should couple with equal strength to the (left-handed) SM electron.
This argument basically excludes the possibility of achieving late kinetic decoupling with 
$\tilde\gamma=\nu_L$ 
in the Lagrangians stated in Eqs.~(\ref{L3_T_SSF}, \ref{L3_T_SVF}, \ref{L3_T_FSF}, \ref{L3_T_FVF}), 
because the coupling of electrons to new light states is generally strongly constrained.
In the case of Eqs.~(\ref{L3_SU_SFF}, \ref{L3_SU_FSF}), on the other hand, the leptons would
couple to {\it heavy} new states -- which is not excluded if the mass scale is high enough
(in supersymmetry, e.g., this would correspond to a coupling between neutralino, lepton and slepton).
Inspecting Eqs.~(\ref{mcutmax3b0RD}, \ref{mcutmax3b2RD}), however, tells us that a large mass 
scale can only be reconciled with late kinetic decoupling for extremely small mass splittings $\delta$.
Again, higher-dimensional operators may potentially allow for qualitatively different options.

%this is sometimes needed in order not to mess up with tables/footnotes on the final page
%\newpage

\bibliography{KD}

\begin{thebibliography}{127}%
\makeatletter
\providecommand \@ifxundefined [1]{%
 \@ifx{#1\undefined}
}%
\providecommand \@ifnum [1]{%
 \ifnum #1\expandafter \@firstoftwo
 \else \expandafter \@secondoftwo
 \fi
}%
\providecommand \@ifx [1]{%
 \ifx #1\expandafter \@firstoftwo
 \else \expandafter \@secondoftwo
 \fi
}%
\providecommand \natexlab [1]{#1}%
\providecommand \enquote  [1]{``#1''}%
\providecommand \bibnamefont  [1]{#1}%
\providecommand \bibfnamefont [1]{#1}%
\providecommand \citenamefont [1]{#1}%
\providecommand \href@noop [0]{\@secondoftwo}%
\providecommand \href [0]{\begingroup \@sanitize@url \@href}%
\providecommand \@href[1]{\@@startlink{#1}\@@href}%
\providecommand \@@href[1]{\endgroup#1\@@endlink}%
\providecommand \@sanitize@url [0]{\catcode `\\12\catcode `\$12\catcode
  `\&12\catcode `\#12\catcode `\^12\catcode `\_12\catcode `\%12\relax}%
\providecommand \@@startlink[1]{}%
\providecommand \@@endlink[0]{}%
\providecommand \url  [0]{\begingroup\@sanitize@url \@url }%
\providecommand \@url [1]{\endgroup\@href {#1}{\urlprefix }}%
\providecommand \urlprefix  [0]{URL }%
\providecommand \Eprint [0]{\href }%
\providecommand \doibase [0]{http://dx.doi.org/}%
\providecommand \selectlanguage [0]{\@gobble}%
\providecommand \bibinfo  [0]{\@secondoftwo}%
\providecommand \bibfield  [0]{\@secondoftwo}%
\providecommand \translation [1]{[#1]}%
\providecommand \BibitemOpen [0]{}%
\providecommand \bibitemStop [0]{}%
\providecommand \bibitemNoStop [0]{.\EOS\space}%
\providecommand \EOS [0]{\spacefactor3000\relax}%
\providecommand \BibitemShut  [1]{\csname bibitem#1\endcsname}%
\let\auto@bib@innerbib\@empty
%</preamble>
\bibitem [{\citenamefont {Ade}\ \emph {et~al.}(2016)\citenamefont {Ade} \emph
  {et~al.}}]{Ade:2015xua}%
  \BibitemOpen
  \bibfield  {author} {\bibinfo {author} {\bibfnamefont {P.~A.~R.}\
  \bibnamefont {Ade}} \emph {et~al.} (\bibinfo {collaboration} {Planck}),\
  }\href {\doibase 10.1051/0004-6361/201525830} {\bibfield  {journal} {\bibinfo
   {journal} {Astron. Astrophys.}\ }\textbf {\bibinfo {volume} {594}},\
  \bibinfo {pages} {A13} (\bibinfo {year} {2016})},\ \Eprint
  {http://arxiv.org/abs/1502.01589} {arXiv:1502.01589 [astro-ph.CO]}
  \BibitemShut {NoStop}%
%%CITATION = ARXIV:1502.01589;%%
\bibitem [{\citenamefont {Springel}(2005)}]{Springel:2005mi}%
  \BibitemOpen
  \bibfield  {author} {\bibinfo {author} {\bibfnamefont {V.}~\bibnamefont
  {Springel}},\ }\href {\doibase 10.1111/j.1365-2966.2005.09655.x} {\bibfield
  {journal} {\bibinfo  {journal} {Mon. Not. Roy. Astron. Soc.}\ }\textbf
  {\bibinfo {volume} {364}},\ \bibinfo {pages} {1105} (\bibinfo {year}
  {2005})},\ \Eprint {http://arxiv.org/abs/astro-ph/0505010}
  {arXiv:astro-ph/0505010} \BibitemShut {NoStop}%
%%CITATION = ASTRO-PH/0505010;%%
\bibitem [{\citenamefont {Vogelsberger}\ \emph {et~al.}(2014)\citenamefont
  {Vogelsberger}, \citenamefont {Genel}, \citenamefont {Springel},
  \citenamefont {Torrey}, \citenamefont {Sijacki}, \citenamefont {Xu},
  \citenamefont {Snyder}, \citenamefont {Bird}, \citenamefont {Nelson},\ and\
  \citenamefont {Hernquist}}]{Vogelsberger:2014kha}%
  \BibitemOpen
  \bibfield  {author} {\bibinfo {author} {\bibfnamefont {M.}~\bibnamefont
  {Vogelsberger}}, \bibinfo {author} {\bibfnamefont {S.}~\bibnamefont {Genel}},
  \bibinfo {author} {\bibfnamefont {V.}~\bibnamefont {Springel}}, \bibinfo
  {author} {\bibfnamefont {P.}~\bibnamefont {Torrey}}, \bibinfo {author}
  {\bibfnamefont {D.}~\bibnamefont {Sijacki}}, \bibinfo {author} {\bibfnamefont
  {D.}~\bibnamefont {Xu}}, \bibinfo {author} {\bibfnamefont {G.~F.}\
  \bibnamefont {Snyder}}, \bibinfo {author} {\bibfnamefont {S.}~\bibnamefont
  {Bird}}, \bibinfo {author} {\bibfnamefont {D.}~\bibnamefont {Nelson}}, \ and\
  \bibinfo {author} {\bibfnamefont {L.}~\bibnamefont {Hernquist}},\ }\href
  {\doibase 10.1038/nature13316} {\bibfield  {journal} {\bibinfo  {journal}
  {Nature}\ }\textbf {\bibinfo {volume} {509}},\ \bibinfo {pages} {177}
  (\bibinfo {year} {2014})},\ \Eprint {http://arxiv.org/abs/1405.1418}
  {arXiv:1405.1418 [astro-ph.CO]} \BibitemShut {NoStop}%
%%CITATION = ARXIV:1405.1418;%%
\bibitem [{\citenamefont {de~Blok}\ and\ \citenamefont
  {McGaugh}(1997)}]{deBlok:1997zlw}%
  \BibitemOpen
  \bibfield  {author} {\bibinfo {author} {\bibfnamefont {W.~J.~G.}\
  \bibnamefont {de~Blok}}\ and\ \bibinfo {author} {\bibfnamefont {S.~S.}\
  \bibnamefont {McGaugh}},\ }\href {\doibase 10.1093/mnras/290.3.533}
  {\bibfield  {journal} {\bibinfo  {journal} {Mon. Not. Roy. Astron. Soc.}\
  }\textbf {\bibinfo {volume} {290}},\ \bibinfo {pages} {533} (\bibinfo {year}
  {1997})},\ \Eprint {http://arxiv.org/abs/astro-ph/9704274}
  {arXiv:astro-ph/9704274} \BibitemShut {NoStop}%
%%CITATION = ASTRO-PH/9704274;%%
\bibitem [{\citenamefont {Klypin}\ \emph {et~al.}(1999)\citenamefont {Klypin},
  \citenamefont {Kravtsov}, \citenamefont {Valenzuela},\ and\ \citenamefont
  {Prada}}]{Klypin:1999uc}%
  \BibitemOpen
  \bibfield  {author} {\bibinfo {author} {\bibfnamefont {A.~A.}\ \bibnamefont
  {Klypin}}, \bibinfo {author} {\bibfnamefont {A.~V.}\ \bibnamefont
  {Kravtsov}}, \bibinfo {author} {\bibfnamefont {O.}~\bibnamefont
  {Valenzuela}}, \ and\ \bibinfo {author} {\bibfnamefont {F.}~\bibnamefont
  {Prada}},\ }\href {\doibase 10.1086/307643} {\bibfield  {journal} {\bibinfo
  {journal} {Astrophys. J.}\ }\textbf {\bibinfo {volume} {522}},\ \bibinfo
  {pages} {82} (\bibinfo {year} {1999})},\ \Eprint
  {http://arxiv.org/abs/astro-ph/9901240} {arXiv:astro-ph/9901240} \BibitemShut
  {NoStop}%
%%CITATION = ASTRO-PH/9901240;%%
\bibitem [{\citenamefont {Moore}\ \emph {et~al.}(1999)\citenamefont {Moore},
  \citenamefont {Ghigna}, \citenamefont {Governato}, \citenamefont {Lake},
  \citenamefont {Quinn}, \citenamefont {Stadel},\ and\ \citenamefont
  {Tozzi}}]{Moore:1999nt}%
  \BibitemOpen
  \bibfield  {author} {\bibinfo {author} {\bibfnamefont {B.}~\bibnamefont
  {Moore}}, \bibinfo {author} {\bibfnamefont {S.}~\bibnamefont {Ghigna}},
  \bibinfo {author} {\bibfnamefont {F.}~\bibnamefont {Governato}}, \bibinfo
  {author} {\bibfnamefont {G.}~\bibnamefont {Lake}}, \bibinfo {author}
  {\bibfnamefont {T.~R.}\ \bibnamefont {Quinn}}, \bibinfo {author}
  {\bibfnamefont {J.}~\bibnamefont {Stadel}}, \ and\ \bibinfo {author}
  {\bibfnamefont {P.}~\bibnamefont {Tozzi}},\ }\href {\doibase 10.1086/312287}
  {\bibfield  {journal} {\bibinfo  {journal} {Astrophys. J.}\ }\textbf
  {\bibinfo {volume} {524}},\ \bibinfo {pages} {L19} (\bibinfo {year}
  {1999})},\ \Eprint {http://arxiv.org/abs/astro-ph/9907411}
  {arXiv:astro-ph/9907411} \BibitemShut {NoStop}%
%%CITATION = ASTRO-PH/9907411;%%
\bibitem [{\citenamefont {Zavala}\ \emph {et~al.}(2009)\citenamefont {Zavala},
  \citenamefont {Jing}, \citenamefont {Faltenbacher}, \citenamefont {Yepes},
  \citenamefont {Hoffman}, \citenamefont {Gottlober},\ and\ \citenamefont
  {Catinella}}]{Zavala:2009ms}%
  \BibitemOpen
  \bibfield  {author} {\bibinfo {author} {\bibfnamefont {J.}~\bibnamefont
  {Zavala}}, \bibinfo {author} {\bibfnamefont {Y.~P.}\ \bibnamefont {Jing}},
  \bibinfo {author} {\bibfnamefont {A.}~\bibnamefont {Faltenbacher}}, \bibinfo
  {author} {\bibfnamefont {G.}~\bibnamefont {Yepes}}, \bibinfo {author}
  {\bibfnamefont {Y.}~\bibnamefont {Hoffman}}, \bibinfo {author} {\bibfnamefont
  {S.}~\bibnamefont {Gottlober}}, \ and\ \bibinfo {author} {\bibfnamefont
  {B.}~\bibnamefont {Catinella}},\ }\href {\doibase
  10.1088/0004-637X/700/2/1779} {\bibfield  {journal} {\bibinfo  {journal}
  {Astrophys. J.}\ }\textbf {\bibinfo {volume} {700}},\ \bibinfo {pages} {1779}
  (\bibinfo {year} {2009})},\ \Eprint {http://arxiv.org/abs/0906.0585}
  {arXiv:0906.0585 [astro-ph.CO]} \BibitemShut {NoStop}%
%%CITATION = ARXIV:0906.0585;%%
\bibitem [{\citenamefont {Oh}\ \emph {et~al.}(2011)\citenamefont {Oh},
  \citenamefont {de~Blok}, \citenamefont {Brinks}, \citenamefont {Walter},\
  and\ \citenamefont {Kennicutt}}]{Oh:2010ea}%
  \BibitemOpen
  \bibfield  {author} {\bibinfo {author} {\bibfnamefont {S.-H.}\ \bibnamefont
  {Oh}}, \bibinfo {author} {\bibfnamefont {W.~J.~G.}\ \bibnamefont {de~Blok}},
  \bibinfo {author} {\bibfnamefont {E.}~\bibnamefont {Brinks}}, \bibinfo
  {author} {\bibfnamefont {F.}~\bibnamefont {Walter}}, \ and\ \bibinfo {author}
  {\bibfnamefont {R.~C.}\ \bibnamefont {Kennicutt}, \bibfnamefont {Jr}},\
  }\href {\doibase 10.1088/0004-6256/141/6/193} {\bibfield  {journal} {\bibinfo
   {journal} {Astron. J.}\ }\textbf {\bibinfo {volume} {141}},\ \bibinfo
  {pages} {193} (\bibinfo {year} {2011})},\ \Eprint
  {http://arxiv.org/abs/1011.0899} {arXiv:1011.0899 [astro-ph.CO]} \BibitemShut
  {NoStop}%
%%CITATION = ARXIV:1011.0899;%%
\bibitem [{\citenamefont {Papastergis}\ \emph {et~al.}(2011)\citenamefont
  {Papastergis}, \citenamefont {Martin}, \citenamefont {Giovanelli},\ and\
  \citenamefont {Haynes}}]{Papastergis:2011xe}%
  \BibitemOpen
  \bibfield  {author} {\bibinfo {author} {\bibfnamefont {E.}~\bibnamefont
  {Papastergis}}, \bibinfo {author} {\bibfnamefont {A.~M.}\ \bibnamefont
  {Martin}}, \bibinfo {author} {\bibfnamefont {R.}~\bibnamefont {Giovanelli}},
  \ and\ \bibinfo {author} {\bibfnamefont {M.~P.}\ \bibnamefont {Haynes}},\
  }\href {\doibase 10.1088/0004-637X/739/1/38} {\bibfield  {journal} {\bibinfo
  {journal} {Astrophys. J.}\ }\textbf {\bibinfo {volume} {739}},\ \bibinfo
  {pages} {38} (\bibinfo {year} {2011})},\ \Eprint
  {http://arxiv.org/abs/1106.0710} {arXiv:1106.0710 [astro-ph.CO]} \BibitemShut
  {NoStop}%
%%CITATION = ARXIV:1106.0710;%%
\bibitem [{\citenamefont {Boylan-Kolchin}\ \emph {et~al.}(2011)\citenamefont
  {Boylan-Kolchin}, \citenamefont {Bullock},\ and\ \citenamefont
  {Kaplinghat}}]{BoylanKolchin:2011de}%
  \BibitemOpen
  \bibfield  {author} {\bibinfo {author} {\bibfnamefont {M.}~\bibnamefont
  {Boylan-Kolchin}}, \bibinfo {author} {\bibfnamefont {J.~S.}\ \bibnamefont
  {Bullock}}, \ and\ \bibinfo {author} {\bibfnamefont {M.}~\bibnamefont
  {Kaplinghat}},\ }\href@noop {} {\bibfield  {journal} {\bibinfo  {journal}
  {Mon. Not. Roy. Astron. Soc.}\ }\textbf {\bibinfo {volume} {415}},\ \bibinfo
  {pages} {L40} (\bibinfo {year} {2011})},\ \Eprint
  {http://arxiv.org/abs/1103.0007} {arXiv:1103.0007 [astro-ph.CO]} \BibitemShut
  {NoStop}%
%%CITATION = ARXIV:1103.0007;%%
\bibitem [{\citenamefont {Walker}\ and\ \citenamefont
  {Penarrubia}(2011)}]{Walker:2011zu}%
  \BibitemOpen
  \bibfield  {author} {\bibinfo {author} {\bibfnamefont {M.~G.}\ \bibnamefont
  {Walker}}\ and\ \bibinfo {author} {\bibfnamefont {J.}~\bibnamefont
  {Penarrubia}},\ }\href {\doibase 10.1088/0004-637X/742/1/20} {\bibfield
  {journal} {\bibinfo  {journal} {Astrophys. J.}\ }\textbf {\bibinfo {volume}
  {742}},\ \bibinfo {pages} {20} (\bibinfo {year} {2011})},\ \Eprint
  {http://arxiv.org/abs/1108.2404} {arXiv:1108.2404 [astro-ph.CO]} \BibitemShut
  {NoStop}%
%%CITATION = ARXIV:1108.2404;%%
\bibitem [{\citenamefont {Pawlowski}\ \emph {et~al.}(2013)\citenamefont
  {Pawlowski}, \citenamefont {Kroupa},\ and\ \citenamefont
  {Jerjen}}]{Pawlowski:2013kpa}%
  \BibitemOpen
  \bibfield  {author} {\bibinfo {author} {\bibfnamefont {M.~S.}\ \bibnamefont
  {Pawlowski}}, \bibinfo {author} {\bibfnamefont {P.}~\bibnamefont {Kroupa}}, \
  and\ \bibinfo {author} {\bibfnamefont {H.}~\bibnamefont {Jerjen}},\ }\href
  {\doibase 10.1093/mnras/stt1384} {\bibfield  {journal} {\bibinfo  {journal}
  {Mon. Not. Roy. Astron. Soc.}\ }\textbf {\bibinfo {volume} {435}},\ \bibinfo
  {pages} {1928} (\bibinfo {year} {2013})},\ \Eprint
  {http://arxiv.org/abs/1307.6210} {arXiv:1307.6210 [astro-ph.CO]} \BibitemShut
  {NoStop}%
%%CITATION = ARXIV:1307.6210;%%
\bibitem [{\citenamefont {Klypin}\ \emph {et~al.}(2015)\citenamefont {Klypin},
  \citenamefont {Karachentsev}, \citenamefont {Makarov},\ and\ \citenamefont
  {Nasonova}}]{Klypin:2014ira}%
  \BibitemOpen
  \bibfield  {author} {\bibinfo {author} {\bibfnamefont {A.}~\bibnamefont
  {Klypin}}, \bibinfo {author} {\bibfnamefont {I.}~\bibnamefont
  {Karachentsev}}, \bibinfo {author} {\bibfnamefont {D.}~\bibnamefont
  {Makarov}}, \ and\ \bibinfo {author} {\bibfnamefont {O.}~\bibnamefont
  {Nasonova}},\ }\href {\doibase 10.1093/mnras/stv2040} {\bibfield  {journal}
  {\bibinfo  {journal} {Mon. Not. Roy. Astron. Soc.}\ }\textbf {\bibinfo
  {volume} {454}},\ \bibinfo {pages} {1798} (\bibinfo {year} {2015})},\ \Eprint
  {http://arxiv.org/abs/1405.4523} {arXiv:1405.4523 [astro-ph.CO]} \BibitemShut
  {NoStop}%
%%CITATION = ARXIV:1405.4523;%%
\bibitem [{\citenamefont {Papastergis}\ \emph {et~al.}(2015)\citenamefont
  {Papastergis}, \citenamefont {Giovanelli}, \citenamefont {Haynes},\ and\
  \citenamefont {Shankar}}]{Papastergis:2014aba}%
  \BibitemOpen
  \bibfield  {author} {\bibinfo {author} {\bibfnamefont {E.}~\bibnamefont
  {Papastergis}}, \bibinfo {author} {\bibfnamefont {R.}~\bibnamefont
  {Giovanelli}}, \bibinfo {author} {\bibfnamefont {M.~P.}\ \bibnamefont
  {Haynes}}, \ and\ \bibinfo {author} {\bibfnamefont {F.}~\bibnamefont
  {Shankar}},\ }\href {\doibase 10.1051/0004-6361/201424909} {\bibfield
  {journal} {\bibinfo  {journal} {Astron. Astrophys.}\ }\textbf {\bibinfo
  {volume} {574}},\ \bibinfo {pages} {A113} (\bibinfo {year} {2015})},\ \Eprint
  {http://arxiv.org/abs/1407.4665} {arXiv:1407.4665 [astro-ph.GA]} \BibitemShut
  {NoStop}%
%%CITATION = ARXIV:1407.4665;%%
\bibitem [{\citenamefont {Oman}\ \emph {et~al.}(2015)\citenamefont {Oman} \emph
  {et~al.}}]{Oman:2015xda}%
  \BibitemOpen
  \bibfield  {author} {\bibinfo {author} {\bibfnamefont {K.~A.}\ \bibnamefont
  {Oman}} \emph {et~al.},\ }\href {\doibase 10.1093/mnras/stv1504} {\bibfield
  {journal} {\bibinfo  {journal} {Mon. Not. Roy. Astron. Soc.}\ }\textbf
  {\bibinfo {volume} {452}},\ \bibinfo {pages} {3650} (\bibinfo {year}
  {2015})},\ \Eprint {http://arxiv.org/abs/1504.01437} {arXiv:1504.01437
  [astro-ph.GA]} \BibitemShut {NoStop}%
%%CITATION = ARXIV:1504.01437;%%
\bibitem [{\citenamefont {Massey}\ \emph {et~al.}(2015)\citenamefont {Massey}
  \emph {et~al.}}]{Massey:2015dkw}%
  \BibitemOpen
  \bibfield  {author} {\bibinfo {author} {\bibfnamefont {R.}~\bibnamefont
  {Massey}} \emph {et~al.},\ }\href {\doibase 10.1093/mnras/stv467} {\bibfield
  {journal} {\bibinfo  {journal} {Mon. Not. Roy. Astron. Soc.}\ }\textbf
  {\bibinfo {volume} {449}},\ \bibinfo {pages} {3393} (\bibinfo {year}
  {2015})},\ \Eprint {http://arxiv.org/abs/1504.03388} {arXiv:1504.03388
  [astro-ph.CO]} \BibitemShut {NoStop}%
%%CITATION = ARXIV:1504.03388;%%
\bibitem [{\citenamefont {Vogelsberger}\ \emph {et~al.}(2016)\citenamefont
  {Vogelsberger}, \citenamefont {Zavala}, \citenamefont {Cyr-Racine},
  \citenamefont {Pfrommer}, \citenamefont {Bringmann},\ and\ \citenamefont
  {Sigurdson}}]{Vogelsberger:2015gpr}%
  \BibitemOpen
  \bibfield  {author} {\bibinfo {author} {\bibfnamefont {M.}~\bibnamefont
  {Vogelsberger}}, \bibinfo {author} {\bibfnamefont {J.}~\bibnamefont
  {Zavala}}, \bibinfo {author} {\bibfnamefont {F.-Y.}\ \bibnamefont
  {Cyr-Racine}}, \bibinfo {author} {\bibfnamefont {C.}~\bibnamefont
  {Pfrommer}}, \bibinfo {author} {\bibfnamefont {T.}~\bibnamefont {Bringmann}},
  \ and\ \bibinfo {author} {\bibfnamefont {K.}~\bibnamefont {Sigurdson}},\
  }\href {\doibase 10.1093/mnras/stw1076} {\bibfield  {journal} {\bibinfo
  {journal} {Mon. Not. Roy. Astron. Soc.}\ }\textbf {\bibinfo {volume} {460}},\
  \bibinfo {pages} {1399} (\bibinfo {year} {2016})},\ \Eprint
  {http://arxiv.org/abs/1512.05349} {arXiv:1512.05349 [astro-ph.CO]}
  \BibitemShut {NoStop}%
%%CITATION = ARXIV:1512.05349;%%
\bibitem [{\citenamefont {Silk}\ \emph {et~al.}(2010)\citenamefont {Silk} \emph
  {et~al.}}]{Bertone:2010zza}%
  \BibitemOpen
  \bibfield  {author} {\bibinfo {author} {\bibfnamefont {J.}~\bibnamefont
  {Silk}} \emph {et~al.},\ }\href
  {http://www.cambridge.org/uk/catalogue/catalogue.asp?isbn=9780521763684}
  {\emph {\bibinfo {title} {{Particle Dark Matter: Observations, Models and
  Searches}}}},\ edited by\ \bibinfo {editor} {\bibfnamefont {G.}~\bibnamefont
  {Bertone}}\ (\bibinfo {year} {2010})\BibitemShut {NoStop}%
%%CITATION = INSPIRE-895273;%%
\bibitem [{\citenamefont {Jungman}\ \emph {et~al.}(1996)\citenamefont
  {Jungman}, \citenamefont {Kamionkowski},\ and\ \citenamefont
  {Griest}}]{Jungman:1995df}%
  \BibitemOpen
  \bibfield  {author} {\bibinfo {author} {\bibfnamefont {G.}~\bibnamefont
  {Jungman}}, \bibinfo {author} {\bibfnamefont {M.}~\bibnamefont
  {Kamionkowski}}, \ and\ \bibinfo {author} {\bibfnamefont {K.}~\bibnamefont
  {Griest}},\ }\href {\doibase 10.1016/0370-1573(95)00058-5} {\bibfield
  {journal} {\bibinfo  {journal} {Phys. Rept.}\ }\textbf {\bibinfo {volume}
  {267}},\ \bibinfo {pages} {195} (\bibinfo {year} {1996})},\ \Eprint
  {http://arxiv.org/abs/hep-ph/9506380} {arXiv:hep-ph/9506380 [hep-ph]}
  \BibitemShut {NoStop}%
%%CITATION = HEP-PH/9506380;%%
\bibitem [{\citenamefont {Hooper}\ and\ \citenamefont
  {Profumo}(2007)}]{Hooper:2007qk}%
  \BibitemOpen
  \bibfield  {author} {\bibinfo {author} {\bibfnamefont {D.}~\bibnamefont
  {Hooper}}\ and\ \bibinfo {author} {\bibfnamefont {S.}~\bibnamefont
  {Profumo}},\ }\href {\doibase 10.1016/j.physrep.2007.09.003} {\bibfield
  {journal} {\bibinfo  {journal} {Phys. Rept.}\ }\textbf {\bibinfo {volume}
  {453}},\ \bibinfo {pages} {29} (\bibinfo {year} {2007})},\ \Eprint
  {http://arxiv.org/abs/hep-ph/0701197} {arXiv:hep-ph/0701197 [hep-ph]}
  \BibitemShut {NoStop}%
%%CITATION = HEP-PH/0701197;%%
\bibitem [{\citenamefont {Cakir}(2015)}]{Cakir:2015gya}%
  \BibitemOpen
  \bibfield  {author} {\bibinfo {author} {\bibfnamefont {A.}~\bibnamefont
  {Cakir}} (\bibinfo {collaboration} {ATLAS, CMS}),\ }\bibfield  {booktitle}
  {\emph {\bibinfo {booktitle} {{Proceedings, 13th Conference on Flavor Physics
  and CP Violation (FPCP 2015)}}},\ }\href@noop {} {\bibfield  {journal}
  {\bibinfo  {journal} {PoS}\ }\textbf {\bibinfo {volume} {FPCP2015}},\
  \bibinfo {pages} {024} (\bibinfo {year} {2015})},\ \Eprint
  {http://arxiv.org/abs/1507.08427} {arXiv:1507.08427 [hep-ph]} \BibitemShut
  {NoStop}%
%%CITATION = ARXIV:1507.08427;%%
\bibitem [{\citenamefont {Aad}\ \emph {et~al.}(2015)\citenamefont {Aad} \emph
  {et~al.}}]{Aad:2015baa}%
  \BibitemOpen
  \bibfield  {author} {\bibinfo {author} {\bibfnamefont {G.}~\bibnamefont
  {Aad}} \emph {et~al.} (\bibinfo {collaboration} {ATLAS}),\ }\href {\doibase
  10.1007/JHEP10(2015)134} {\bibfield  {journal} {\bibinfo  {journal} {JHEP}\
  }\textbf {\bibinfo {volume} {10}},\ \bibinfo {pages} {134} (\bibinfo {year}
  {2015})},\ \Eprint {http://arxiv.org/abs/1508.06608} {arXiv:1508.06608
  [hep-ex]} \BibitemShut {NoStop}%
%%CITATION = ARXIV:1508.06608;%%
\bibitem [{\citenamefont {Aprile}\ \emph {et~al.}(2012)\citenamefont {Aprile}
  \emph {et~al.}}]{Aprile:2012nq}%
  \BibitemOpen
  \bibfield  {author} {\bibinfo {author} {\bibfnamefont {E.}~\bibnamefont
  {Aprile}} \emph {et~al.} (\bibinfo {collaboration} {XENON100}),\ }\href
  {\doibase 10.1103/PhysRevLett.109.181301} {\bibfield  {journal} {\bibinfo
  {journal} {Phys. Rev. Lett.}\ }\textbf {\bibinfo {volume} {109}},\ \bibinfo
  {pages} {181301} (\bibinfo {year} {2012})},\ \Eprint
  {http://arxiv.org/abs/1207.5988} {arXiv:1207.5988 [astro-ph.CO]} \BibitemShut
  {NoStop}%
%%CITATION = ARXIV:1207.5988;%%
\bibitem [{\citenamefont {Akerib}\ \emph {et~al.}(2014)\citenamefont {Akerib}
  \emph {et~al.}}]{Akerib:2013tjd}%
  \BibitemOpen
  \bibfield  {author} {\bibinfo {author} {\bibfnamefont {D.~S.}\ \bibnamefont
  {Akerib}} \emph {et~al.} (\bibinfo {collaboration} {LUX}),\ }\href {\doibase
  10.1103/PhysRevLett.112.091303} {\bibfield  {journal} {\bibinfo  {journal}
  {Phys. Rev. Lett.}\ }\textbf {\bibinfo {volume} {112}},\ \bibinfo {pages}
  {091303} (\bibinfo {year} {2014})},\ \Eprint {http://arxiv.org/abs/1310.8214}
  {arXiv:1310.8214 [astro-ph.CO]} \BibitemShut {NoStop}%
%%CITATION = ARXIV:1310.8214;%%
\bibitem [{\citenamefont {Steffen}(2009)}]{Steffen:2008qp}%
  \BibitemOpen
  \bibfield  {author} {\bibinfo {author} {\bibfnamefont {F.~D.}\ \bibnamefont
  {Steffen}},\ }\href {\doibase 10.1140/epjc/s10052-008-0830-0} {\bibfield
  {journal} {\bibinfo  {journal} {Eur. Phys. J.}\ }\textbf {\bibinfo {volume}
  {C59}},\ \bibinfo {pages} {557} (\bibinfo {year} {2009})},\ \Eprint
  {http://arxiv.org/abs/0811.3347} {arXiv:0811.3347 [hep-ph]} \BibitemShut
  {NoStop}%
%%CITATION = ARXIV:0811.3347;%%
\bibitem [{\citenamefont {Feng}(2010)}]{Feng:2010gw}%
  \BibitemOpen
  \bibfield  {author} {\bibinfo {author} {\bibfnamefont {J.~L.}\ \bibnamefont
  {Feng}},\ }\href {\doibase 10.1146/annurev-astro-082708-101659} {\bibfield
  {journal} {\bibinfo  {journal} {Ann. Rev. Astron. Astrophys.}\ }\textbf
  {\bibinfo {volume} {48}},\ \bibinfo {pages} {495} (\bibinfo {year} {2010})},\
  \Eprint {http://arxiv.org/abs/1003.0904} {arXiv:1003.0904 [astro-ph.CO]}
  \BibitemShut {NoStop}%
%%CITATION = ARXIV:1003.0904;%%
\bibitem [{\citenamefont {Baer}\ \emph {et~al.}(2015)\citenamefont {Baer},
  \citenamefont {Choi}, \citenamefont {Kim},\ and\ \citenamefont
  {Roszkowski}}]{Baer:2014eja}%
  \BibitemOpen
  \bibfield  {author} {\bibinfo {author} {\bibfnamefont {H.}~\bibnamefont
  {Baer}}, \bibinfo {author} {\bibfnamefont {K.-Y.}\ \bibnamefont {Choi}},
  \bibinfo {author} {\bibfnamefont {J.~E.}\ \bibnamefont {Kim}}, \ and\
  \bibinfo {author} {\bibfnamefont {L.}~\bibnamefont {Roszkowski}},\ }\href
  {\doibase 10.1016/j.physrep.2014.10.002} {\bibfield  {journal} {\bibinfo
  {journal} {Phys. Rept.}\ }\textbf {\bibinfo {volume} {555}},\ \bibinfo
  {pages} {1} (\bibinfo {year} {2015})},\ \Eprint
  {http://arxiv.org/abs/1407.0017} {arXiv:1407.0017 [hep-ph]} \BibitemShut
  {NoStop}%
%%CITATION = ARXIV:1407.0017;%%
\bibitem [{\citenamefont {Adhikari}\ \emph {et~al.}(2016)\citenamefont
  {Adhikari} \emph {et~al.}}]{Adhikari:2016bei}%
  \BibitemOpen
  \bibfield  {author} {\bibinfo {author} {\bibfnamefont {R.}~\bibnamefont
  {Adhikari}} \emph {et~al.},\ }\href@noop {} {\  (\bibinfo {year} {2016})},\
  \Eprint {http://arxiv.org/abs/1602.04816} {arXiv:1602.04816 [hep-ph]}
  \BibitemShut {NoStop}%
%%CITATION = ARXIV:1602.04816;%%
\bibitem [{\citenamefont {Goldberg}\ and\ \citenamefont
  {Hall}(1986)}]{Goldberg:1986nk}%
  \BibitemOpen
  \bibfield  {author} {\bibinfo {author} {\bibfnamefont {H.}~\bibnamefont
  {Goldberg}}\ and\ \bibinfo {author} {\bibfnamefont {L.~J.}\ \bibnamefont
  {Hall}},\ }\href {\doibase 10.1016/0370-2693(86)90731-8} {\bibfield
  {journal} {\bibinfo  {journal} {Phys. Lett.}\ }\textbf {\bibinfo {volume}
  {B174}},\ \bibinfo {pages} {151} (\bibinfo {year} {1986})}\BibitemShut
  {NoStop}%
%%CITATION = PHLTA,B174,151;%%
\bibitem [{\citenamefont {Gradwohl}\ and\ \citenamefont
  {Frieman}(1992)}]{Gradwohl:1992ue}%
  \BibitemOpen
  \bibfield  {author} {\bibinfo {author} {\bibfnamefont {B.-A.}\ \bibnamefont
  {Gradwohl}}\ and\ \bibinfo {author} {\bibfnamefont {J.~A.}\ \bibnamefont
  {Frieman}},\ }\href {\doibase 10.1086/171865} {\bibfield  {journal} {\bibinfo
   {journal} {Astrophys. J.}\ }\textbf {\bibinfo {volume} {398}},\ \bibinfo
  {pages} {407} (\bibinfo {year} {1992})}\BibitemShut {NoStop}%
%%CITATION = ASJOA,398,407;%%
\bibitem [{\citenamefont {Bode}\ \emph {et~al.}(2001)\citenamefont {Bode},
  \citenamefont {Ostriker},\ and\ \citenamefont {Turok}}]{Bode:2000gq}%
  \BibitemOpen
  \bibfield  {author} {\bibinfo {author} {\bibfnamefont {P.}~\bibnamefont
  {Bode}}, \bibinfo {author} {\bibfnamefont {J.~P.}\ \bibnamefont {Ostriker}},
  \ and\ \bibinfo {author} {\bibfnamefont {N.}~\bibnamefont {Turok}},\ }\href
  {\doibase 10.1086/321541} {\bibfield  {journal} {\bibinfo  {journal}
  {Astrophys. J.}\ }\textbf {\bibinfo {volume} {556}},\ \bibinfo {pages} {93}
  (\bibinfo {year} {2001})},\ \Eprint {http://arxiv.org/abs/astro-ph/0010389}
  {arXiv:astro-ph/0010389} \BibitemShut {NoStop}%
%%CITATION = ASTRO-PH/0010389;%%
\bibitem [{\citenamefont {Hannestad}\ and\ \citenamefont
  {Scherrer}(2000)}]{Hannestad:2000gt}%
  \BibitemOpen
  \bibfield  {author} {\bibinfo {author} {\bibfnamefont {S.}~\bibnamefont
  {Hannestad}}\ and\ \bibinfo {author} {\bibfnamefont {R.~J.}\ \bibnamefont
  {Scherrer}},\ }\href {\doibase 10.1103/PhysRevD.62.043522} {\bibfield
  {journal} {\bibinfo  {journal} {Phys. Rev.}\ }\textbf {\bibinfo {volume}
  {D62}},\ \bibinfo {pages} {043522} (\bibinfo {year} {2000})},\ \Eprint
  {http://arxiv.org/abs/astro-ph/0003046} {arXiv:astro-ph/0003046} \BibitemShut
  {NoStop}%
%%CITATION = ASTRO-PH/0003046;%%
\bibitem [{\citenamefont {Boehm}\ \emph {et~al.}(2001)\citenamefont {Boehm},
  \citenamefont {Fayet},\ and\ \citenamefont {Schaeffer}}]{Boehm:2000gq}%
  \BibitemOpen
  \bibfield  {author} {\bibinfo {author} {\bibfnamefont {C.}~\bibnamefont
  {Boehm}}, \bibinfo {author} {\bibfnamefont {P.}~\bibnamefont {Fayet}}, \ and\
  \bibinfo {author} {\bibfnamefont {R.}~\bibnamefont {Schaeffer}},\ }\href
  {\doibase 10.1016/S0370-2693(01)01060-7} {\bibfield  {journal} {\bibinfo
  {journal} {Phys. Lett.}\ }\textbf {\bibinfo {volume} {B518}},\ \bibinfo
  {pages} {8} (\bibinfo {year} {2001})},\ \Eprint
  {http://arxiv.org/abs/astro-ph/0012504} {arXiv:astro-ph/0012504} \BibitemShut
  {NoStop}%
%%CITATION = ASTRO-PH/0012504;%%
\bibitem [{\citenamefont {Boehm}\ \emph {et~al.}(2002)\citenamefont {Boehm},
  \citenamefont {Riazuelo}, \citenamefont {Hansen},\ and\ \citenamefont
  {Schaeffer}}]{Boehm:2001hm}%
  \BibitemOpen
  \bibfield  {author} {\bibinfo {author} {\bibfnamefont {C.}~\bibnamefont
  {Boehm}}, \bibinfo {author} {\bibfnamefont {A.}~\bibnamefont {Riazuelo}},
  \bibinfo {author} {\bibfnamefont {S.~H.}\ \bibnamefont {Hansen}}, \ and\
  \bibinfo {author} {\bibfnamefont {R.}~\bibnamefont {Schaeffer}},\ }\href
  {\doibase 10.1103/PhysRevD.66.083505} {\bibfield  {journal} {\bibinfo
  {journal} {Phys. Rev.}\ }\textbf {\bibinfo {volume} {D66}},\ \bibinfo {pages}
  {083505} (\bibinfo {year} {2002})},\ \Eprint
  {http://arxiv.org/abs/astro-ph/0112522} {arXiv:astro-ph/0112522} \BibitemShut
  {NoStop}%
%%CITATION = ASTRO-PH/0112522;%%
\bibitem [{\citenamefont {Boehm}\ and\ \citenamefont
  {Schaeffer}(2005)}]{Boehm:2004th}%
  \BibitemOpen
  \bibfield  {author} {\bibinfo {author} {\bibfnamefont {C.}~\bibnamefont
  {Boehm}}\ and\ \bibinfo {author} {\bibfnamefont {R.}~\bibnamefont
  {Schaeffer}},\ }\href {\doibase 10.1051/0004-6361:20042238} {\bibfield
  {journal} {\bibinfo  {journal} {Astron. Astrophys.}\ }\textbf {\bibinfo
  {volume} {438}},\ \bibinfo {pages} {419} (\bibinfo {year} {2005})},\ \Eprint
  {http://arxiv.org/abs/astro-ph/0410591} {arXiv:astro-ph/0410591} \BibitemShut
  {NoStop}%
%%CITATION = ASTRO-PH/0410591;%%
\bibitem [{\citenamefont {Cembranos}\ \emph {et~al.}(2005)\citenamefont
  {Cembranos}, \citenamefont {Feng}, \citenamefont {Rajaraman},\ and\
  \citenamefont {Takayama}}]{Cembranos:2005us}%
  \BibitemOpen
  \bibfield  {author} {\bibinfo {author} {\bibfnamefont {J.~A.~R.}\
  \bibnamefont {Cembranos}}, \bibinfo {author} {\bibfnamefont {J.~L.}\
  \bibnamefont {Feng}}, \bibinfo {author} {\bibfnamefont {A.}~\bibnamefont
  {Rajaraman}}, \ and\ \bibinfo {author} {\bibfnamefont {F.}~\bibnamefont
  {Takayama}},\ }\href {\doibase 10.1103/PhysRevLett.95.181301} {\bibfield
  {journal} {\bibinfo  {journal} {Phys. Rev. Lett.}\ }\textbf {\bibinfo
  {volume} {95}},\ \bibinfo {pages} {181301} (\bibinfo {year} {2005})},\
  \Eprint {http://arxiv.org/abs/hep-ph/0507150} {arXiv:hep-ph/0507150 [hep-ph]}
  \BibitemShut {NoStop}%
%%CITATION = HEP-PH/0507150;%%
\bibitem [{\citenamefont {Hooper}\ \emph {et~al.}(2007)\citenamefont {Hooper},
  \citenamefont {Kaplinghat}, \citenamefont {Strigari},\ and\ \citenamefont
  {Zurek}}]{Hooper:2007tu}%
  \BibitemOpen
  \bibfield  {author} {\bibinfo {author} {\bibfnamefont {D.}~\bibnamefont
  {Hooper}}, \bibinfo {author} {\bibfnamefont {M.}~\bibnamefont {Kaplinghat}},
  \bibinfo {author} {\bibfnamefont {L.~E.}\ \bibnamefont {Strigari}}, \ and\
  \bibinfo {author} {\bibfnamefont {K.~M.}\ \bibnamefont {Zurek}},\ }\href
  {\doibase 10.1103/PhysRevD.76.103515} {\bibfield  {journal} {\bibinfo
  {journal} {Phys. Rev.}\ }\textbf {\bibinfo {volume} {D76}},\ \bibinfo {pages}
  {103515} (\bibinfo {year} {2007})},\ \Eprint {http://arxiv.org/abs/0704.2558}
  {arXiv:0704.2558 [astro-ph]} \BibitemShut {NoStop}%
%%CITATION = ARXIV:0704.2558;%%
\bibitem [{\citenamefont {Feng}\ \emph {et~al.}(2009)\citenamefont {Feng},
  \citenamefont {Kaplinghat}, \citenamefont {Tu},\ and\ \citenamefont
  {Yu}}]{Feng:2009mn}%
  \BibitemOpen
  \bibfield  {author} {\bibinfo {author} {\bibfnamefont {J.~L.}\ \bibnamefont
  {Feng}}, \bibinfo {author} {\bibfnamefont {M.}~\bibnamefont {Kaplinghat}},
  \bibinfo {author} {\bibfnamefont {H.}~\bibnamefont {Tu}}, \ and\ \bibinfo
  {author} {\bibfnamefont {H.-B.}\ \bibnamefont {Yu}},\ }\href {\doibase
  10.1088/1475-7516/2009/07/004} {\bibfield  {journal} {\bibinfo  {journal}
  {JCAP}\ }\textbf {\bibinfo {volume} {0907}},\ \bibinfo {pages} {004}
  (\bibinfo {year} {2009})},\ \Eprint {http://arxiv.org/abs/0905.3039}
  {arXiv:0905.3039 [hep-ph]} \BibitemShut {NoStop}%
%%CITATION = ARXIV:0905.3039;%%
\bibitem [{\citenamefont {Kaplan}\ \emph {et~al.}(2010)\citenamefont {Kaplan},
  \citenamefont {Krnjaic}, \citenamefont {Rehermann},\ and\ \citenamefont
  {Wells}}]{Kaplan:2009de}%
  \BibitemOpen
  \bibfield  {author} {\bibinfo {author} {\bibfnamefont {D.~E.}\ \bibnamefont
  {Kaplan}}, \bibinfo {author} {\bibfnamefont {G.~Z.}\ \bibnamefont {Krnjaic}},
  \bibinfo {author} {\bibfnamefont {K.~R.}\ \bibnamefont {Rehermann}}, \ and\
  \bibinfo {author} {\bibfnamefont {C.~M.}\ \bibnamefont {Wells}},\ }\href
  {\doibase 10.1088/1475-7516/2010/05/021} {\bibfield  {journal} {\bibinfo
  {journal} {JCAP}\ }\textbf {\bibinfo {volume} {1005}},\ \bibinfo {pages}
  {021} (\bibinfo {year} {2010})},\ \Eprint {http://arxiv.org/abs/0909.0753}
  {arXiv:0909.0753 [hep-ph]} \BibitemShut {NoStop}%
%%CITATION = ARXIV:0909.0753;%%
\bibitem [{\citenamefont {Ackerman}\ \emph {et~al.}(2009)\citenamefont
  {Ackerman}, \citenamefont {Buckley}, \citenamefont {Carroll},\ and\
  \citenamefont {Kamionkowski}}]{Ackerman:mha}%
  \BibitemOpen
  \bibfield  {author} {\bibinfo {author} {\bibfnamefont {L.}~\bibnamefont
  {Ackerman}}, \bibinfo {author} {\bibfnamefont {M.~R.}\ \bibnamefont
  {Buckley}}, \bibinfo {author} {\bibfnamefont {S.~M.}\ \bibnamefont
  {Carroll}}, \ and\ \bibinfo {author} {\bibfnamefont {M.}~\bibnamefont
  {Kamionkowski}},\ }\bibfield  {booktitle} {\emph {\bibinfo {booktitle}
  {{Proceedings, 7th International Heidelberg Conference on Dark Matter in
  Astro and Particle Physics (DARK 2009)}}},\ }\href {\doibase
  10.1103/PhysRevD.79.023519, 10.1142/9789814293792_0021} {\bibfield  {journal}
  {\bibinfo  {journal} {Phys. Rev.}\ }\textbf {\bibinfo {volume} {D79}},\
  \bibinfo {pages} {023519} (\bibinfo {year} {2009})},\ \bibinfo {note}
  {[,277(2008)]},\ \Eprint {http://arxiv.org/abs/0810.5126} {arXiv:0810.5126
  [hep-ph]} \BibitemShut {NoStop}%
%%CITATION = ARXIV:0810.5126;%%
\bibitem [{\citenamefont {Kaplan}\ \emph {et~al.}(2011)\citenamefont {Kaplan},
  \citenamefont {Krnjaic}, \citenamefont {Rehermann},\ and\ \citenamefont
  {Wells}}]{Kaplan:2011yj}%
  \BibitemOpen
  \bibfield  {author} {\bibinfo {author} {\bibfnamefont {D.~E.}\ \bibnamefont
  {Kaplan}}, \bibinfo {author} {\bibfnamefont {G.~Z.}\ \bibnamefont {Krnjaic}},
  \bibinfo {author} {\bibfnamefont {K.~R.}\ \bibnamefont {Rehermann}}, \ and\
  \bibinfo {author} {\bibfnamefont {C.~M.}\ \bibnamefont {Wells}},\ }\href
  {\doibase 10.1088/1475-7516/2011/10/011} {\bibfield  {journal} {\bibinfo
  {journal} {JCAP}\ }\textbf {\bibinfo {volume} {1110}},\ \bibinfo {pages}
  {011} (\bibinfo {year} {2011})},\ \Eprint {http://arxiv.org/abs/1105.2073}
  {arXiv:1105.2073 [hep-ph]} \BibitemShut {NoStop}%
%%CITATION = ARXIV:1105.2073;%%
\bibitem [{\citenamefont {van~den Aarssen}\ \emph
  {et~al.}(2012{\natexlab{a}})\citenamefont {van~den Aarssen}, \citenamefont
  {Bringmann},\ and\ \citenamefont {Goedecke}}]{vandenAarssen:2012ag}%
  \BibitemOpen
  \bibfield  {author} {\bibinfo {author} {\bibfnamefont {L.~G.}\ \bibnamefont
  {van~den Aarssen}}, \bibinfo {author} {\bibfnamefont {T.}~\bibnamefont
  {Bringmann}}, \ and\ \bibinfo {author} {\bibfnamefont {Y.~C.}\ \bibnamefont
  {Goedecke}},\ }\href {\doibase 10.1103/PhysRevD.85.123512} {\bibfield
  {journal} {\bibinfo  {journal} {Phys. Rev.}\ }\textbf {\bibinfo {volume}
  {D85}},\ \bibinfo {pages} {123512} (\bibinfo {year} {2012}{\natexlab{a}})},\
  \Eprint {http://arxiv.org/abs/1202.5456} {arXiv:1202.5456 [hep-ph]}
  \BibitemShut {NoStop}%
%%CITATION = ARXIV:1202.5456;%%
\bibitem [{\citenamefont {van~den Aarssen}\ \emph
  {et~al.}(2012{\natexlab{b}})\citenamefont {van~den Aarssen}, \citenamefont
  {Bringmann},\ and\ \citenamefont {Pfrommer}}]{Aarssen:2012fx}%
  \BibitemOpen
  \bibfield  {author} {\bibinfo {author} {\bibfnamefont {L.~G.}\ \bibnamefont
  {van~den Aarssen}}, \bibinfo {author} {\bibfnamefont {T.}~\bibnamefont
  {Bringmann}}, \ and\ \bibinfo {author} {\bibfnamefont {C.}~\bibnamefont
  {Pfrommer}},\ }\href {\doibase 10.1103/PhysRevLett.109.231301} {\bibfield
  {journal} {\bibinfo  {journal} {Phys. Rev. Lett.}\ }\textbf {\bibinfo
  {volume} {109}},\ \bibinfo {pages} {231301} (\bibinfo {year}
  {2012}{\natexlab{b}})},\ \Eprint {http://arxiv.org/abs/1205.5809}
  {arXiv:1205.5809 [astro-ph.CO]} \BibitemShut {NoStop}%
%%CITATION = ARXIV:1205.5809;%%
\bibitem [{\citenamefont {Cline}\ \emph {et~al.}(2012)\citenamefont {Cline},
  \citenamefont {Liu},\ and\ \citenamefont {Xue}}]{Cline:2012is}%
  \BibitemOpen
  \bibfield  {author} {\bibinfo {author} {\bibfnamefont {J.~M.}\ \bibnamefont
  {Cline}}, \bibinfo {author} {\bibfnamefont {Z.}~\bibnamefont {Liu}}, \ and\
  \bibinfo {author} {\bibfnamefont {W.}~\bibnamefont {Xue}},\ }\href {\doibase
  10.1103/PhysRevD.85.101302} {\bibfield  {journal} {\bibinfo  {journal} {Phys.
  Rev.}\ }\textbf {\bibinfo {volume} {D85}},\ \bibinfo {pages} {101302}
  (\bibinfo {year} {2012})},\ \Eprint {http://arxiv.org/abs/1201.4858}
  {arXiv:1201.4858 [hep-ph]} \BibitemShut {NoStop}%
%%CITATION = ARXIV:1201.4858;%%
\bibitem [{\citenamefont {Cyr-Racine}\ and\ \citenamefont
  {Sigurdson}(2013)}]{CyrRacine:2012fz}%
  \BibitemOpen
  \bibfield  {author} {\bibinfo {author} {\bibfnamefont {F.-Y.}\ \bibnamefont
  {Cyr-Racine}}\ and\ \bibinfo {author} {\bibfnamefont {K.}~\bibnamefont
  {Sigurdson}},\ }\href {\doibase 10.1103/PhysRevD.87.103515} {\bibfield
  {journal} {\bibinfo  {journal} {Phys. Rev.}\ }\textbf {\bibinfo {volume}
  {D87}},\ \bibinfo {pages} {103515} (\bibinfo {year} {2013})},\ \Eprint
  {http://arxiv.org/abs/1209.5752} {arXiv:1209.5752 [astro-ph.CO]} \BibitemShut
  {NoStop}%
%%CITATION = ARXIV:1209.5752;%%
\bibitem [{\citenamefont {Tulin}\ \emph
  {et~al.}(2013{\natexlab{a}})\citenamefont {Tulin}, \citenamefont {Yu},\ and\
  \citenamefont {Zurek}}]{Tulin:2012wi}%
  \BibitemOpen
  \bibfield  {author} {\bibinfo {author} {\bibfnamefont {S.}~\bibnamefont
  {Tulin}}, \bibinfo {author} {\bibfnamefont {H.-B.}\ \bibnamefont {Yu}}, \
  and\ \bibinfo {author} {\bibfnamefont {K.~M.}\ \bibnamefont {Zurek}},\ }\href
  {\doibase 10.1103/PhysRevLett.110.111301} {\bibfield  {journal} {\bibinfo
  {journal} {Phys. Rev. Lett.}\ }\textbf {\bibinfo {volume} {110}},\ \bibinfo
  {pages} {111301} (\bibinfo {year} {2013}{\natexlab{a}})},\ \Eprint
  {http://arxiv.org/abs/1210.0900} {arXiv:1210.0900 [hep-ph]} \BibitemShut
  {NoStop}%
%%CITATION = ARXIV:1210.0900;%%
\bibitem [{\citenamefont {Bringmann}\ \emph {et~al.}(2014)\citenamefont
  {Bringmann}, \citenamefont {Hasenkamp},\ and\ \citenamefont
  {Kersten}}]{Bringmann:2013vra}%
  \BibitemOpen
  \bibfield  {author} {\bibinfo {author} {\bibfnamefont {T.}~\bibnamefont
  {Bringmann}}, \bibinfo {author} {\bibfnamefont {J.}~\bibnamefont
  {Hasenkamp}}, \ and\ \bibinfo {author} {\bibfnamefont {J.}~\bibnamefont
  {Kersten}},\ }\href {\doibase 10.1088/1475-7516/2014/07/042} {\bibfield
  {journal} {\bibinfo  {journal} {JCAP}\ }\textbf {\bibinfo {volume} {1407}},\
  \bibinfo {pages} {042} (\bibinfo {year} {2014})},\ \Eprint
  {http://arxiv.org/abs/1312.4947} {arXiv:1312.4947 [hep-ph]} \BibitemShut
  {NoStop}%
%%CITATION = ARXIV:1312.4947;%%
\bibitem [{\citenamefont {Dasgupta}\ and\ \citenamefont
  {Kopp}(2014)}]{Dasgupta:2013zpn}%
  \BibitemOpen
  \bibfield  {author} {\bibinfo {author} {\bibfnamefont {B.}~\bibnamefont
  {Dasgupta}}\ and\ \bibinfo {author} {\bibfnamefont {J.}~\bibnamefont
  {Kopp}},\ }\href {\doibase 10.1103/PhysRevLett.112.031803} {\bibfield
  {journal} {\bibinfo  {journal} {Phys. Rev. Lett.}\ }\textbf {\bibinfo
  {volume} {112}},\ \bibinfo {pages} {031803} (\bibinfo {year} {2014})},\
  \Eprint {http://arxiv.org/abs/1310.6337} {arXiv:1310.6337 [hep-ph]}
  \BibitemShut {NoStop}%
%%CITATION = ARXIV:1310.6337;%%
\bibitem [{\citenamefont {Tulin}\ \emph
  {et~al.}(2013{\natexlab{b}})\citenamefont {Tulin}, \citenamefont {Yu},\ and\
  \citenamefont {Zurek}}]{Tulin:2013teo}%
  \BibitemOpen
  \bibfield  {author} {\bibinfo {author} {\bibfnamefont {S.}~\bibnamefont
  {Tulin}}, \bibinfo {author} {\bibfnamefont {H.-B.}\ \bibnamefont {Yu}}, \
  and\ \bibinfo {author} {\bibfnamefont {K.~M.}\ \bibnamefont {Zurek}},\ }\href
  {\doibase 10.1103/PhysRevD.87.115007} {\bibfield  {journal} {\bibinfo
  {journal} {Phys. Rev.}\ }\textbf {\bibinfo {volume} {D87}},\ \bibinfo {pages}
  {115007} (\bibinfo {year} {2013}{\natexlab{b}})},\ \Eprint
  {http://arxiv.org/abs/1302.3898} {arXiv:1302.3898 [hep-ph]} \BibitemShut
  {NoStop}%
%%CITATION = ARXIV:1302.3898;%%
\bibitem [{\citenamefont {Cyr-Racine}\ \emph {et~al.}(2014)\citenamefont
  {Cyr-Racine}, \citenamefont {de~Putter}, \citenamefont {Raccanelli},\ and\
  \citenamefont {Sigurdson}}]{Cyr-Racine:2013fsa}%
  \BibitemOpen
  \bibfield  {author} {\bibinfo {author} {\bibfnamefont {F.-Y.}\ \bibnamefont
  {Cyr-Racine}}, \bibinfo {author} {\bibfnamefont {R.}~\bibnamefont
  {de~Putter}}, \bibinfo {author} {\bibfnamefont {A.}~\bibnamefont
  {Raccanelli}}, \ and\ \bibinfo {author} {\bibfnamefont {K.}~\bibnamefont
  {Sigurdson}},\ }\href {\doibase 10.1103/PhysRevD.89.063517} {\bibfield
  {journal} {\bibinfo  {journal} {Phys. Rev.}\ }\textbf {\bibinfo {volume}
  {D89}},\ \bibinfo {pages} {063517} (\bibinfo {year} {2014})},\ \Eprint
  {http://arxiv.org/abs/1310.3278} {arXiv:1310.3278 [astro-ph.CO]} \BibitemShut
  {NoStop}%
%%CITATION = ARXIV:1310.3278;%%
\bibitem [{\citenamefont {Shoemaker}(2013)}]{Shoemaker:2013tda}%
  \BibitemOpen
  \bibfield  {author} {\bibinfo {author} {\bibfnamefont {I.~M.}\ \bibnamefont
  {Shoemaker}},\ }\href {\doibase 10.1016/j.dark.2013.07.002} {\bibfield
  {journal} {\bibinfo  {journal} {Phys. Dark Univ.}\ }\textbf {\bibinfo
  {volume} {2}},\ \bibinfo {pages} {157} (\bibinfo {year} {2013})},\ \Eprint
  {http://arxiv.org/abs/1305.1936} {arXiv:1305.1936 [hep-ph]} \BibitemShut
  {NoStop}%
%%CITATION = ARXIV:1305.1936;%%
\bibitem [{\citenamefont {Cline}\ \emph {et~al.}(2014)\citenamefont {Cline},
  \citenamefont {Liu}, \citenamefont {Moore},\ and\ \citenamefont
  {Xue}}]{Cline:2013zca}%
  \BibitemOpen
  \bibfield  {author} {\bibinfo {author} {\bibfnamefont {J.~M.}\ \bibnamefont
  {Cline}}, \bibinfo {author} {\bibfnamefont {Z.}~\bibnamefont {Liu}}, \bibinfo
  {author} {\bibfnamefont {G.}~\bibnamefont {Moore}}, \ and\ \bibinfo {author}
  {\bibfnamefont {W.}~\bibnamefont {Xue}},\ }\href {\doibase
  10.1103/PhysRevD.90.015023} {\bibfield  {journal} {\bibinfo  {journal} {Phys.
  Rev.}\ }\textbf {\bibinfo {volume} {D90}},\ \bibinfo {pages} {015023}
  (\bibinfo {year} {2014})},\ \Eprint {http://arxiv.org/abs/1312.3325}
  {arXiv:1312.3325 [hep-ph]} \BibitemShut {NoStop}%
%%CITATION = ARXIV:1312.3325;%%
\bibitem [{\citenamefont {Fan}\ \emph {et~al.}(2013)\citenamefont {Fan},
  \citenamefont {Katz}, \citenamefont {Randall},\ and\ \citenamefont
  {Reece}}]{Fan:2013yva}%
  \BibitemOpen
  \bibfield  {author} {\bibinfo {author} {\bibfnamefont {J.}~\bibnamefont
  {Fan}}, \bibinfo {author} {\bibfnamefont {A.}~\bibnamefont {Katz}}, \bibinfo
  {author} {\bibfnamefont {L.}~\bibnamefont {Randall}}, \ and\ \bibinfo
  {author} {\bibfnamefont {M.}~\bibnamefont {Reece}},\ }\href {\doibase
  10.1016/j.dark.2013.07.001} {\bibfield  {journal} {\bibinfo  {journal} {Phys.
  Dark Univ.}\ }\textbf {\bibinfo {volume} {2}},\ \bibinfo {pages} {139}
  (\bibinfo {year} {2013})},\ \Eprint {http://arxiv.org/abs/1303.1521}
  {arXiv:1303.1521 [astro-ph.CO]} \BibitemShut {NoStop}%
%%CITATION = ARXIV:1303.1521;%%
\bibitem [{\citenamefont {Chu}\ and\ \citenamefont
  {Dasgupta}(2014)}]{Chu:2014lja}%
  \BibitemOpen
  \bibfield  {author} {\bibinfo {author} {\bibfnamefont {X.}~\bibnamefont
  {Chu}}\ and\ \bibinfo {author} {\bibfnamefont {B.}~\bibnamefont {Dasgupta}},\
  }\href {\doibase 10.1103/PhysRevLett.113.161301} {\bibfield  {journal}
  {\bibinfo  {journal} {Phys. Rev. Lett.}\ }\textbf {\bibinfo {volume} {113}},\
  \bibinfo {pages} {161301} (\bibinfo {year} {2014})},\ \Eprint
  {http://arxiv.org/abs/1404.6127} {arXiv:1404.6127 [hep-ph]} \BibitemShut
  {NoStop}%
%%CITATION = ARXIV:1404.6127;%%
\bibitem [{\citenamefont {Ko}\ and\ \citenamefont {Tang}(2014)}]{Ko:2014bka}%
  \BibitemOpen
  \bibfield  {author} {\bibinfo {author} {\bibfnamefont {P.}~\bibnamefont
  {Ko}}\ and\ \bibinfo {author} {\bibfnamefont {Y.}~\bibnamefont {Tang}},\
  }\href {\doibase 10.1016/j.physletb.2014.10.035} {\bibfield  {journal}
  {\bibinfo  {journal} {Phys. Lett.}\ }\textbf {\bibinfo {volume} {B739}},\
  \bibinfo {pages} {62} (\bibinfo {year} {2014})},\ \Eprint
  {http://arxiv.org/abs/1404.0236} {arXiv:1404.0236 [hep-ph]} \BibitemShut
  {NoStop}%
%%CITATION = ARXIV:1404.0236;%%
\bibitem [{\citenamefont {Foot}\ and\ \citenamefont
  {Vagnozzi}(2015)}]{Foot:2014uba}%
  \BibitemOpen
  \bibfield  {author} {\bibinfo {author} {\bibfnamefont {R.}~\bibnamefont
  {Foot}}\ and\ \bibinfo {author} {\bibfnamefont {S.}~\bibnamefont
  {Vagnozzi}},\ }\href {\doibase 10.1103/PhysRevD.91.023512} {\bibfield
  {journal} {\bibinfo  {journal} {Phys. Rev.}\ }\textbf {\bibinfo {volume}
  {D91}},\ \bibinfo {pages} {023512} (\bibinfo {year} {2015})},\ \Eprint
  {http://arxiv.org/abs/1409.7174} {arXiv:1409.7174 [hep-ph]} \BibitemShut
  {NoStop}%
%%CITATION = ARXIV:1409.7174;%%
\bibitem [{\citenamefont {Cherry}\ \emph {et~al.}(2014)\citenamefont {Cherry},
  \citenamefont {Friedland},\ and\ \citenamefont {Shoemaker}}]{Cherry:2014xra}%
  \BibitemOpen
  \bibfield  {author} {\bibinfo {author} {\bibfnamefont {J.~F.}\ \bibnamefont
  {Cherry}}, \bibinfo {author} {\bibfnamefont {A.}~\bibnamefont {Friedland}}, \
  and\ \bibinfo {author} {\bibfnamefont {I.~M.}\ \bibnamefont {Shoemaker}},\
  }\href@noop {} {\  (\bibinfo {year} {2014})},\ \Eprint
  {http://arxiv.org/abs/1411.1071} {arXiv:1411.1071 [hep-ph]} \BibitemShut
  {NoStop}%
%%CITATION = ARXIV:1411.1071;%%
\bibitem [{\citenamefont {Buckley}\ \emph {et~al.}(2014)\citenamefont
  {Buckley}, \citenamefont {Zavala}, \citenamefont {Cyr-Racine}, \citenamefont
  {Sigurdson},\ and\ \citenamefont {Vogelsberger}}]{Buckley:2014hja}%
  \BibitemOpen
  \bibfield  {author} {\bibinfo {author} {\bibfnamefont {M.~R.}\ \bibnamefont
  {Buckley}}, \bibinfo {author} {\bibfnamefont {J.}~\bibnamefont {Zavala}},
  \bibinfo {author} {\bibfnamefont {F.-Y.}\ \bibnamefont {Cyr-Racine}},
  \bibinfo {author} {\bibfnamefont {K.}~\bibnamefont {Sigurdson}}, \ and\
  \bibinfo {author} {\bibfnamefont {M.}~\bibnamefont {Vogelsberger}},\ }\href
  {\doibase 10.1103/PhysRevD.90.043524} {\bibfield  {journal} {\bibinfo
  {journal} {Phys. Rev.}\ }\textbf {\bibinfo {volume} {D90}},\ \bibinfo {pages}
  {043524} (\bibinfo {year} {2014})},\ \Eprint {http://arxiv.org/abs/1405.2075}
  {arXiv:1405.2075 [astro-ph.CO]} \BibitemShut {NoStop}%
%%CITATION = ARXIV:1405.2075;%%
\bibitem [{\citenamefont {Boehm}\ \emph {et~al.}(2014)\citenamefont {Boehm},
  \citenamefont {Schewtschenko}, \citenamefont {Wilkinson}, \citenamefont
  {Baugh},\ and\ \citenamefont {Pascoli}}]{Boehm:2014vja}%
  \BibitemOpen
  \bibfield  {author} {\bibinfo {author} {\bibfnamefont {C.}~\bibnamefont
  {Boehm}}, \bibinfo {author} {\bibfnamefont {J.~A.}\ \bibnamefont
  {Schewtschenko}}, \bibinfo {author} {\bibfnamefont {R.~J.}\ \bibnamefont
  {Wilkinson}}, \bibinfo {author} {\bibfnamefont {C.~M.}\ \bibnamefont
  {Baugh}}, \ and\ \bibinfo {author} {\bibfnamefont {S.}~\bibnamefont
  {Pascoli}},\ }\href {\doibase 10.1093/mnrasl/slu115} {\bibfield  {journal}
  {\bibinfo  {journal} {Mon. Not. Roy. Astron. Soc.}\ }\textbf {\bibinfo
  {volume} {445}},\ \bibinfo {pages} {L31} (\bibinfo {year} {2014})},\ \Eprint
  {http://arxiv.org/abs/1404.7012} {arXiv:1404.7012 [astro-ph.CO]} \BibitemShut
  {NoStop}%
%%CITATION = ARXIV:1404.7012;%%
\bibitem [{\citenamefont {Buen-Abad}\ \emph {et~al.}(2015)\citenamefont
  {Buen-Abad}, \citenamefont {Marques-Tavares},\ and\ \citenamefont
  {Schmaltz}}]{Buen-Abad:2015ova}%
  \BibitemOpen
  \bibfield  {author} {\bibinfo {author} {\bibfnamefont {M.~A.}\ \bibnamefont
  {Buen-Abad}}, \bibinfo {author} {\bibfnamefont {G.}~\bibnamefont
  {Marques-Tavares}}, \ and\ \bibinfo {author} {\bibfnamefont {M.}~\bibnamefont
  {Schmaltz}},\ }\href {\doibase 10.1103/PhysRevD.92.023531} {\bibfield
  {journal} {\bibinfo  {journal} {Phys. Rev.}\ }\textbf {\bibinfo {volume}
  {D92}},\ \bibinfo {pages} {023531} (\bibinfo {year} {2015})},\ \Eprint
  {http://arxiv.org/abs/1505.03542} {arXiv:1505.03542 [hep-ph]} \BibitemShut
  {NoStop}%
%%CITATION = ARXIV:1505.03542;%%
\bibitem [{\citenamefont {Lesgourgues}\ \emph {et~al.}(2016)\citenamefont
  {Lesgourgues}, \citenamefont {Marques-Tavares},\ and\ \citenamefont
  {Schmaltz}}]{Lesgourgues:2015wza}%
  \BibitemOpen
  \bibfield  {author} {\bibinfo {author} {\bibfnamefont {J.}~\bibnamefont
  {Lesgourgues}}, \bibinfo {author} {\bibfnamefont {G.}~\bibnamefont
  {Marques-Tavares}}, \ and\ \bibinfo {author} {\bibfnamefont {M.}~\bibnamefont
  {Schmaltz}},\ }\href {\doibase 10.1088/1475-7516/2016/02/037} {\bibfield
  {journal} {\bibinfo  {journal} {JCAP}\ }\textbf {\bibinfo {volume} {1602}},\
  \bibinfo {pages} {037} (\bibinfo {year} {2016})},\ \Eprint
  {http://arxiv.org/abs/1507.04351} {arXiv:1507.04351 [astro-ph.CO]}
  \BibitemShut {NoStop}%
%%CITATION = ARXIV:1507.04351;%%
\bibitem [{\citenamefont {Foot}\ and\ \citenamefont
  {Vagnozzi}(2016)}]{Foot:2016wvj}%
  \BibitemOpen
  \bibfield  {author} {\bibinfo {author} {\bibfnamefont {R.}~\bibnamefont
  {Foot}}\ and\ \bibinfo {author} {\bibfnamefont {S.}~\bibnamefont
  {Vagnozzi}},\ }\href {\doibase 10.1088/1475-7516/2016/07/013} {\bibfield
  {journal} {\bibinfo  {journal} {JCAP}\ }\textbf {\bibinfo {volume} {1607}},\
  \bibinfo {pages} {013} (\bibinfo {year} {2016})},\ \Eprint
  {http://arxiv.org/abs/1602.02467} {arXiv:1602.02467 [astro-ph.CO]}
  \BibitemShut {NoStop}%
%%CITATION = ARXIV:1602.02467;%%
\bibitem [{\citenamefont {Cyr-Racine}\ \emph {et~al.}(2016)\citenamefont
  {Cyr-Racine}, \citenamefont {Sigurdson}, \citenamefont {Zavala},
  \citenamefont {Bringmann}, \citenamefont {Vogelsberger},\ and\ \citenamefont
  {Pfrommer}}]{Cyr-Racine:2015ihg}%
  \BibitemOpen
  \bibfield  {author} {\bibinfo {author} {\bibfnamefont {F.-Y.}\ \bibnamefont
  {Cyr-Racine}}, \bibinfo {author} {\bibfnamefont {K.}~\bibnamefont
  {Sigurdson}}, \bibinfo {author} {\bibfnamefont {J.}~\bibnamefont {Zavala}},
  \bibinfo {author} {\bibfnamefont {T.}~\bibnamefont {Bringmann}}, \bibinfo
  {author} {\bibfnamefont {M.}~\bibnamefont {Vogelsberger}}, \ and\ \bibinfo
  {author} {\bibfnamefont {C.}~\bibnamefont {Pfrommer}},\ }\href {\doibase
  10.1103/PhysRevD.93.123527} {\bibfield  {journal} {\bibinfo  {journal} {Phys.
  Rev.}\ }\textbf {\bibinfo {volume} {D93}},\ \bibinfo {pages} {123527}
  (\bibinfo {year} {2016})},\ \Eprint {http://arxiv.org/abs/1512.05344}
  {arXiv:1512.05344 [astro-ph.CO]} \BibitemShut {NoStop}%
%%CITATION = ARXIV:1512.05344;%%
\bibitem [{\citenamefont {Viel}\ \emph {et~al.}(2013)\citenamefont {Viel},
  \citenamefont {Becker}, \citenamefont {Bolton},\ and\ \citenamefont
  {Haehnelt}}]{Viel:2013apy}%
  \BibitemOpen
  \bibfield  {author} {\bibinfo {author} {\bibfnamefont {M.}~\bibnamefont
  {Viel}}, \bibinfo {author} {\bibfnamefont {G.~D.}\ \bibnamefont {Becker}},
  \bibinfo {author} {\bibfnamefont {J.~S.}\ \bibnamefont {Bolton}}, \ and\
  \bibinfo {author} {\bibfnamefont {M.~G.}\ \bibnamefont {Haehnelt}},\ }\href
  {\doibase 10.1103/PhysRevD.88.043502} {\bibfield  {journal} {\bibinfo
  {journal} {Phys. Rev.}\ }\textbf {\bibinfo {volume} {D88}},\ \bibinfo {pages}
  {043502} (\bibinfo {year} {2013})},\ \Eprint {http://arxiv.org/abs/1306.2314}
  {arXiv:1306.2314 [astro-ph.CO]} \BibitemShut {NoStop}%
%%CITATION = ARXIV:1306.2314;%%
\bibitem [{\citenamefont {Baur}\ \emph {et~al.}(2015)\citenamefont {Baur},
  \citenamefont {Palanque-Delabrouille}, \citenamefont {Yèche}, \citenamefont
  {Magneville},\ and\ \citenamefont {Viel}}]{Baur:2015jsy}%
  \BibitemOpen
  \bibfield  {author} {\bibinfo {author} {\bibfnamefont {J.}~\bibnamefont
  {Baur}}, \bibinfo {author} {\bibfnamefont {N.}~\bibnamefont
  {Palanque-Delabrouille}}, \bibinfo {author} {\bibfnamefont {C.}~\bibnamefont
  {Yèche}}, \bibinfo {author} {\bibfnamefont {C.}~\bibnamefont {Magneville}},
  \ and\ \bibinfo {author} {\bibfnamefont {M.}~\bibnamefont {Viel}},\ }in\
  \href {http://inspirehep.net/record/1408473/files/arXiv:1512.01981.pdf}
  {\emph {\bibinfo {booktitle} {{SDSS-IV Collaboration Meeting, July 20-23,
  2015}}}}\ (\bibinfo {year} {2015})\ \Eprint {http://arxiv.org/abs/1512.01981}
  {arXiv:1512.01981 [astro-ph.CO]} \BibitemShut {NoStop}%
%%CITATION = ARXIV:1512.01981;%%
\bibitem [{\citenamefont {Garzilli}\ \emph {et~al.}(2015)\citenamefont
  {Garzilli}, \citenamefont {Boyarsky},\ and\ \citenamefont
  {Ruchayskiy}}]{Garzilli:2015iwa}%
  \BibitemOpen
  \bibfield  {author} {\bibinfo {author} {\bibfnamefont {A.}~\bibnamefont
  {Garzilli}}, \bibinfo {author} {\bibfnamefont {A.}~\bibnamefont {Boyarsky}},
  \ and\ \bibinfo {author} {\bibfnamefont {O.}~\bibnamefont {Ruchayskiy}},\
  }\href@noop {} {\  (\bibinfo {year} {2015})},\ \Eprint
  {http://arxiv.org/abs/1510.07006} {arXiv:1510.07006 [astro-ph.CO]}
  \BibitemShut {NoStop}%
%%CITATION = ARXIV:1510.07006;%%
\bibitem [{\citenamefont {Horiuchi}\ \emph {et~al.}(2014)\citenamefont
  {Horiuchi}, \citenamefont {Humphrey}, \citenamefont {Onorbe}, \citenamefont
  {Abazajian}, \citenamefont {Kaplinghat},\ and\ \citenamefont
  {Garrison-Kimmel}}]{Horiuchi:2013noa}%
  \BibitemOpen
  \bibfield  {author} {\bibinfo {author} {\bibfnamefont {S.}~\bibnamefont
  {Horiuchi}}, \bibinfo {author} {\bibfnamefont {P.~J.}\ \bibnamefont
  {Humphrey}}, \bibinfo {author} {\bibfnamefont {J.}~\bibnamefont {Onorbe}},
  \bibinfo {author} {\bibfnamefont {K.~N.}\ \bibnamefont {Abazajian}}, \bibinfo
  {author} {\bibfnamefont {M.}~\bibnamefont {Kaplinghat}}, \ and\ \bibinfo
  {author} {\bibfnamefont {S.}~\bibnamefont {Garrison-Kimmel}},\ }\href
  {\doibase 10.1103/PhysRevD.89.025017} {\bibfield  {journal} {\bibinfo
  {journal} {Phys. Rev.}\ }\textbf {\bibinfo {volume} {D89}},\ \bibinfo {pages}
  {025017} (\bibinfo {year} {2014})},\ \Eprint {http://arxiv.org/abs/1311.0282}
  {arXiv:1311.0282 [astro-ph.CO]} \BibitemShut {NoStop}%
%%CITATION = ARXIV:1311.0282;%%
\bibitem [{\citenamefont {Inoue}\ \emph {et~al.}(2015)\citenamefont {Inoue},
  \citenamefont {Takahashi}, \citenamefont {Takahashi},\ and\ \citenamefont
  {Ishiyama}}]{Inoue:2014jka}%
  \BibitemOpen
  \bibfield  {author} {\bibinfo {author} {\bibfnamefont {K.~T.}\ \bibnamefont
  {Inoue}}, \bibinfo {author} {\bibfnamefont {R.}~\bibnamefont {Takahashi}},
  \bibinfo {author} {\bibfnamefont {T.}~\bibnamefont {Takahashi}}, \ and\
  \bibinfo {author} {\bibfnamefont {T.}~\bibnamefont {Ishiyama}},\ }\href
  {\doibase 10.1093/mnras/stv194} {\bibfield  {journal} {\bibinfo  {journal}
  {Mon. Not. Roy. Astron. Soc.}\ }\textbf {\bibinfo {volume} {448}},\ \bibinfo
  {pages} {2704} (\bibinfo {year} {2015})},\ \Eprint
  {http://arxiv.org/abs/1409.1326} {arXiv:1409.1326 [astro-ph.CO]} \BibitemShut
  {NoStop}%
%%CITATION = ARXIV:1409.1326;%%
\bibitem [{\citenamefont {Schneider}\ \emph {et~al.}(2014)\citenamefont
  {Schneider}, \citenamefont {Anderhalden}, \citenamefont {Maccio},\ and\
  \citenamefont {Diemand}}]{Schneider:2013wwa}%
  \BibitemOpen
  \bibfield  {author} {\bibinfo {author} {\bibfnamefont {A.}~\bibnamefont
  {Schneider}}, \bibinfo {author} {\bibfnamefont {D.}~\bibnamefont
  {Anderhalden}}, \bibinfo {author} {\bibfnamefont {A.}~\bibnamefont {Maccio}},
  \ and\ \bibinfo {author} {\bibfnamefont {J.}~\bibnamefont {Diemand}},\ }\href
  {\doibase 10.1093/mnrasl/slu034} {\bibfield  {journal} {\bibinfo  {journal}
  {Mon. Not. Roy. Astron. Soc.}\ }\textbf {\bibinfo {volume} {441}},\ \bibinfo
  {pages} {6} (\bibinfo {year} {2014})},\ \Eprint
  {http://arxiv.org/abs/1309.5960} {arXiv:1309.5960 [astro-ph.CO]} \BibitemShut
  {NoStop}%
%%CITATION = ARXIV:1309.5960;%%
\bibitem [{\citenamefont {Macciò}\ and\ \citenamefont
  {Fontanot}(2010)}]{Maccio':2009rx}%
  \BibitemOpen
  \bibfield  {author} {\bibinfo {author} {\bibfnamefont {A.~V.}\ \bibnamefont
  {Macciò}}\ and\ \bibinfo {author} {\bibfnamefont {F.}~\bibnamefont
  {Fontanot}},\ }\href {\doibase 10.1111/j.1745-3933.2010.00825.x} {\bibfield
  {journal} {\bibinfo  {journal} {Mon. Not. Roy. Astron. Soc.}\ }\textbf
  {\bibinfo {volume} {404}},\ \bibinfo {pages} {16} (\bibinfo {year} {2010})},\
  \Eprint {http://arxiv.org/abs/0910.2460} {arXiv:0910.2460 [astro-ph.CO]}
  \BibitemShut {NoStop}%
%%CITATION = ARXIV:0910.2460;%%
\bibitem [{\citenamefont {Macciò}\ \emph {et~al.}(2012)\citenamefont
  {Macciò}, \citenamefont {Paduroiu}, \citenamefont {Anderhalden},
  \citenamefont {Schneider},\ and\ \citenamefont {Moore}}]{Maccio:2012qf}%
  \BibitemOpen
  \bibfield  {author} {\bibinfo {author} {\bibfnamefont {A.~V.}\ \bibnamefont
  {Macciò}}, \bibinfo {author} {\bibfnamefont {S.}~\bibnamefont {Paduroiu}},
  \bibinfo {author} {\bibfnamefont {D.}~\bibnamefont {Anderhalden}}, \bibinfo
  {author} {\bibfnamefont {A.}~\bibnamefont {Schneider}}, \ and\ \bibinfo
  {author} {\bibfnamefont {B.}~\bibnamefont {Moore}},\ }\href {\doibase
  10.1111/j.1365-2966.2012.21284.x} {\bibfield  {journal} {\bibinfo  {journal}
  {Mon. Not. Roy. Astron. Soc.}\ }\textbf {\bibinfo {volume} {424}},\ \bibinfo
  {pages} {1105} (\bibinfo {year} {2012})},\ \bibinfo {note} {erratum ibid.\
  \textbf{428}, 3715 (2013)},\ \Eprint {http://arxiv.org/abs/1202.1282}
  {arXiv:1202.1282 [astro-ph.CO]} \BibitemShut {NoStop}%
%%CITATION = ARXIV:1202.1282;%%
\bibitem [{\citenamefont {Shao}\ \emph {et~al.}(2013)\citenamefont {Shao},
  \citenamefont {Gao}, \citenamefont {Theuns},\ and\ \citenamefont
  {Frenk}}]{Shao:2012cg}%
  \BibitemOpen
  \bibfield  {author} {\bibinfo {author} {\bibfnamefont {S.}~\bibnamefont
  {Shao}}, \bibinfo {author} {\bibfnamefont {L.}~\bibnamefont {Gao}}, \bibinfo
  {author} {\bibfnamefont {T.}~\bibnamefont {Theuns}}, \ and\ \bibinfo {author}
  {\bibfnamefont {C.~S.}\ \bibnamefont {Frenk}},\ }\href {\doibase
  10.1093/mnras/stt053} {\bibfield  {journal} {\bibinfo  {journal} {Mon. Not.
  Roy. Astron. Soc.}\ }\textbf {\bibinfo {volume} {430}},\ \bibinfo {pages}
  {2346} (\bibinfo {year} {2013})},\ \Eprint {http://arxiv.org/abs/1209.5563}
  {arXiv:1209.5563 [astro-ph.CO]} \BibitemShut {NoStop}%
%%CITATION = ARXIV:1209.5563;%%
\bibitem [{\citenamefont {Chen}\ \emph {et~al.}(2001)\citenamefont {Chen},
  \citenamefont {Kamionkowski},\ and\ \citenamefont {Zhang}}]{Chen:2001jz}%
  \BibitemOpen
  \bibfield  {author} {\bibinfo {author} {\bibfnamefont {X.-l.}\ \bibnamefont
  {Chen}}, \bibinfo {author} {\bibfnamefont {M.}~\bibnamefont {Kamionkowski}},
  \ and\ \bibinfo {author} {\bibfnamefont {X.-m.}\ \bibnamefont {Zhang}},\
  }\href {\doibase 10.1103/PhysRevD.64.021302} {\bibfield  {journal} {\bibinfo
  {journal} {Phys. Rev.}\ }\textbf {\bibinfo {volume} {D64}},\ \bibinfo {pages}
  {021302} (\bibinfo {year} {2001})},\ \Eprint
  {http://arxiv.org/abs/astro-ph/0103452} {arXiv:astro-ph/0103452} \BibitemShut
  {NoStop}%
%%CITATION = ASTRO-PH/0103452;%%
\bibitem [{\citenamefont {Bringmann}(2009)}]{Bringmann:2009vf}%
  \BibitemOpen
  \bibfield  {author} {\bibinfo {author} {\bibfnamefont {T.}~\bibnamefont
  {Bringmann}},\ }\href {\doibase 10.1088/1367-2630/11/10/105027} {\bibfield
  {journal} {\bibinfo  {journal} {New J. Phys.}\ }\textbf {\bibinfo {volume}
  {11}},\ \bibinfo {pages} {105027} (\bibinfo {year} {2009})},\ \Eprint
  {http://arxiv.org/abs/0903.0189} {arXiv:0903.0189 [astro-ph.CO]} \BibitemShut
  {NoStop}%
%%CITATION = ARXIV:0903.0189;%%
\bibitem [{\citenamefont {Loeb}\ and\ \citenamefont
  {Zaldarriaga}(2005)}]{Loeb:2005pm}%
  \BibitemOpen
  \bibfield  {author} {\bibinfo {author} {\bibfnamefont {A.}~\bibnamefont
  {Loeb}}\ and\ \bibinfo {author} {\bibfnamefont {M.}~\bibnamefont
  {Zaldarriaga}},\ }\href {\doibase 10.1103/PhysRevD.71.103520} {\bibfield
  {journal} {\bibinfo  {journal} {Phys. Rev.}\ }\textbf {\bibinfo {volume}
  {D71}},\ \bibinfo {pages} {103520} (\bibinfo {year} {2005})},\ \Eprint
  {http://arxiv.org/abs/astro-ph/0504112} {arXiv:astro-ph/0504112} \BibitemShut
  {NoStop}%
%%CITATION = ASTRO-PH/0504112;%%
\bibitem [{\citenamefont {Bertschinger}(2006)}]{Bertschinger:2006nq}%
  \BibitemOpen
  \bibfield  {author} {\bibinfo {author} {\bibfnamefont {E.}~\bibnamefont
  {Bertschinger}},\ }\href {\doibase 10.1103/PhysRevD.74.063509} {\bibfield
  {journal} {\bibinfo  {journal} {Phys. Rev.}\ }\textbf {\bibinfo {volume}
  {D74}},\ \bibinfo {pages} {063509} (\bibinfo {year} {2006})},\ \Eprint
  {http://arxiv.org/abs/astro-ph/0607319} {arXiv:astro-ph/0607319} \BibitemShut
  {NoStop}%
%%CITATION = ASTRO-PH/0607319;%%
\bibitem [{\citenamefont {Schewtschenko}\ \emph {et~al.}(2016)\citenamefont
  {Schewtschenko}, \citenamefont {Baugh}, \citenamefont {Wilkinson},
  \citenamefont {Boehm}, \citenamefont {Pascoli},\ and\ \citenamefont
  {Sawala}}]{Schewtschenko:2015rno}%
  \BibitemOpen
  \bibfield  {author} {\bibinfo {author} {\bibfnamefont {J.~A.}\ \bibnamefont
  {Schewtschenko}}, \bibinfo {author} {\bibfnamefont {C.~M.}\ \bibnamefont
  {Baugh}}, \bibinfo {author} {\bibfnamefont {R.~J.}\ \bibnamefont
  {Wilkinson}}, \bibinfo {author} {\bibfnamefont {C.}~\bibnamefont {Boehm}},
  \bibinfo {author} {\bibfnamefont {S.}~\bibnamefont {Pascoli}}, \ and\
  \bibinfo {author} {\bibfnamefont {T.}~\bibnamefont {Sawala}},\ }\href
  {\doibase 10.1093/mnras/stw1078} {\bibfield  {journal} {\bibinfo  {journal}
  {Mon. Not. Roy. Astron. Soc.}\ }\textbf {\bibinfo {volume} {461}},\ \bibinfo
  {pages} {2282} (\bibinfo {year} {2016})},\ \Eprint
  {http://arxiv.org/abs/1512.06774} {arXiv:1512.06774 [astro-ph.CO]}
  \BibitemShut {NoStop}%
%%CITATION = ARXIV:1512.06774;%%
\bibitem [{\citenamefont {Bertoni}\ \emph {et~al.}(2015)\citenamefont
  {Bertoni}, \citenamefont {Ipek}, \citenamefont {McKeen},\ and\ \citenamefont
  {Nelson}}]{Bertoni:2014mva}%
  \BibitemOpen
  \bibfield  {author} {\bibinfo {author} {\bibfnamefont {B.}~\bibnamefont
  {Bertoni}}, \bibinfo {author} {\bibfnamefont {S.}~\bibnamefont {Ipek}},
  \bibinfo {author} {\bibfnamefont {D.}~\bibnamefont {McKeen}}, \ and\ \bibinfo
  {author} {\bibfnamefont {A.~E.}\ \bibnamefont {Nelson}},\ }\href {\doibase
  10.1007/JHEP04(2015)170} {\bibfield  {journal} {\bibinfo  {journal} {JHEP}\
  }\textbf {\bibinfo {volume} {04}},\ \bibinfo {pages} {170} (\bibinfo {year}
  {2015})},\ \Eprint {http://arxiv.org/abs/1412.3113} {arXiv:1412.3113
  [hep-ph]} \BibitemShut {NoStop}%
%%CITATION = ARXIV:1412.3113;%%
\bibitem [{\citenamefont {Brust}\ \emph {et~al.}(2013)\citenamefont {Brust},
  \citenamefont {Kaplan},\ and\ \citenamefont {Walters}}]{Brust:2013xpv}%
  \BibitemOpen
  \bibfield  {author} {\bibinfo {author} {\bibfnamefont {C.}~\bibnamefont
  {Brust}}, \bibinfo {author} {\bibfnamefont {D.~E.}\ \bibnamefont {Kaplan}}, \
  and\ \bibinfo {author} {\bibfnamefont {M.~T.}\ \bibnamefont {Walters}},\
  }\href {\doibase 10.1007/JHEP12(2013)058} {\bibfield  {journal} {\bibinfo
  {journal} {JHEP}\ }\textbf {\bibinfo {volume} {12}},\ \bibinfo {pages} {058}
  (\bibinfo {year} {2013})},\ \Eprint {http://arxiv.org/abs/1303.5379}
  {arXiv:1303.5379 [hep-ph]} \BibitemShut {NoStop}%
%%CITATION = ARXIV:1303.5379;%%
\bibitem [{\citenamefont {Archidiacono}\ \emph {et~al.}(2013)\citenamefont
  {Archidiacono}, \citenamefont {Giusarma}, \citenamefont {Hannestad},\ and\
  \citenamefont {Mena}}]{Archidiacono:2013fha}%
  \BibitemOpen
  \bibfield  {author} {\bibinfo {author} {\bibfnamefont {M.}~\bibnamefont
  {Archidiacono}}, \bibinfo {author} {\bibfnamefont {E.}~\bibnamefont
  {Giusarma}}, \bibinfo {author} {\bibfnamefont {S.}~\bibnamefont {Hannestad}},
  \ and\ \bibinfo {author} {\bibfnamefont {O.}~\bibnamefont {Mena}},\ }\href
  {\doibase 10.1155/2013/191047} {\bibfield  {journal} {\bibinfo  {journal}
  {Adv. High Energy Phys.}\ }\textbf {\bibinfo {volume} {2013}},\ \bibinfo
  {pages} {191047} (\bibinfo {year} {2013})},\ \Eprint
  {http://arxiv.org/abs/1307.0637} {arXiv:1307.0637 [astro-ph.CO]} \BibitemShut
  {NoStop}%
%%CITATION = ARXIV:1307.0637;%%
\bibitem [{\citenamefont {Wyman}\ \emph {et~al.}(2014)\citenamefont {Wyman},
  \citenamefont {Rudd}, \citenamefont {Vanderveld},\ and\ \citenamefont
  {Hu}}]{Wyman:2013lza}%
  \BibitemOpen
  \bibfield  {author} {\bibinfo {author} {\bibfnamefont {M.}~\bibnamefont
  {Wyman}}, \bibinfo {author} {\bibfnamefont {D.~H.}\ \bibnamefont {Rudd}},
  \bibinfo {author} {\bibfnamefont {R.~A.}\ \bibnamefont {Vanderveld}}, \ and\
  \bibinfo {author} {\bibfnamefont {W.}~\bibnamefont {Hu}},\ }\href {\doibase
  10.1103/PhysRevLett.112.051302} {\bibfield  {journal} {\bibinfo  {journal}
  {Phys. Rev. Lett.}\ }\textbf {\bibinfo {volume} {112}},\ \bibinfo {pages}
  {051302} (\bibinfo {year} {2014})},\ \Eprint {http://arxiv.org/abs/1307.7715}
  {arXiv:1307.7715 [astro-ph.CO]} \BibitemShut {NoStop}%
%%CITATION = ARXIV:1307.7715;%%
\bibitem [{\citenamefont {Hamann}\ and\ \citenamefont
  {Hasenkamp}(2013)}]{Hamann:2013iba}%
  \BibitemOpen
  \bibfield  {author} {\bibinfo {author} {\bibfnamefont {J.}~\bibnamefont
  {Hamann}}\ and\ \bibinfo {author} {\bibfnamefont {J.}~\bibnamefont
  {Hasenkamp}},\ }\href {\doibase 10.1088/1475-7516/2013/10/044} {\bibfield
  {journal} {\bibinfo  {journal} {JCAP}\ }\textbf {\bibinfo {volume} {1310}},\
  \bibinfo {pages} {044} (\bibinfo {year} {2013})},\ \Eprint
  {http://arxiv.org/abs/1308.3255} {arXiv:1308.3255 [astro-ph.CO]} \BibitemShut
  {NoStop}%
%%CITATION = ARXIV:1308.3255;%%
\bibitem [{\citenamefont {Battye}\ and\ \citenamefont
  {Moss}(2014)}]{Battye:2013xqa}%
  \BibitemOpen
  \bibfield  {author} {\bibinfo {author} {\bibfnamefont {R.~A.}\ \bibnamefont
  {Battye}}\ and\ \bibinfo {author} {\bibfnamefont {A.}~\bibnamefont {Moss}},\
  }\href {\doibase 10.1103/PhysRevLett.112.051303} {\bibfield  {journal}
  {\bibinfo  {journal} {Phys. Rev. Lett.}\ }\textbf {\bibinfo {volume} {112}},\
  \bibinfo {pages} {051303} (\bibinfo {year} {2014})},\ \Eprint
  {http://arxiv.org/abs/1308.5870} {arXiv:1308.5870 [astro-ph.CO]} \BibitemShut
  {NoStop}%
%%CITATION = ARXIV:1308.5870;%%
\bibitem [{\citenamefont {Gariazzo}\ \emph {et~al.}(2013)\citenamefont
  {Gariazzo}, \citenamefont {Giunti},\ and\ \citenamefont
  {Laveder}}]{Gariazzo:2013gua}%
  \BibitemOpen
  \bibfield  {author} {\bibinfo {author} {\bibfnamefont {S.}~\bibnamefont
  {Gariazzo}}, \bibinfo {author} {\bibfnamefont {C.}~\bibnamefont {Giunti}}, \
  and\ \bibinfo {author} {\bibfnamefont {M.}~\bibnamefont {Laveder}},\ }\href
  {\doibase 10.1007/JHEP11(2013)211} {\bibfield  {journal} {\bibinfo  {journal}
  {JHEP}\ }\textbf {\bibinfo {volume} {1311}},\ \bibinfo {pages} {211}
  (\bibinfo {year} {2013})},\ \Eprint {http://arxiv.org/abs/1309.3192}
  {arXiv:1309.3192 [hep-ph]} \BibitemShut {NoStop}%
%%CITATION = ARXIV:1309.3192;%%
\bibitem [{\citenamefont {Battye}\ \emph {et~al.}(2015)\citenamefont {Battye},
  \citenamefont {Charnock},\ and\ \citenamefont {Moss}}]{Battye:2014qga}%
  \BibitemOpen
  \bibfield  {author} {\bibinfo {author} {\bibfnamefont {R.~A.}\ \bibnamefont
  {Battye}}, \bibinfo {author} {\bibfnamefont {T.}~\bibnamefont {Charnock}}, \
  and\ \bibinfo {author} {\bibfnamefont {A.}~\bibnamefont {Moss}},\ }\href
  {\doibase 10.1103/PhysRevD.91.103508} {\bibfield  {journal} {\bibinfo
  {journal} {Phys. Rev.}\ }\textbf {\bibinfo {volume} {D91}},\ \bibinfo {pages}
  {103508} (\bibinfo {year} {2015})},\ \Eprint {http://arxiv.org/abs/1409.2769}
  {arXiv:1409.2769 [astro-ph.CO]} \BibitemShut {NoStop}%
%%CITATION = ARXIV:1409.2769;%%
\bibitem [{\citenamefont {Green}\ \emph {et~al.}(2005)\citenamefont {Green},
  \citenamefont {Hofmann},\ and\ \citenamefont {Schwarz}}]{Green:2005fa}%
  \BibitemOpen
  \bibfield  {author} {\bibinfo {author} {\bibfnamefont {A.~M.}\ \bibnamefont
  {Green}}, \bibinfo {author} {\bibfnamefont {S.}~\bibnamefont {Hofmann}}, \
  and\ \bibinfo {author} {\bibfnamefont {D.~J.}\ \bibnamefont {Schwarz}},\
  }\href {\doibase 10.1088/1475-7516/2005/08/003} {\bibfield  {journal}
  {\bibinfo  {journal} {JCAP}\ }\textbf {\bibinfo {volume} {0508}},\ \bibinfo
  {pages} {003} (\bibinfo {year} {2005})},\ \Eprint
  {http://arxiv.org/abs/astro-ph/0503387} {arXiv:astro-ph/0503387 [astro-ph]}
  \BibitemShut {NoStop}%
%%CITATION = ASTRO-PH/0503387;%%
\bibitem [{\citenamefont {Nollett}\ and\ \citenamefont
  {Steigman}(2015)}]{Nollett:2014lwa}%
  \BibitemOpen
  \bibfield  {author} {\bibinfo {author} {\bibfnamefont {K.~M.}\ \bibnamefont
  {Nollett}}\ and\ \bibinfo {author} {\bibfnamefont {G.}~\bibnamefont
  {Steigman}},\ }\href {\doibase 10.1103/PhysRevD.91.083505} {\bibfield
  {journal} {\bibinfo  {journal} {Phys. Rev.}\ }\textbf {\bibinfo {volume}
  {D91}},\ \bibinfo {pages} {083505} (\bibinfo {year} {2015})},\ \Eprint
  {http://arxiv.org/abs/1411.6005} {arXiv:1411.6005 [astro-ph.CO]} \BibitemShut
  {NoStop}%
%%CITATION = ARXIV:1411.6005;%%
\bibitem [{\citenamefont {Kahlhoefer}\ \emph {et~al.}(2016)\citenamefont
  {Kahlhoefer}, \citenamefont {Schmidt-Hoberg}, \citenamefont {Schwetz},\ and\
  \citenamefont {Vogl}}]{Kahlhoefer:2015bea}%
  \BibitemOpen
  \bibfield  {author} {\bibinfo {author} {\bibfnamefont {F.}~\bibnamefont
  {Kahlhoefer}}, \bibinfo {author} {\bibfnamefont {K.}~\bibnamefont
  {Schmidt-Hoberg}}, \bibinfo {author} {\bibfnamefont {T.}~\bibnamefont
  {Schwetz}}, \ and\ \bibinfo {author} {\bibfnamefont {S.}~\bibnamefont
  {Vogl}},\ }\href {\doibase 10.1007/JHEP02(2016)016} {\bibfield  {journal}
  {\bibinfo  {journal} {JHEP}\ }\textbf {\bibinfo {volume} {02}},\ \bibinfo
  {pages} {016} (\bibinfo {year} {2016})},\ \Eprint
  {http://arxiv.org/abs/1510.02110} {arXiv:1510.02110 [hep-ph]} \BibitemShut
  {NoStop}%
%%CITATION = ARXIV:1510.02110;%%
\bibitem [{\citenamefont {Boehm}\ \emph {et~al.}(2013)\citenamefont {Boehm},
  \citenamefont {Dolan},\ and\ \citenamefont {McCabe}}]{Boehm:2013jpa}%
  \BibitemOpen
  \bibfield  {author} {\bibinfo {author} {\bibfnamefont {C.}~\bibnamefont
  {Boehm}}, \bibinfo {author} {\bibfnamefont {M.~J.}\ \bibnamefont {Dolan}}, \
  and\ \bibinfo {author} {\bibfnamefont {C.}~\bibnamefont {McCabe}},\ }\href
  {\doibase 10.1088/1475-7516/2013/08/041} {\bibfield  {journal} {\bibinfo
  {journal} {JCAP}\ }\textbf {\bibinfo {volume} {1308}},\ \bibinfo {pages}
  {041} (\bibinfo {year} {2013})},\ \Eprint {http://arxiv.org/abs/1303.6270}
  {arXiv:1303.6270 [hep-ph]} \BibitemShut {NoStop}%
%%CITATION = ARXIV:1303.6270;%%
\bibitem [{\citenamefont {Gondolo}\ and\ \citenamefont
  {Gelmini}(1991)}]{Gondolo:1990dk}%
  \BibitemOpen
  \bibfield  {author} {\bibinfo {author} {\bibfnamefont {P.}~\bibnamefont
  {Gondolo}}\ and\ \bibinfo {author} {\bibfnamefont {G.}~\bibnamefont
  {Gelmini}},\ }\href {\doibase 10.1016/0550-3213(91)90438-4} {\bibfield
  {journal} {\bibinfo  {journal} {Nucl. Phys.}\ }\textbf {\bibinfo {volume}
  {B360}},\ \bibinfo {pages} {145} (\bibinfo {year} {1991})}\BibitemShut
  {NoStop}%
%%CITATION = NUPHA,B360,145;%%
\bibitem [{\citenamefont {Feng}\ and\ \citenamefont
  {Kumar}(2008)}]{Feng:2008ya}%
  \BibitemOpen
  \bibfield  {author} {\bibinfo {author} {\bibfnamefont {J.~L.}\ \bibnamefont
  {Feng}}\ and\ \bibinfo {author} {\bibfnamefont {J.}~\bibnamefont {Kumar}},\
  }\href {\doibase 10.1103/PhysRevLett.101.231301} {\bibfield  {journal}
  {\bibinfo  {journal} {Phys. Rev. Lett.}\ }\textbf {\bibinfo {volume} {101}},\
  \bibinfo {pages} {231301} (\bibinfo {year} {2008})},\ \Eprint
  {http://arxiv.org/abs/0803.4196} {arXiv:0803.4196 [hep-ph]} \BibitemShut
  {NoStop}%
%%CITATION = ARXIV:0803.4196;%%
\bibitem [{\citenamefont {Bellazzini}\ \emph {et~al.}(2013)\citenamefont
  {Bellazzini}, \citenamefont {Cliche},\ and\ \citenamefont
  {Tanedo}}]{Bellazzini:2013foa}%
  \BibitemOpen
  \bibfield  {author} {\bibinfo {author} {\bibfnamefont {B.}~\bibnamefont
  {Bellazzini}}, \bibinfo {author} {\bibfnamefont {M.}~\bibnamefont {Cliche}},
  \ and\ \bibinfo {author} {\bibfnamefont {P.}~\bibnamefont {Tanedo}},\ }\href
  {\doibase 10.1103/PhysRevD.88.083506} {\bibfield  {journal} {\bibinfo
  {journal} {Phys. Rev.}\ }\textbf {\bibinfo {volume} {D88}},\ \bibinfo {pages}
  {083506} (\bibinfo {year} {2013})},\ \Eprint {http://arxiv.org/abs/1307.1129}
  {arXiv:1307.1129} \BibitemShut {NoStop}%
%%CITATION = ARXIV:1307.1129;%%
\bibitem [{\citenamefont {Archidiacono}\ \emph {et~al.}(2015)\citenamefont
  {Archidiacono}, \citenamefont {Hannestad}, \citenamefont {Hansen},\ and\
  \citenamefont {Tram}}]{Archidiacono:2014nda}%
  \BibitemOpen
  \bibfield  {author} {\bibinfo {author} {\bibfnamefont {M.}~\bibnamefont
  {Archidiacono}}, \bibinfo {author} {\bibfnamefont {S.}~\bibnamefont
  {Hannestad}}, \bibinfo {author} {\bibfnamefont {R.~S.}\ \bibnamefont
  {Hansen}}, \ and\ \bibinfo {author} {\bibfnamefont {T.}~\bibnamefont
  {Tram}},\ }\href {\doibase 10.1103/PhysRevD.91.065021} {\bibfield  {journal}
  {\bibinfo  {journal} {Phys. Rev.}\ }\textbf {\bibinfo {volume} {D91}},\
  \bibinfo {pages} {065021} (\bibinfo {year} {2015})},\ \Eprint
  {http://arxiv.org/abs/1404.5915} {arXiv:1404.5915 [astro-ph.CO]} \BibitemShut
  {NoStop}%
%%CITATION = ARXIV:1404.5915;%%
\bibitem [{\citenamefont {Bedaque}\ \emph {et~al.}(2009)\citenamefont
  {Bedaque}, \citenamefont {Buchoff},\ and\ \citenamefont
  {Mishra}}]{Bedaque:2009ri}%
  \BibitemOpen
  \bibfield  {author} {\bibinfo {author} {\bibfnamefont {P.~F.}\ \bibnamefont
  {Bedaque}}, \bibinfo {author} {\bibfnamefont {M.~I.}\ \bibnamefont
  {Buchoff}}, \ and\ \bibinfo {author} {\bibfnamefont {R.~K.}\ \bibnamefont
  {Mishra}},\ }\href {\doibase 10.1088/1126-6708/2009/11/046} {\bibfield
  {journal} {\bibinfo  {journal} {JHEP}\ }\textbf {\bibinfo {volume} {11}},\
  \bibinfo {pages} {046} (\bibinfo {year} {2009})},\ \Eprint
  {http://arxiv.org/abs/0907.0235} {arXiv:0907.0235 [hep-ph]} \BibitemShut
  {NoStop}%
%%CITATION = ARXIV:0907.0235;%%
\bibitem [{\citenamefont {Dolan}\ \emph {et~al.}(2015)\citenamefont {Dolan},
  \citenamefont {Kahlhoefer}, \citenamefont {McCabe},\ and\ \citenamefont
  {Schmidt-Hoberg}}]{Dolan:2014ska}%
  \BibitemOpen
  \bibfield  {author} {\bibinfo {author} {\bibfnamefont {M.~J.}\ \bibnamefont
  {Dolan}}, \bibinfo {author} {\bibfnamefont {F.}~\bibnamefont {Kahlhoefer}},
  \bibinfo {author} {\bibfnamefont {C.}~\bibnamefont {McCabe}}, \ and\ \bibinfo
  {author} {\bibfnamefont {K.}~\bibnamefont {Schmidt-Hoberg}},\ }\href
  {\doibase 10.1007/JHEP07(2015)103, 10.1007/JHEP03(2015)171} {\bibfield
  {journal} {\bibinfo  {journal} {JHEP}\ }\textbf {\bibinfo {volume} {03}},\
  \bibinfo {pages} {171} (\bibinfo {year} {2015})},\ \bibinfo {note} {erratum
  ibid.\ \textbf{07}, 103 (2015)},\ \Eprint {http://arxiv.org/abs/1412.5174}
  {arXiv:1412.5174 [hep-ph]} \BibitemShut {NoStop}%
%%CITATION = ARXIV:1412.5174;%%
\bibitem [{\citenamefont {Feng}\ \emph
  {et~al.}(2010{\natexlab{a}})\citenamefont {Feng}, \citenamefont
  {Kaplinghat},\ and\ \citenamefont {Yu}}]{Feng:2009hw}%
  \BibitemOpen
  \bibfield  {author} {\bibinfo {author} {\bibfnamefont {J.~L.}\ \bibnamefont
  {Feng}}, \bibinfo {author} {\bibfnamefont {M.}~\bibnamefont {Kaplinghat}}, \
  and\ \bibinfo {author} {\bibfnamefont {H.-B.}\ \bibnamefont {Yu}},\ }\href
  {\doibase 10.1103/PhysRevLett.104.151301} {\bibfield  {journal} {\bibinfo
  {journal} {Phys. Rev. Lett.}\ }\textbf {\bibinfo {volume} {104}},\ \bibinfo
  {pages} {151301} (\bibinfo {year} {2010}{\natexlab{a}})},\ \Eprint
  {http://arxiv.org/abs/0911.0422} {arXiv:0911.0422 [hep-ph]} \BibitemShut
  {NoStop}%
%%CITATION = ARXIV:0911.0422;%%
\bibitem [{\citenamefont {Wandelt}\ \emph {et~al.}(2000)\citenamefont
  {Wandelt}, \citenamefont {Dave}, \citenamefont {Farrar}, \citenamefont
  {McGuire}, \citenamefont {Spergel},\ and\ \citenamefont
  {Steinhardt}}]{Wandelt:2000ad}%
  \BibitemOpen
  \bibfield  {author} {\bibinfo {author} {\bibfnamefont {B.~D.}\ \bibnamefont
  {Wandelt}}, \bibinfo {author} {\bibfnamefont {R.}~\bibnamefont {Dave}},
  \bibinfo {author} {\bibfnamefont {G.~R.}\ \bibnamefont {Farrar}}, \bibinfo
  {author} {\bibfnamefont {P.~C.}\ \bibnamefont {McGuire}}, \bibinfo {author}
  {\bibfnamefont {D.~N.}\ \bibnamefont {Spergel}}, \ and\ \bibinfo {author}
  {\bibfnamefont {P.~J.}\ \bibnamefont {Steinhardt}},\ }in\ \href
  {http://www.slac.stanford.edu/spires/find/books/www?cl=QB461:I57:2000} {\emph
  {\bibinfo {booktitle} {{Sources and detection of dark matter and dark energy
  in the universe. Proceedings, 4th International Symposium, DM 2000, Marina
  del Rey, USA, February 23-25, 2000}}}}\ (\bibinfo {year} {2000})\ pp.\
  \bibinfo {pages} {263--274},\ \Eprint {http://arxiv.org/abs/astro-ph/0006344}
  {arXiv:astro-ph/0006344} \BibitemShut {NoStop}%
%%CITATION = ASTRO-PH/0006344;%%
\bibitem [{\citenamefont {Spergel}\ and\ \citenamefont
  {Steinhardt}(2000)}]{Spergel:1999mh}%
  \BibitemOpen
  \bibfield  {author} {\bibinfo {author} {\bibfnamefont {D.~N.}\ \bibnamefont
  {Spergel}}\ and\ \bibinfo {author} {\bibfnamefont {P.~J.}\ \bibnamefont
  {Steinhardt}},\ }\href {\doibase 10.1103/PhysRevLett.84.3760} {\bibfield
  {journal} {\bibinfo  {journal} {Phys. Rev. Lett.}\ }\textbf {\bibinfo
  {volume} {84}},\ \bibinfo {pages} {3760} (\bibinfo {year} {2000})},\ \Eprint
  {http://arxiv.org/abs/astro-ph/9909386} {arXiv:astro-ph/9909386} \BibitemShut
  {NoStop}%
%%CITATION = ASTRO-PH/9909386;%%
\bibitem [{\citenamefont {Rocha}\ \emph {et~al.}(2013)\citenamefont {Rocha},
  \citenamefont {Peter}, \citenamefont {Bullock}, \citenamefont {Kaplinghat},
  \citenamefont {Garrison-Kimmel}, \citenamefont {Onorbe},\ and\ \citenamefont
  {Moustakas}}]{Rocha:2012jg}%
  \BibitemOpen
  \bibfield  {author} {\bibinfo {author} {\bibfnamefont {M.}~\bibnamefont
  {Rocha}}, \bibinfo {author} {\bibfnamefont {A.~H.~G.}\ \bibnamefont {Peter}},
  \bibinfo {author} {\bibfnamefont {J.~S.}\ \bibnamefont {Bullock}}, \bibinfo
  {author} {\bibfnamefont {M.}~\bibnamefont {Kaplinghat}}, \bibinfo {author}
  {\bibfnamefont {S.}~\bibnamefont {Garrison-Kimmel}}, \bibinfo {author}
  {\bibfnamefont {J.}~\bibnamefont {Onorbe}}, \ and\ \bibinfo {author}
  {\bibfnamefont {L.~A.}\ \bibnamefont {Moustakas}},\ }\href {\doibase
  10.1093/mnras/sts514} {\bibfield  {journal} {\bibinfo  {journal} {Mon. Not.
  Roy. Astron. Soc.}\ }\textbf {\bibinfo {volume} {430}},\ \bibinfo {pages}
  {81} (\bibinfo {year} {2013})},\ \Eprint {http://arxiv.org/abs/1208.3025}
  {arXiv:1208.3025 [astro-ph.CO]} \BibitemShut {NoStop}%
%%CITATION = ARXIV:1208.3025;%%
\bibitem [{\citenamefont {Zavala}\ \emph {et~al.}(2013)\citenamefont {Zavala},
  \citenamefont {Vogelsberger},\ and\ \citenamefont {Walker}}]{Zavala:2012us}%
  \BibitemOpen
  \bibfield  {author} {\bibinfo {author} {\bibfnamefont {J.}~\bibnamefont
  {Zavala}}, \bibinfo {author} {\bibfnamefont {M.}~\bibnamefont
  {Vogelsberger}}, \ and\ \bibinfo {author} {\bibfnamefont {M.~G.}\
  \bibnamefont {Walker}},\ }\href {\doibase 10.1093/mnrasl/sls053} {\bibfield
  {journal} {\bibinfo  {journal} {Mon. Not. Roy. Astron. Soc.}\ }\textbf
  {\bibinfo {volume} {431}},\ \bibinfo {pages} {L20} (\bibinfo {year}
  {2013})},\ \Eprint {http://arxiv.org/abs/1211.6426} {arXiv:1211.6426
  [astro-ph.CO]} \BibitemShut {NoStop}%
%%CITATION = ARXIV:1211.6426;%%
\bibitem [{\citenamefont {Kaplinghat}\ \emph {et~al.}(2016)\citenamefont
  {Kaplinghat}, \citenamefont {Tulin},\ and\ \citenamefont
  {Yu}}]{Kaplinghat:2015aga}%
  \BibitemOpen
  \bibfield  {author} {\bibinfo {author} {\bibfnamefont {M.}~\bibnamefont
  {Kaplinghat}}, \bibinfo {author} {\bibfnamefont {S.}~\bibnamefont {Tulin}}, \
  and\ \bibinfo {author} {\bibfnamefont {H.-B.}\ \bibnamefont {Yu}},\ }\href
  {\doibase 10.1103/PhysRevLett.116.041302} {\bibfield  {journal} {\bibinfo
  {journal} {Phys. Rev. Lett.}\ }\textbf {\bibinfo {volume} {116}},\ \bibinfo
  {pages} {041302} (\bibinfo {year} {2016})},\ \Eprint
  {http://arxiv.org/abs/1508.03339} {arXiv:1508.03339 [astro-ph.CO]}
  \BibitemShut {NoStop}%
%%CITATION = ARXIV:1508.03339;%%
\bibitem [{\citenamefont {Markevitch}\ \emph {et~al.}(2004)\citenamefont
  {Markevitch}, \citenamefont {Gonzalez}, \citenamefont {Clowe}, \citenamefont
  {Vikhlinin}, \citenamefont {David}, \citenamefont {Forman}, \citenamefont
  {Jones}, \citenamefont {Murray},\ and\ \citenamefont
  {Tucker}}]{Markevitch:2003at}%
  \BibitemOpen
  \bibfield  {author} {\bibinfo {author} {\bibfnamefont {M.}~\bibnamefont
  {Markevitch}}, \bibinfo {author} {\bibfnamefont {A.~H.}\ \bibnamefont
  {Gonzalez}}, \bibinfo {author} {\bibfnamefont {D.}~\bibnamefont {Clowe}},
  \bibinfo {author} {\bibfnamefont {A.}~\bibnamefont {Vikhlinin}}, \bibinfo
  {author} {\bibfnamefont {L.}~\bibnamefont {David}}, \bibinfo {author}
  {\bibfnamefont {W.}~\bibnamefont {Forman}}, \bibinfo {author} {\bibfnamefont
  {C.}~\bibnamefont {Jones}}, \bibinfo {author} {\bibfnamefont
  {S.}~\bibnamefont {Murray}}, \ and\ \bibinfo {author} {\bibfnamefont
  {W.}~\bibnamefont {Tucker}},\ }\href {\doibase 10.1086/383178} {\bibfield
  {journal} {\bibinfo  {journal} {Astrophys. J.}\ }\textbf {\bibinfo {volume}
  {606}},\ \bibinfo {pages} {819} (\bibinfo {year} {2004})},\ \Eprint
  {http://arxiv.org/abs/astro-ph/0309303} {arXiv:astro-ph/0309303} \BibitemShut
  {NoStop}%
%%CITATION = ASTRO-PH/0309303;%%
\bibitem [{\citenamefont {Randall}\ \emph {et~al.}(2008)\citenamefont
  {Randall}, \citenamefont {Markevitch}, \citenamefont {Clowe}, \citenamefont
  {Gonzalez},\ and\ \citenamefont {Bradac}}]{Randall:2007ph}%
  \BibitemOpen
  \bibfield  {author} {\bibinfo {author} {\bibfnamefont {S.~W.}\ \bibnamefont
  {Randall}}, \bibinfo {author} {\bibfnamefont {M.}~\bibnamefont {Markevitch}},
  \bibinfo {author} {\bibfnamefont {D.}~\bibnamefont {Clowe}}, \bibinfo
  {author} {\bibfnamefont {A.~H.}\ \bibnamefont {Gonzalez}}, \ and\ \bibinfo
  {author} {\bibfnamefont {M.}~\bibnamefont {Bradac}},\ }\href {\doibase
  10.1086/587859} {\bibfield  {journal} {\bibinfo  {journal} {Astrophys. J.}\
  }\textbf {\bibinfo {volume} {679}},\ \bibinfo {pages} {1173} (\bibinfo {year}
  {2008})},\ \Eprint {http://arxiv.org/abs/0704.0261} {arXiv:0704.0261
  [astro-ph]} \BibitemShut {NoStop}%
%%CITATION = ARXIV:0704.0261;%%
\bibitem [{\citenamefont {Buckley}\ and\ \citenamefont
  {Fox}(2010)}]{Buckley:2009in}%
  \BibitemOpen
  \bibfield  {author} {\bibinfo {author} {\bibfnamefont {M.~R.}\ \bibnamefont
  {Buckley}}\ and\ \bibinfo {author} {\bibfnamefont {P.~J.}\ \bibnamefont
  {Fox}},\ }\href {\doibase 10.1103/PhysRevD.81.083522} {\bibfield  {journal}
  {\bibinfo  {journal} {Phys. Rev.}\ }\textbf {\bibinfo {volume} {D81}},\
  \bibinfo {pages} {083522} (\bibinfo {year} {2010})},\ \Eprint
  {http://arxiv.org/abs/0911.3898} {arXiv:0911.3898 [hep-ph]} \BibitemShut
  {NoStop}%
%%CITATION = ARXIV:0911.3898;%%
\bibitem [{\citenamefont {Loeb}\ and\ \citenamefont
  {Weiner}(2011)}]{Loeb:2010gj}%
  \BibitemOpen
  \bibfield  {author} {\bibinfo {author} {\bibfnamefont {A.}~\bibnamefont
  {Loeb}}\ and\ \bibinfo {author} {\bibfnamefont {N.}~\bibnamefont {Weiner}},\
  }\href {\doibase 10.1103/PhysRevLett.106.171302} {\bibfield  {journal}
  {\bibinfo  {journal} {Phys. Rev. Lett.}\ }\textbf {\bibinfo {volume} {106}},\
  \bibinfo {pages} {171302} (\bibinfo {year} {2011})},\ \Eprint
  {http://arxiv.org/abs/1011.6374} {arXiv:1011.6374 [astro-ph.CO]} \BibitemShut
  {NoStop}%
%%CITATION = ARXIV:1011.6374;%%
\bibitem [{\citenamefont {Vogelsberger}\ \emph {et~al.}(2012)\citenamefont
  {Vogelsberger}, \citenamefont {Zavala},\ and\ \citenamefont
  {Loeb}}]{Vogelsberger:2012ku}%
  \BibitemOpen
  \bibfield  {author} {\bibinfo {author} {\bibfnamefont {M.}~\bibnamefont
  {Vogelsberger}}, \bibinfo {author} {\bibfnamefont {J.}~\bibnamefont
  {Zavala}}, \ and\ \bibinfo {author} {\bibfnamefont {A.}~\bibnamefont
  {Loeb}},\ }\href {\doibase 10.1111/j.1365-2966.2012.21182.x} {\bibfield
  {journal} {\bibinfo  {journal} {Mon. Not. Roy. Astron. Soc.}\ }\textbf
  {\bibinfo {volume} {423}},\ \bibinfo {pages} {3740} (\bibinfo {year}
  {2012})},\ \Eprint {http://arxiv.org/abs/1201.5892} {arXiv:1201.5892
  [astro-ph.CO]} \BibitemShut {NoStop}%
%%CITATION = ARXIV:1201.5892;%%
\bibitem [{\citenamefont {Kahlhoefer}\ \emph {et~al.}(2015)\citenamefont
  {Kahlhoefer}, \citenamefont {Schmidt-Hoberg}, \citenamefont {Kummer},\ and\
  \citenamefont {Sarkar}}]{Kahlhoefer:2015vua}%
  \BibitemOpen
  \bibfield  {author} {\bibinfo {author} {\bibfnamefont {F.}~\bibnamefont
  {Kahlhoefer}}, \bibinfo {author} {\bibfnamefont {K.}~\bibnamefont
  {Schmidt-Hoberg}}, \bibinfo {author} {\bibfnamefont {J.}~\bibnamefont
  {Kummer}}, \ and\ \bibinfo {author} {\bibfnamefont {S.}~\bibnamefont
  {Sarkar}},\ }\href {\doibase 10.1093/mnrasl/slv088} {\bibfield  {journal}
  {\bibinfo  {journal} {Mon. Not. Roy. Astron. Soc.}\ }\textbf {\bibinfo
  {volume} {452}},\ \bibinfo {pages} {L54} (\bibinfo {year} {2015})},\ \Eprint
  {http://arxiv.org/abs/1504.06576} {arXiv:1504.06576 [astro-ph.CO]}
  \BibitemShut {NoStop}%
%%CITATION = ARXIV:1504.06576;%%
\bibitem [{\citenamefont {Lebedev}\ \emph {et~al.}(2012)\citenamefont
  {Lebedev}, \citenamefont {Lee},\ and\ \citenamefont
  {Mambrini}}]{Lebedev:2011iq}%
  \BibitemOpen
  \bibfield  {author} {\bibinfo {author} {\bibfnamefont {O.}~\bibnamefont
  {Lebedev}}, \bibinfo {author} {\bibfnamefont {H.~M.}\ \bibnamefont {Lee}}, \
  and\ \bibinfo {author} {\bibfnamefont {Y.}~\bibnamefont {Mambrini}},\ }\href
  {\doibase 10.1016/j.physletb.2012.01.029} {\bibfield  {journal} {\bibinfo
  {journal} {Phys. Lett.}\ }\textbf {\bibinfo {volume} {B707}},\ \bibinfo
  {pages} {570} (\bibinfo {year} {2012})},\ \Eprint
  {http://arxiv.org/abs/1111.4482} {arXiv:1111.4482 [hep-ph]} \BibitemShut
  {NoStop}%
%%CITATION = ARXIV:1111.4482;%%
\bibitem [{\citenamefont {Gross}\ \emph {et~al.}(2015)\citenamefont {Gross},
  \citenamefont {Lebedev},\ and\ \citenamefont {Mambrini}}]{Gross:2015cwa}%
  \BibitemOpen
  \bibfield  {author} {\bibinfo {author} {\bibfnamefont {C.}~\bibnamefont
  {Gross}}, \bibinfo {author} {\bibfnamefont {O.}~\bibnamefont {Lebedev}}, \
  and\ \bibinfo {author} {\bibfnamefont {Y.}~\bibnamefont {Mambrini}},\ }\href
  {\doibase 10.1007/JHEP08(2015)158} {\bibfield  {journal} {\bibinfo  {journal}
  {JHEP}\ }\textbf {\bibinfo {volume} {08}},\ \bibinfo {pages} {158} (\bibinfo
  {year} {2015})},\ \Eprint {http://arxiv.org/abs/1505.07480} {arXiv:1505.07480
  [hep-ph]} \BibitemShut {NoStop}%
%%CITATION = ARXIV:1505.07480;%%
\bibitem [{\citenamefont {Ko}\ and\ \citenamefont {Tang}(2016)}]{Ko:2016fcd}%
  \BibitemOpen
  \bibfield  {author} {\bibinfo {author} {\bibfnamefont {P.}~\bibnamefont
  {Ko}}\ and\ \bibinfo {author} {\bibfnamefont {Y.}~\bibnamefont {Tang}},\
  }\href@noop {} {\  (\bibinfo {year} {2016})},\ \Eprint
  {http://arxiv.org/abs/1609.02307} {arXiv:1609.02307 [hep-ph]} \BibitemShut
  {NoStop}%
%%CITATION = ARXIV:1609.02307;%%
\bibitem [{\citenamefont {Arnold}\ and\ \citenamefont
  {Yaffe}(1995)}]{Arnold:1995bh}%
  \BibitemOpen
  \bibfield  {author} {\bibinfo {author} {\bibfnamefont {P.~B.}\ \bibnamefont
  {Arnold}}\ and\ \bibinfo {author} {\bibfnamefont {L.~G.}\ \bibnamefont
  {Yaffe}},\ }\href {\doibase 10.1103/PhysRevD.52.7208} {\bibfield  {journal}
  {\bibinfo  {journal} {Phys. Rev.}\ }\textbf {\bibinfo {volume} {D52}},\
  \bibinfo {pages} {7208} (\bibinfo {year} {1995})},\ \Eprint
  {http://arxiv.org/abs/hep-ph/9508280} {arXiv:hep-ph/9508280} \BibitemShut
  {NoStop}%
%%CITATION = HEP-PH/9508280;%%
\bibitem [{\citenamefont {Beutler}\ \emph {et~al.}(2014)\citenamefont {Beutler}
  \emph {et~al.}}]{Beutler:2014yhv}%
  \BibitemOpen
  \bibfield  {author} {\bibinfo {author} {\bibfnamefont {F.}~\bibnamefont
  {Beutler}} \emph {et~al.} (\bibinfo {collaboration} {BOSS}),\ }\href
  {\doibase 10.1093/mnras/stu1702} {\bibfield  {journal} {\bibinfo  {journal}
  {Mon. Not. Roy. Astron. Soc.}\ }\textbf {\bibinfo {volume} {444}},\ \bibinfo
  {pages} {3501} (\bibinfo {year} {2014})},\ \Eprint
  {http://arxiv.org/abs/1403.4599} {arXiv:1403.4599 [astro-ph.CO]} \BibitemShut
  {NoStop}%
%%CITATION = ARXIV:1403.4599;%%
\bibitem [{\citenamefont {Edsj{\"o}}\ and\ \citenamefont
  {Gondolo}(1997)}]{Edsjo:1997bg}%
  \BibitemOpen
  \bibfield  {author} {\bibinfo {author} {\bibfnamefont {J.}~\bibnamefont
  {Edsj{\"o}}}\ and\ \bibinfo {author} {\bibfnamefont {P.}~\bibnamefont
  {Gondolo}},\ }\href {\doibase 10.1103/PhysRevD.56.1879} {\bibfield  {journal}
  {\bibinfo  {journal} {Phys. Rev.}\ }\textbf {\bibinfo {volume} {D56}},\
  \bibinfo {pages} {1879} (\bibinfo {year} {1997})},\ \Eprint
  {http://arxiv.org/abs/hep-ph/9704361} {arXiv:hep-ph/9704361} \BibitemShut
  {NoStop}%
%%CITATION = HEP-PH/9704361;%%
\bibitem [{\citenamefont {Feng}\ \emph
  {et~al.}(2010{\natexlab{b}})\citenamefont {Feng}, \citenamefont
  {Kaplinghat},\ and\ \citenamefont {Yu}}]{Feng:2010zp}%
  \BibitemOpen
  \bibfield  {author} {\bibinfo {author} {\bibfnamefont {J.~L.}\ \bibnamefont
  {Feng}}, \bibinfo {author} {\bibfnamefont {M.}~\bibnamefont {Kaplinghat}}, \
  and\ \bibinfo {author} {\bibfnamefont {H.-B.}\ \bibnamefont {Yu}},\ }\href
  {\doibase 10.1103/PhysRevD.82.083525} {\bibfield  {journal} {\bibinfo
  {journal} {Phys. Rev.}\ }\textbf {\bibinfo {volume} {D82}},\ \bibinfo {pages}
  {083525} (\bibinfo {year} {2010}{\natexlab{b}})},\ \Eprint
  {http://arxiv.org/abs/1005.4678} {arXiv:1005.4678 [hep-ph]} \BibitemShut
  {NoStop}%
%%CITATION = ARXIV:1005.4678;%%
\bibitem [{\citenamefont {Hahn}\ and\ \citenamefont
  {Perez-Victoria}(1999)}]{Hahn:1998yk}%
  \BibitemOpen
  \bibfield  {author} {\bibinfo {author} {\bibfnamefont {T.}~\bibnamefont
  {Hahn}}\ and\ \bibinfo {author} {\bibfnamefont {M.}~\bibnamefont
  {Perez-Victoria}},\ }\href {\doibase 10.1016/S0010-4655(98)00173-8}
  {\bibfield  {journal} {\bibinfo  {journal} {Comput. Phys. Commun.}\ }\textbf
  {\bibinfo {volume} {118}},\ \bibinfo {pages} {153} (\bibinfo {year}
  {1999})},\ \Eprint {http://arxiv.org/abs/hep-ph/9807565}
  {arXiv:hep-ph/9807565} \BibitemShut {NoStop}%
%%CITATION = HEP-PH/9807565;%%
\bibitem [{\citenamefont {Hisano}\ \emph {et~al.}(2005)\citenamefont {Hisano},
  \citenamefont {Matsumoto}, \citenamefont {Nojiri},\ and\ \citenamefont
  {Saito}}]{Hisano:2004ds}%
  \BibitemOpen
  \bibfield  {author} {\bibinfo {author} {\bibfnamefont {J.}~\bibnamefont
  {Hisano}}, \bibinfo {author} {\bibfnamefont {S.}~\bibnamefont {Matsumoto}},
  \bibinfo {author} {\bibfnamefont {M.~M.}\ \bibnamefont {Nojiri}}, \ and\
  \bibinfo {author} {\bibfnamefont {O.}~\bibnamefont {Saito}},\ }\href
  {\doibase 10.1103/PhysRevD.71.063528} {\bibfield  {journal} {\bibinfo
  {journal} {Phys. Rev.}\ }\textbf {\bibinfo {volume} {D71}},\ \bibinfo {pages}
  {063528} (\bibinfo {year} {2005})},\ \Eprint
  {http://arxiv.org/abs/hep-ph/0412403} {arXiv:hep-ph/0412403} \BibitemShut
  {NoStop}%
%%CITATION = HEP-PH/0412403;%%
\bibitem [{\citenamefont {Iengo}(2009)}]{Iengo:2009ni}%
  \BibitemOpen
  \bibfield  {author} {\bibinfo {author} {\bibfnamefont {R.}~\bibnamefont
  {Iengo}},\ }\href {\doibase 10.1088/1126-6708/2009/05/024} {\bibfield
  {journal} {\bibinfo  {journal} {JHEP}\ }\textbf {\bibinfo {volume} {05}},\
  \bibinfo {pages} {024} (\bibinfo {year} {2009})},\ \Eprint
  {http://arxiv.org/abs/0902.0688} {arXiv:0902.0688 [hep-ph]} \BibitemShut
  {NoStop}%
%%CITATION = ARXIV:0902.0688;%%
\bibitem [{\citenamefont {Arkani-Hamed}\ \emph {et~al.}(2009)\citenamefont
  {Arkani-Hamed}, \citenamefont {Finkbeiner}, \citenamefont {Slatyer},\ and\
  \citenamefont {Weiner}}]{ArkaniHamed:2008qn}%
  \BibitemOpen
  \bibfield  {author} {\bibinfo {author} {\bibfnamefont {N.}~\bibnamefont
  {Arkani-Hamed}}, \bibinfo {author} {\bibfnamefont {D.~P.}\ \bibnamefont
  {Finkbeiner}}, \bibinfo {author} {\bibfnamefont {T.~R.}\ \bibnamefont
  {Slatyer}}, \ and\ \bibinfo {author} {\bibfnamefont {N.}~\bibnamefont
  {Weiner}},\ }\href {\doibase 10.1103/PhysRevD.79.015014} {\bibfield
  {journal} {\bibinfo  {journal} {Phys. Rev.}\ }\textbf {\bibinfo {volume}
  {D79}},\ \bibinfo {pages} {015014} (\bibinfo {year} {2009})},\ \Eprint
  {http://arxiv.org/abs/0810.0713} {arXiv:0810.0713 [hep-ph]} \BibitemShut
  {NoStop}%
%%CITATION = ARXIV:0810.0713;%%
\bibitem [{\citenamefont {Binder}\ \emph {et~al.}(2016)\citenamefont {Binder},
  \citenamefont {Covi}, \citenamefont {Kamada}, \citenamefont {Murayama},
  \citenamefont {Takahashi},\ and\ \citenamefont {Yoshida}}]{Binder:2016pnr}%
  \BibitemOpen
  \bibfield  {author} {\bibinfo {author} {\bibfnamefont {T.}~\bibnamefont
  {Binder}}, \bibinfo {author} {\bibfnamefont {L.}~\bibnamefont {Covi}},
  \bibinfo {author} {\bibfnamefont {A.}~\bibnamefont {Kamada}}, \bibinfo
  {author} {\bibfnamefont {H.}~\bibnamefont {Murayama}}, \bibinfo {author}
  {\bibfnamefont {T.}~\bibnamefont {Takahashi}}, \ and\ \bibinfo {author}
  {\bibfnamefont {N.}~\bibnamefont {Yoshida}},\ }\href {\doibase
  10.1088/1475-7516/2016/11/043} {\bibfield  {journal} {\bibinfo  {journal}
  {JCAP}\ }\textbf {\bibinfo {volume} {1611}},\ \bibinfo {pages} {043}
  (\bibinfo {year} {2016})},\ \Eprint {http://arxiv.org/abs/1602.07624}
  {arXiv:1602.07624 [hep-ph]} \BibitemShut {NoStop}%
%%CITATION = ARXIV:1602.07624;%%
\bibitem [{\citenamefont {Tang}(2016)}]{Tang:2016mot}%
  \BibitemOpen
  \bibfield  {author} {\bibinfo {author} {\bibfnamefont {Y.}~\bibnamefont
  {Tang}},\ }\href {\doibase 10.1016/j.physletb.2016.04.026} {\bibfield
  {journal} {\bibinfo  {journal} {Phys. Lett.}\ }\textbf {\bibinfo {volume}
  {B757}},\ \bibinfo {pages} {387} (\bibinfo {year} {2016})},\ \Eprint
  {http://arxiv.org/abs/1603.00165} {arXiv:1603.00165 [astro-ph.CO]}
  \BibitemShut {NoStop}%
%%CITATION = ARXIV:1603.00165;%%
\bibitem [{\citenamefont {Bringmann}\ and\ \citenamefont
  {Hofmann}(2007)}]{Bringmann:2006mu}%
  \BibitemOpen
  \bibfield  {author} {\bibinfo {author} {\bibfnamefont {T.}~\bibnamefont
  {Bringmann}}\ and\ \bibinfo {author} {\bibfnamefont {S.}~\bibnamefont
  {Hofmann}},\ }\href {\doibase 10.1088/1475-7516/2007/04/016} {\bibfield
  {journal} {\bibinfo  {journal} {JCAP}\ }\textbf {\bibinfo {volume} {0704}},\
  \bibinfo {pages} {016} (\bibinfo {year} {2007})},\ \Eprint
  {http://arxiv.org/abs/hep-ph/0612238} {arXiv:hep-ph/0612238} \BibitemShut
  {NoStop}%
%%CITATION = HEP-PH/0612238;%%
\bibitem [{\citenamefont {Kasahara}(2009)}]{KasaharaPHD}%
  \BibitemOpen
  \bibfield  {author} {\bibinfo {author} {\bibfnamefont {J.}~\bibnamefont
  {Kasahara}},\ }\emph {\bibinfo {title} {Neutralino Dark Matter: The Mass of
  the smallest Halo and the golden Region}},\ \href@noop {} {Ph.D. thesis},\
  \bibinfo  {school} {University of Utah} (\bibinfo {year} {2009})\BibitemShut
  {NoStop}%
\bibitem [{\citenamefont {Gondolo}\ \emph {et~al.}(2012)\citenamefont
  {Gondolo}, \citenamefont {Hisano},\ and\ \citenamefont
  {Kadota}}]{Gondolo:2012vh}%
  \BibitemOpen
  \bibfield  {author} {\bibinfo {author} {\bibfnamefont {P.}~\bibnamefont
  {Gondolo}}, \bibinfo {author} {\bibfnamefont {J.}~\bibnamefont {Hisano}}, \
  and\ \bibinfo {author} {\bibfnamefont {K.}~\bibnamefont {Kadota}},\ }\href
  {\doibase 10.1103/PhysRevD.86.083523} {\bibfield  {journal} {\bibinfo
  {journal} {Phys. Rev.}\ }\textbf {\bibinfo {volume} {D86}},\ \bibinfo {pages}
  {083523} (\bibinfo {year} {2012})},\ \Eprint {http://arxiv.org/abs/1205.1914}
  {arXiv:1205.1914 [hep-ph]} \BibitemShut {NoStop}%
%%CITATION = ARXIV:1205.1914;%%
\bibitem [{\citenamefont {Peskin}\ and\ \citenamefont
  {Schroeder}(1995)}]{Peskin:1995ev}%
  \BibitemOpen
  \bibfield  {author} {\bibinfo {author} {\bibfnamefont {M.~E.}\ \bibnamefont
  {Peskin}}\ and\ \bibinfo {author} {\bibfnamefont {D.~V.}\ \bibnamefont
  {Schroeder}},\ }\href
  {http://www.slac.stanford.edu/spires/find/books/www?cl=QC174.45%3AP4} {\emph
  {\bibinfo {title} {{An Introduction to quantum field theory}}}}\ (\bibinfo
  {year} {1995})\BibitemShut {NoStop}%
%%CITATION = ISBN-9780201503975;%%
\bibitem [{\citenamefont {Profumo}\ \emph {et~al.}(2006)\citenamefont
  {Profumo}, \citenamefont {Sigurdson},\ and\ \citenamefont
  {Kamionkowski}}]{Profumo:2006bv}%
  \BibitemOpen
  \bibfield  {author} {\bibinfo {author} {\bibfnamefont {S.}~\bibnamefont
  {Profumo}}, \bibinfo {author} {\bibfnamefont {K.}~\bibnamefont {Sigurdson}},
  \ and\ \bibinfo {author} {\bibfnamefont {M.}~\bibnamefont {Kamionkowski}},\
  }\href {\doibase 10.1103/PhysRevLett.97.031301} {\bibfield  {journal}
  {\bibinfo  {journal} {Phys. Rev. Lett.}\ }\textbf {\bibinfo {volume} {97}},\
  \bibinfo {pages} {031301} (\bibinfo {year} {2006})},\ \Eprint
  {http://arxiv.org/abs/astro-ph/0603373} {arXiv:astro-ph/0603373} \BibitemShut
  {NoStop}%
%%CITATION = ASTRO-PH/0603373;%%
\bibitem [{\citenamefont {Visinelli}\ and\ \citenamefont
  {Gondolo}(2015)}]{Visinelli:2015eka}%
  \BibitemOpen
  \bibfield  {author} {\bibinfo {author} {\bibfnamefont {L.}~\bibnamefont
  {Visinelli}}\ and\ \bibinfo {author} {\bibfnamefont {P.}~\bibnamefont
  {Gondolo}},\ }\href {\doibase 10.1103/PhysRevD.91.083526} {\bibfield
  {journal} {\bibinfo  {journal} {Phys. Rev.}\ }\textbf {\bibinfo {volume}
  {D91}},\ \bibinfo {pages} {083526} (\bibinfo {year} {2015})},\ \Eprint
  {http://arxiv.org/abs/1501.02233} {arXiv:1501.02233 [astro-ph.CO]}
  \BibitemShut {NoStop}%
%%CITATION = ARXIV:1501.02233;%%
\bibitem [{\citenamefont {Gondolo}\ \emph {et~al.}()\citenamefont {Gondolo},
  \citenamefont {Edsj\"o}, \citenamefont {Ullio}, \citenamefont {Bergstr\"om},
  \citenamefont {Schelke}, \citenamefont {Baltz}, \citenamefont {Bringmann},\
  and\ \citenamefont {Duda}}]{DSweb}%
  \BibitemOpen
  \bibfield  {author} {\bibinfo {author} {\bibfnamefont {P.}~\bibnamefont
  {Gondolo}}, \bibinfo {author} {\bibfnamefont {J.}~\bibnamefont {Edsj\"o}},
  \bibinfo {author} {\bibfnamefont {P.}~\bibnamefont {Ullio}}, \bibinfo
  {author} {\bibfnamefont {L.}~\bibnamefont {Bergstr\"om}}, \bibinfo {author}
  {\bibfnamefont {M.}~\bibnamefont {Schelke}}, \bibinfo {author} {\bibfnamefont
  {E.}~\bibnamefont {Baltz}}, \bibinfo {author} {\bibfnamefont
  {T.}~\bibnamefont {Bringmann}}, \ and\ \bibinfo {author} {\bibfnamefont
  {G.}~\bibnamefont {Duda}},\ }\href@noop {} {\enquote {\bibinfo {title} {{\tt
  DarkSUSY}},}\ }\bibinfo {note}
  {\href{http://www.darksusy.org}{http://www.darksusy.org} {}}\BibitemShut
  {NoStop}%
\end{thebibliography}%

\end{document}